\documentclass[lettersize,journal]{IEEEtran}
\usepackage{verbatim}
\usepackage{graphicx, amsbsy, amsmath, cuted}
 
\usepackage{amssymb}
\usepackage{multirow}
\usepackage{booktabs}
\usepackage{threeparttable}
\usepackage{enumitem}
\usepackage{ragged2e}
\usepackage{subfig}
\usepackage{hyperref}
\usepackage{float}
\usepackage{xcolor, soul}
\usepackage{subcaption}
\usepackage{orcidlink}
\usepackage{flushend}
\usepackage{balance}
\usepackage{framed}

\graphicspath{{./Fig/}}
\colorlet{shadecolor}{yellow!50} 

\makeatletter
\newcommand*\titleheader[1]{\gdef\@titleheader{#1}}
\AtBeginDocument{%
  \let\st@red@title\@title
  \def\@title{%
    \bgroup\normalfont\large\centering\@titleheader\par\egroup
    \vskip0.5em\st@red@title}
}
\makeatother

\title{Regional Weather Variable Predictions by Machine Learning with Near-Surface Observational and Atmospheric Numerical Data\textsuperscript{*}}
\titleheader{Accepted for publication in IEEE Transactions on Geoscience and Remote Sensing (TGRS)}

\author{
    Yihe Zhang\orcidlink{0000-0003-3540-8172}$^{1\dagger}$,
    Bryce Turney\orcidlink{0009-0004-8422-1931}$^{1\dagger}$,
    Purushottam Sigdel\orcidlink{0000-0002-4410-793X}$^{2}$,
    Xu Yuan\orcidlink{0000-0003-3775-3033}$^{3}$,
    Eric Rappin\orcidlink{0000-0002-8966-7076}$^{4}$,
    Adrian L. Lago\orcidlink{0009-0002-5072-6814}$^{5}$,
    Sytske Kimball\orcidlink{0000-0002-8467-1017}$^{6}$,
    Li Chen\orcidlink{0000-0002-2300-6996}$^{1}$, 
    Paul Darby\orcidlink{0000-0001-8395-0711}$^{1}$,
    Lu Peng\orcidlink{0000-0003-3545-286X}$^{7}$,
    Sercan Aygun\orcidlink{0000-0002-4615-7914}$^{1}$,
    Yazhou Tu\orcidlink{0000-0001-7640-1829}$^{8}$,
    M. Hassan Najafi\orcidlink{0000-0002-4655-6229}$^{1}$,
    and
    Nian-Feng Tzeng\orcidlink{0000-0002-8357-6632}$^{1}$

    \thanks{
      $^{1}$University of Louisiana at Lafayette,
      $^{2}$Intel Corp.,
      $^{3}$University of Delaware,
      $^{4}$Western Kentucky University,
      $^{5}$ National Weather Office, Paducah, Kentucky, 
      $^{6}$University of South Alabama,
      $^{7}$Tulane University,
      and $^{8}$Auburn University

      \vspace{0.5em}
      \noindent $^\dagger$ Authors with major contributions equally.
      
      \vspace{0.5em}
      \noindent $^*$ A preliminary version of this manuscript was presented at 2021 European Conference on Machine Learning and Principles and Practice of Knowledge Discovery (ECML-PKDD), September 2021~\cite{original_mima55}.
    }
}

\begin{document}
  
\maketitle

\begin{abstract}
Accurate and timely regional weather prediction is vital for sectors dependent on weather-related decisions. Traditional prediction methods, based on atmospheric equations, often struggle with coarse temporal resolutions and inaccuracies. This paper presents a novel machine learning (ML) model, called MiMa (short for \underline{Mi}cro-\underline{Ma}cro), that integrates both near-surface observational data from Kentucky Mesonet stations (collected every five minutes, known as Micro data) and hourly atmospheric numerical outputs (termed as Macro data) for fine-resolution weather forecasting. The MiMa model employs an encoder-decoder transformer structure, with two encoders for processing multivariate data from both datasets and a decoder for forecasting weather variables over short time horizons. Each instance of the MiMa model, called a modelet, predicts the values of a specific weather parameter at an individual Mesonet station. The approach is extended with Re-MiMa modelets, which are designed to predict weather variables at ungauged locations by training on multivariate data from a few representative stations in a region, tagged with their elevations. Re-MiMa (short for \underline{Re}gional-MiMa) can provide highly accurate predictions across an entire region, even in areas without observational stations. Experimental results show that MiMa significantly outperforms current models, with Re-MiMa offering precise short-term forecasts for ungauged locations, marking a significant advancement in weather forecasting accuracy and applicability.
\end{abstract}

	\section{Introduction}
\label{sec:Intro}

Accurate short-term weather predictions with fine temporal resolutions are crucial for sectors that depend on real-time weather-related decision-making, such as transportation, emergency response, solar farm operations, etc. However, current forecasting models, such as the Weather Research and Forecasting (WRF) model with High-Resolution Rapid Refresh (HRRR) \cite{wrf}, fall short of meeting these demands due to their coarse hourly outputs and high computational complexity. These models generate around 148 weather parameter values (i.e., variables) per hour over large geo-grids (e.g., 3 km × 3 km), with coarse temporal granularity (hourly forecasts) often deemed insufficient for applications requiring predictions in the interval of 5 or 15 minutes \cite{earth}. Additionally, a lack of near-surface observational data at tactical locations limits their accuracy.

\begin{figure}
	\centering
    \includegraphics[width=8.8cm, keepaspectratio]{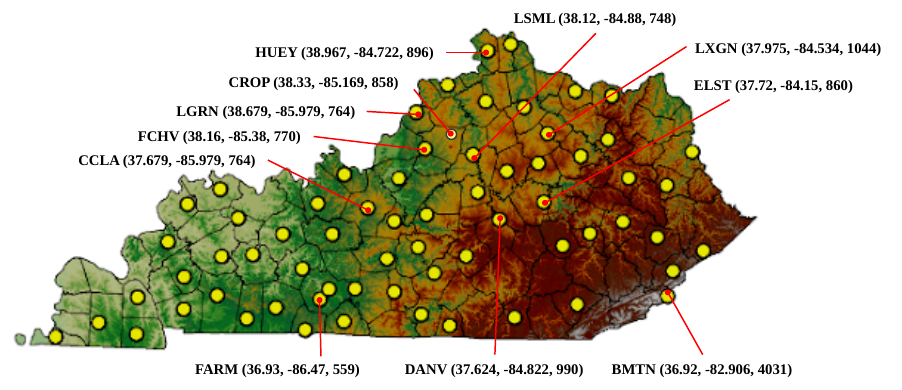}
	\caption{\small  Kentucky Mesonet weather observational stations denoted by yellow circles, with those stations chosen for MiMa model evaluation and pointed by red line segments tagged with their latitudes, longitudes, and elevations.}
	\label{fig:KentuckyMesonet}
\end{figure}

On the other hand, regional Mesonet networks, such as the Kentucky Mesonet~\cite{KAmesonet}, provide real-time, location-specific weather data with fine temporal granularity (e.g., every five minutes). These networks operate under the U.S. National Mesonet Program~\cite{mesonet_org} and consist of strategically located observational stations. The Kentucky Mesonet, for example, comprises over 70 stations for collecting values of some 22 weather parameters, including temperature, humidity, wind speed, pressure, and precipitation. This high-resolution dataset provides valuable micro-level data that can be leveraged to improve prediction accuracy (see Fig.~\ref{fig:KentuckyMesonet} for station distribution).

\begin{figure*}
	\centering
    \includegraphics[width=18cm, height=7cm]{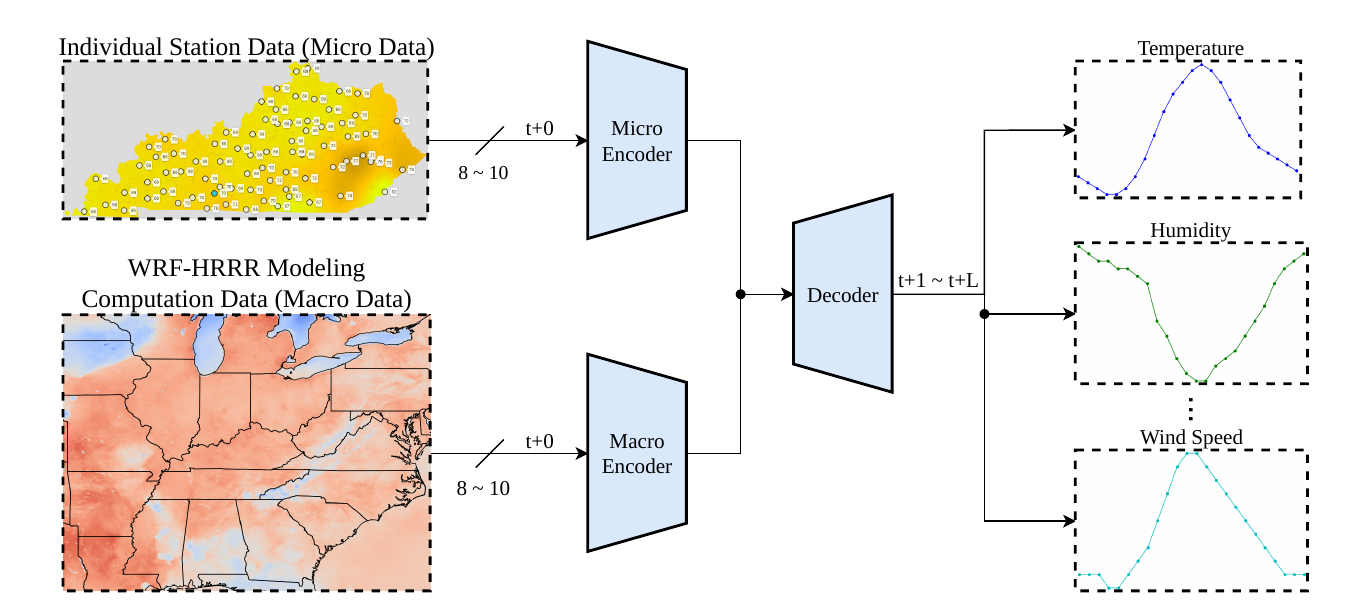}
	\caption{\small  Overview of the MiMa (short for Micro-Macro) model inputted with data from both an individual station and WRF-HRRR modeling computation to yield the weather variable predictions.}
	\label{fig:intro_mima}
\end{figure*}

Recent advances in machine learning (ML) technology have propelled weather forecasting into a new era. Numerous studies have explored ML-centric techniques for weather forecasting, yielding promising results. These techniques include the deep neural networks (DNN), convolutional neural networks (CNN), long short-term memory networks (LSTM)~\cite{hochreiter97:NC:long}, generative adversarial networks (GAN)~\cite{goodfellow14:NIPS:generative}, among others, for predicting values of such parameters as wind speed and direction~\cite{dalto15:ICIT:deep, hu16:RE:transfer, alemany19:AAAI:predicting, pandit20::data}, solar radiation~\cite{gensler16:SMC:solor, arshi19:IJCNN:weather}, precipitation~\cite{xingjian15::convolutional, hernandez16:HAIS:rainfall, pan2019:WRR:improving, kim22:AR:near, zhang21:JH:evaluation}, air quality~\cite{yi18:KDD:deep}, weather changes~\cite{grover15:KDD:5deep, weyn20::improving, yang21:JH:large, yang20:JWRPM:simulating}, etc. However, existing ML forecasting models have not yet achieved accurate predictions of weather variables at fine temporal resolutions (as seen with recent transformer-variant forecasters like \cite{autoformer}, \cite{fedformer}, \cite{interpretable_weather_forecasting}, \cite{fourcastnet}). This shortcoming is largely due to the lack of geo-location-aligned ground observational data, which is essential for enhancing model training and prediction accuracy. Many models rely solely on satellite or radar images~\cite{ravuri21:Nature:skillful}, limiting their precision. Additionally, most neural network-based models are designed to predict values of specific parameters, lacking a generalizable framework that can be readily adapted to forecast all parameters of interest.

To address these shortcomings, this paper introduces the MiMa model, a novel ML-based approach that integrates fine-grained observational data (Micro data) from regional Mesonet stations with larger-scale numerical outputs (macro data) from the WRF-HRRR model~\cite{earth}. This integration enables accurate weather predictions over short time horizons (in minutes) for a region of interest. The MiMa model employs an encoder-decoder transformer architecture, where two encoders process multivariate sequences from the micro and the macro datasets, and a decoder forecasts the values of multiple weather parameters across a sequence of time points from $t+1$ to $t+L$ (see Fig.~\ref{fig:intro_mima}).

Each instance of the MiMa model is referred to as a modelet, dedicated to predicting a specific weather variable for an individual Mesonet station. To further enhance its utility, the MiMa model is extended to become Regional MiMa (Re-MiMa) modelets, which can predict weather variables at ungauged locations (where no observational stations exist). By using data from several representative stations in a region of interest, typically 3 or 4, tagged with their elevations, Re-MiMa modelets can generalize predictions across an entire region, providing accurate forecasts at both gauged and ungauged locations. This extension addresses a long-standing challenge in meteorological forecasting: accurate predictions at locations without direct observational data.

We conducted experiments on weather forecasting at various Kentucky Mesonet stations, focusing on four key weather parameters: air temperature, relative humidity, wind speed, and atmospheric pressure, across eleven stations (Figure \ref{fig:KentuckyMesonet}). The MiMa modelets consistently outperform all comparative techniques, achieving the lowest RMSE (Root Mean Squared Error) values in 39 out of 44 forecasting instances (Table~\ref{model_errors}. Additionally, Re-MiMa modelets, trained using data from three representative stations in eastern Kentucky (BMTN, FARM, and DANV), were tested at eight ungauged stations. The results show that Re-MiMa modelets predicted weather variables with accuracy comparable to, or better than, location-specific MiMa modelets, with 22 out of 32 parameters at those ungauged locations predicted more accurately (see Table~\ref{Re-MiMa_errors}). This demonstrates the effectiveness of Re-MiMa in avoiding the need for multiple modelets for predicting a given variable while maintaining high accuracy across the entire region.

In summary, the MiMa model integrates micro and macro data to deliver precise weather predictions at fine temporal resolutions. The Re-MiMa extension further enhances its regional forecasting capability, making it a versatile tool for applications that require high-accuracy predictions in real-time. The MiMa model code, documentation, and datasets are made publicly available at~\cite{MiMaModelets} for further research and applications to other regions.

\vspace{1em}

\noindent This paper makes several key contributions:
\begin{itemize}
    \item \textbf{Novel Weather Prediction Model (MiMa)}. We introduce the MiMa model, a machine learning framework designed to predict weather parameters at fine temporal resolutions accurately. The model integrates high-frequency observational data (Micro data) with geo-aligned atmospheric numerical outputs (macro data) to provide accurate short-term weather forecasts.
    \item \textbf{Adaptable Prediction for Arbitrary Lead Times}. The MiMa model employs an encoder-decoder architecture with LSTM units, allowing it to handle arbitrary lead times and forecast weather variables with fine temporal granularity, such as 5- or 15-minute intervals, meeting real-world demands for high-resolution weather forecasts.
    \item \textbf{Regional MiMa (Re-MiMa)}. We extend the MiMa model by introducing Re-MiMa modelets, which enable accurate weather forecasting at ungauged locations (where no observational stations exist). Re-MiMa uses observational data from a small number of representative stations, avoiding the need for location-specific models while maintaining high prediction accuracy across a region.
    \item \textbf{Reduction in Modelet Count}. By utilizing transfer learning and data from representative stations, Re-MiMa reduces the number of required modelets, achieving accurate regional forecasts using one single modelet per weather variable, compared to traditional approaches requiring individual modelets for each location.
    \item \textbf{Comprehensive Evaluation}. Our experimental evaluation across multiple Kentucky Mesonet stations demonstrates that MiMa and Re-MiMa significantly outperform their counterparts in forecasting accuracy for multiple weather parameters, including temperature, humidity, wind speed, and pressure. Re-MiMa models achieve high accuracy at ungauged locations, further validating their effectiveness and boosting their usability.
    \item \textbf{Data and Code Availability}. To facilitate further research and reproducibility, we have made the MiMa model code, documentation, and datasets available at~\cite{MiMaModelets}, enabling other researchers to apply our approach to different regions and weather forecasting tasks.
\end{itemize}

	\section {Related Work}

Abundant applications of ML techniques for weather forecasting exist.  
This section reviews the recent advances in such applications, which mostly follow three lines of work.

\subsection{Neural Networks for Simulating Atmospheric Systems}
The first line aims to explore whether NNs can simulate the physical principles of atmosphere systems.  In particular, a Global NN and a Local NN are employed in~\cite{dueben18:GMD:challenges} to simulate the dynamics of a simple global atmosphere model at $500$ hPa geopotential. The results conclude that prediction outcomes by the NN models can be better than those of the coarse-resolution atmosphere models over a short duration with the 1-hour time scale.  Scher~\cite{scher18::toward} applied the CNN structure with an AutoEncoder to learn the simplified general circulation models (GCMs), which can predict the weather variables for up to 14 days. The CNN incorporating LSTM components was leveraged in~\cite{weyn19::can} to achieve  14-day lead time forecasting as well. Vlachas {\em et al.}~\cite{vlachas18::data} employed the LSTM model to reduce the order space of a chaotic system. However, known solutions along this line of work all focused on developing prediction models for simulated or simplified climate environments, without taking into account real-world conditions like observed weather parameters at a region of interest. Their applicability and effectiveness in real environments are questionable, given the complex real-world conditions in practice. For example, the actual measurements from Mesonet stations are highly dependent on local conditions. In addition, their solutions cannot make accurate fine-grained forecasts (e.g., in the 5- or 15-minute resolution) over short horizons (for one or two hours) flexibly.

\subsection{Neural Networks for Real-World Weather Prediction}
The second line of work pursues new NN models for the real-world weather parameters prediction. For example, the LSTM and fully connected NNs are leveraged in~\cite{pandit20::data} to predict the wind speed at an offshore site, by capturing its rapidly changing features. Grover {\em et al.}~\cite{grover15:KDD:5deep} combined the discriminatively trained predictive models with a DNN (Deep NN) to predict the atmospheric pressure, temperature, wind speed, and dew point. A convolutional LSTM model was adopted in~\cite{xingjian15::convolutional} to predict precipitation, whereas the CNN with a stack of delicately selected frames was employed in~\cite{pan2019:WRR:improving} for precipitation forecasting. In addition, a model with the AutoEncoder structure was proposed to predict rain-fall~\cite{hernandez16:HAIS:rainfall}. Forecasting the hurricane trajectories via a Recurrent NN structure was considered in~\cite{alemany19:AAAI:predicting}. The LSTM structures were employed in \cite{gensler16:SMC:solor} and \cite{arshi19:IJCNN:weather} to predict solar radiation and photovoltaic energy, respectively. \cite{yi18:KDD:deep} proposed a deep fusion network to predict air quality. \cite{veillette20:NIPS:sevir} crafted a storm event imagery dataset while leveraging the VGG16 model to analyze storm events. U-net models~\cite{ronneberger15:ICMICC:u-net} were considered in~\cite{pan21:GRL:improving, li21:GMD:msdm} for fine-grained radar nowcasting. A deep CNN model was developed over a cubed sphere~\cite{weyn20::improving} for predicting several basic atmospheric variables on a global grid. In \cite{kumar21:KDD:micro}, the DeepMC model with attention mechanisms was proposed to predict micro climate. A near real-time hurricane rainfall forecasting model was proposed in~\cite{kim22:AR:near}, where a basic CNN model inputted with the integrated IMERG dataset was leveraged. In~\cite{yang20:JWRPM:simulating},  basic machine learning and data mining algorithms were developed for forecasting the reservoir release. Recently, a lightweight model inputted with satellite and radar images for real-world storm prediction has been treated~\cite{transformer-based_storm}. Meanwhile, a nonlinear nowcasting model under a neural network framework has been proposed for precipitation forecasting based on composite radar observations to exhibit more accurate and instructive outcomes than other deep-learning methods \cite{nowcastnet}.

\subsection{Transformer-based Models for Long-term Weather Forecasting}
As the third line of work, various transformer-based solutions have been developed for long time-series predictions, including weather forecasting but in coarse resolutions. Specifically, Autoformer \cite{autoformer} performs weather forecasting in the daily resolution, whereas FEDformer \cite{fedformer} evaluates the predictions of weather time-series data for the hourly resolution. Likewise, the Corrformer model \cite{interpretable_weather_forecasting} forecasts weather conditions over a large number of stations in coarse granularity temporally, FourCastNet \cite{fourcastnet} considers weather predictions at the temporal resolution of 6 hours, and the iTransformer model \cite{liu2024itransformerinvertedtransformerseffective} makes long-term weather forecasting in the range of 96 hours to 720 hours at the hourly resolution.  Lately, general circulation models for weather and climate by combining atmospheric physics with machine learning have aimed at the daily (or longer) resolution coarsely over medium range (1-14 days) time horizons~\cite{neural_generation_circulation}, which are also the target of the recently published global weather forecasting benchmark~\cite{wang2019deep57}.

\subsection{Station Forecasting}
The fourth line of work includes weather forecasting via (1) data from a large number of weather stations \cite{han2024weather}, (2) station downscaling \cite{liu2024deriving}, (3) a physical-ML hybrid model \cite{li2024deepphysinet}, (4) data from dense and sparse sensors \cite{deeplearning_day_forecasts}, and (5) GNN-based method \cite{GNN_predictions}. Specifically, multiple critical weather variables are forecast using a comprehensive collection of over 5,000 weather stations, called the Weather-5k dataset \cite{han2024weather}. Leveraging the dataset like Weather-5k, recent pursuits improve weather predictions at specific station locations through advanced downscaling techniques. For instance, station-scale downscaling \cite{liu2024deriving} accurately derives meteorological conditions at station locations from coarse-resolution meteorological fields. Meanwhile, the hybrid DeepPhysiNet approach \cite{li2024deepphysinet} integrates physical laws into deep learning models to enhance prediction accuracy. Another study \cite{deeplearning_day_forecasts} utilizes both dense and sparse sensor data to make predictions for lead times up to 24 hours, with its focus on extending the lead time from 12 to 24 hours. However, the study relies solely on ground-based sensor data without valuable computational numerical data, such as WRF-HRRR, which would enhance accuracy. A GNN-based method \cite{GNN_predictions} has also been proposed for downscaling global grids to off-grid locations of interest. However, like the previous study, it does not leverage computational numerical data, such as WRF-HRRR, to improve predictions for these off-grid locations. While all aforementioned solutions help to advance weather prediction, they are not meant to predict weather parameters accurately in fine temporal granularity (in minutes) over flexible time horizons and lead times, hence calling for accurate weather forecasting with fine-grained temporal resolutions. 

\subsection{MiMa and Re-MiMa: Fine-Grained Weather Prediction}
With encoder-decoder transformer-variant structures, our MiMa and Re-MiMa modelets, for the first time, achieve accurate predictions with short time horizons in fine temporal resolutions on all weather variables at locations in a target region, realized by (1) tailoring a modelet for one variable prediction per location (or per region) under MiMa (or Re-MiMa), and (2) taking both near-surface observational and atmospheric numerical multi-variate data as their inputs, and (3) letting modelets' encoders input suitable data (including predicted outcomes) for encoding adaptively.

	\section{Pertinent Background}

This section first explains near-surface observations conducted by Mesonet stations \cite{KAmesonet}, followed by describing the WRF-HRRR (Weather Research and Forecasting with High-Resolution Rapid Refresh) computational model \cite{hrrr}. The limitations of applying such datasets for weather forecasting are then stated.

\vspace{1em}

\noindent{\bf \em Kentucky Mesonet}.  Under the U.S. National Mesonet Program, this Mesonet comprises a set of automated weather stations (towers) located at specific locations in the State of Kentucky, as marked by yellow circles in Fig.~\ref{fig:KentuckyMesonet}. Its towers aim to gather real-time meteorological and soil measurements relevant to local weather phenomena, involving tens of meteorological measurements, such as air temperature, relative humidity, wind speed, atmospheric pressure, and precipitation, among others, periodically \cite{KAmesonet}. 
Meteorological measurements are gathered once in five minutes, whereas soil measurements are taken once in 15 or 30 minutes. 

\vspace{1em}

\begin{table*}[htbp]
    \caption{\small RMSE (Root Mean Squared Error) values of WRF-HRRR outputs versus Mesonet observations over three months at each of those 11 Kentucky Mesonet stations marked in Fig.~\ref{fig:KentuckyMesonet}, with TEMP, HUMI, WSPD, and PRES denoting air temperature, relative humidity, wind speed, and atmospheric pressure, respectively}
    \centering
    \Huge
    \renewcommand{\arraystretch}{1.5} 
    \resizebox{\textwidth}{!}{
        \begin{tabular}{l|lllllllllll}
            \toprule
            \hspace{2em} & \textbf{BMTN} \hspace{2em} & \textbf{CCLA} \hspace{2em} & \textbf{CROP} \hspace{1.5em} & \textbf{DANV} \hspace{1.5em} & \textbf{ELST} \hspace{1.5em} & \textbf{FARM} \hspace{1.5em} & \textbf{FCHV} \hspace{2em} & \textbf{HUEY} \hspace{2em} & \textbf{LGRN }\hspace{1.5em} & \textbf{LSML} \hspace{1.5em} & \textbf{LXGN} \hspace{1.5em} \\ \hline
            TEMP & 1.79 & 1.78 & 1.65 & 1.36 & 1.46 & 1.18 & 1.63 & 1.00 & 1.41 & 1.49 & 1.12 \\
            HUMI & 13.95 & 12.80 & 10.49 & 9.07 & 10.91 & 6.45 & 10.88 & 7.54 & 9.41 & 9.71 & 5.16 \\
            WSPD & 4.05 & 2.97 & 2.77 & 2.79 & 2.69 & 2.82 & 2.78 & 3.45 & 2.44 & 2.90 & 3.83 \\
            PRES & 24.12 & 0.30 & 1.45 & 0.38 & 0.51 & 0.78 & 0.32 & 0.50 & 6.81 & 2.52 & 1.09 \\ 
            \bottomrule
        \end{tabular}
    }
    \vspace{0.5em}
    \label{wrf_var}
\end{table*}

\noindent{\bf \em WRF with HRRR modeling}. The WRF model takes actual atmospheric conditions (mainly from satellite, ground radar imagery,  METAR, SYNOP, Sonde, etc) as the input of atmosphere physical equations to calculate numerical outputs that serve a wide range of meteorological applications across the nation. The WRF-HRRR model is the ARW core~\cite{skamarock19:ARW:description} simulation results of the WRF model~\cite{WRFARW} initialized by the HRRR assimilating system~\cite{hrrr}. It takes multiple sources as inputs, including radar reflectivity and observations~\cite{benjamin16:MWR:north} related to rawinsonde, boundary layer, cloud, precipitation processes, etc. It computes up to $148$ weather parameters over the 18-hour time horizon in hourly increments with the spatial resolution of $3$-km and across $50$ vertical levels. In this work, we take the HRRR assimilated results archived in the University of Utah for public use, and those results cover the whole United States continent with a total of $1059 \times 1799$  geo-grids sized 3km $\times$ 3km~\cite{wrf}. On July 12, 2018, the HRRR implementation Version 2 was upgraded to Version 3, with some changes to parameters; please refer to~\cite{HRRR_impl} for more details. Every parameter selected for our evaluation exists in both versions. To obtain the WRF-HRRR data that are geo-aligned with ground observational stations (in a mesonet) for MiMa modelet training, each involved hourly WRF-HRRR data file (sized  $~120 MB$) has to be preprocessed, given that those hourly parameter values of all geo-grids over the US continent are compressed to one single file for efficient transfer and storage. Preprocessing an hourly data file takes about 2 minutes by one Dell server in our lab and modelet training needs WRF-HRRR computed parameters held in thousands of such files, deemed a rather time-consuming task, as detailed under ``WRF-HRRR Data Preprocessing" in Section~\ref{subsec:setting}. 
\vspace{1em}

\noindent{\bf \em Limitations.} Both Mesonet and WRF-HRRR datasets have their respective limitations. Specifically, the mesonet dataset contains near-surface weather measurements gathered continuously by stations with various sensors and devices in minutes. However, it does not provide forecasting results and involves only tens of observation parameters. It can serve as the ground truth for ML model training but is unable to reveal future weather parameter values by itself. The WRF-HRRR numerical outputs cover the whole US at hourly granularity, but they usually suffer from considerable inaccuracy at geo-grids of interest. For example, Table~\ref{wrf_var} presents the RMSE (Root Mean Squared Error) values of WRF-HRRR outputs over Mesonet observations in three months of 2018 at eleven stations. Besides, its hourly prediction scale limits its suitability for meteorological applications that require high temporal resolutions (in minutes) in support of real-time decision-making. With affluent weather parameters (i.e., 148 or 192), the WRF-HRRR data can be inputted into our prediction modelets for complementing Mesonet observational data.

Our developed prediction modelets take multivariate time series systematically chosen from both datasets as their inputs to complement each other for accurate prediction in fine temporal resolutions. As such, WRF-HRRR can provide affluent weather condition information while Mesonet stations gather accurate ground observations. Utilizing both of them (in our developed modelets) properly enables precise weather forecasting in fine-grained temporal resolutions.

\section {ML-based Models for Weather Forecasting}
\vspace{0.25em}

The MiMa meteorological model utilizes Meteo modelets to accurately and concurrently predict weather variables with fine temporal resolution. These modelets are fed with minute-level near-surface observational data (the micro dataset) and hourly atmospheric numerical outputs from WRF-HRRR modeling (the macro dataset). Each Meteo modelet is specifically designed to predict a single weather parameter at a location where observational data is available. For each predicted parameter, two subsets of input parameters are selected: one from the micro dataset and the other from the macro dataset. These subsets are chosen based on their relevance levels with respect to the weather parameter being predicted, as detailed in Section ~\ref{subsec:setting}. The primary objective of the modelet design is to extract temporal variation features from relevant sequences of previous measurements to accurately predict weather parameter values at multiple consecutive future time points (say, in T minutes, 2T minutes, 3T minutes, etc.). This is achieved by leveraging advanced ML techniques to learn temporal sequence patterns from both datasets, capturing weather-situational variations essential for predicting specific parameters. The Meteo modelets deliver precise weather forecasts for a target region at desirable temporal resolutions. Before detailing the configuration of the MiMa modelets, we first describe a micro model that relies solely on the micro dataset for its predictions, as below.

\subsection {Micro Model}

\begin{figure}
    \centering
    \includegraphics[width=8.8cm, keepaspectratio]{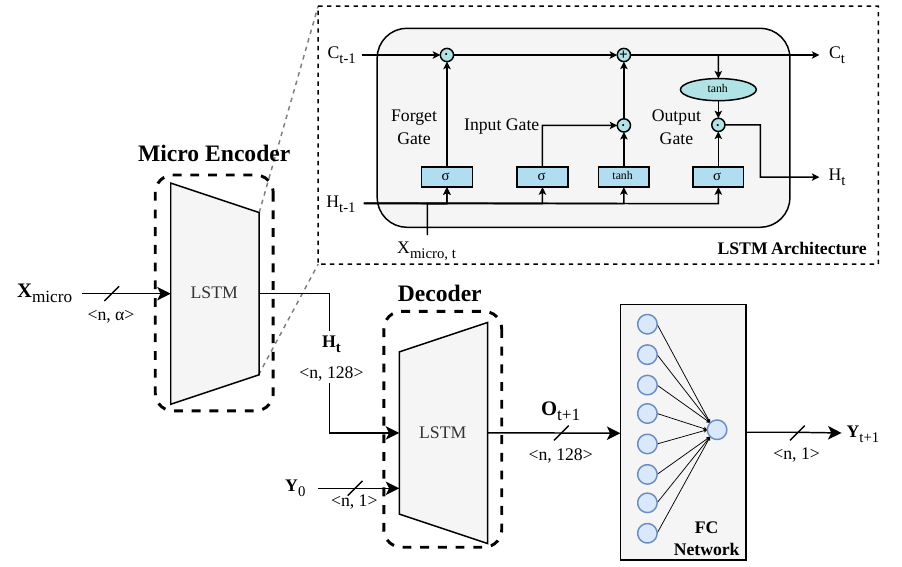}
    \vspace{-2em}
    \caption{\small
        Structure of the Micro model, with the hidden state $\textbf{H}_{t}$ obtained by inputting $\textbf{X}_{\text{micro}}$ to an encoder and the output $\textbf{O}_{t+1}$ obtained by inputting $\textbf{H}_{t}$ plus $\textbf{Y}_{0}$ to a decoder. Output $\textbf{O}_{t+1}$ is then passed to a fully connected layer which generates the predicted parameter value $\textbf{Y}_{t+1}$ via a fully connected network.}
    \label{fig:MiED}
\end{figure}

Most atmospheric data exhibit noticeable temporal sequences and periodic patterns, with weather conditions continuously changing over time. To capture these patterns for forecasting in consecutive future time points, an Encoder-Decoder structure with the Long Short-Term Memory (LSTM) network \cite{hochreiter97:NC:long} as its building block to capture the temporal and periodic patterns, as depicted in Fig. \ref{fig:MiED}. Although the encoder-decoder LSTM model has been widely applied to sequence tasks such as language translation and question answering, the physical meaning of each element in the input vectors is not well-explored. Hence, the encoder's LSTM is detailed next, enabling it to keep rich element-wise features when encoding all features into a dense vector. 

\vspace{1em}

{\noindent \bf \em Micro Encoder.} \
The Micro Encoder consists of an LSTM unit designed to encode appropriate multivariate time series of data over a specific period into a single dense vector, representing their temporal feature variations. The input to this encoder, $\bf{X}_{micro}$, is a matrix of the most relevant parameter values at each timestamp, defined as follows:
\[
\mathbf{X}_{\text{micro}} = \begin{pmatrix}
    P_{1}^{1} & P_{2}^{1} & P_{3}^{1} & \cdots & P_{\alpha}^{1} \\[0.3em]
    P_{1}^{2} & P_{2}^{2} & P_{3}^{2} & \cdots & P_{\alpha}^{2} \\[0.3em]
    \vdots & \vdots & \vdots & \ddots & \vdots \\[0.3em]
    P_{1}^{n} & P_{2}^{n} & P_{3}^{n} & \cdots & P_{\alpha}^{n}
\end{pmatrix},
\]
where the $i^{th}$ row, for $1 \leq i \leq n$, represents those $\alpha$ most relevant parameters chosen from the micro dataset at the $i^{th}$ timestep of the multivariate time series data observed at a station to train the model for the station's location. These $n$ time steps constitute the \textit{lookback window} for predicting the results of future time points over a horizon, where the time gap between the lookback window and the prediction horizon is known as the lead time. The past $T\times n$-minute surface observational data points from the $\alpha$ time series gathered by the station are taken as a data frame ($\mathbf{X}_{\text{micro}}$) representing an observed weather snapshot as the model input. In the ablation study (Section \ref{subsec:ablation}), results under different lookback windows are provided and discussed. After inputting the data of $\textbf X_{\text{micro}}$ to the Micro Encoder, a hidden state vector of size $n \times 128$ is obtained. During the training, the exact hidden state vector size is $128 \times 128$, under the mini-batch size of 64. Concatenation of the hidden and cell states lead to the resulting hidden state vector size of $128 \times 128$. The LSTM unit learns key features and updates its associated hidden state vector. This vector, along with the next data frame, is input to the same LSTM unit to update the hidden state vector $\bf{H}_t$, expressed by:
\begin{equation}
     {\bf H}_{t} = \sigma(\bf{X}_{micro}W_{xo} + \bf{H}_{t-1}\bf{W}_{ho} + \bf{b}_o) \times \tanh{(\bf{C}_t)},
     \label{eqn:Micro Hidden State}
\end{equation}
where $\bf{W}_{xo}$ are the output weights for the input $\bf{X}_{micro}$, $\bf{H}_{t-1}$ is the hidden state vector from the previous timestamp, $\bf{W}_{ho}$ are the output weights for the hidden state, $\bf{b}_o$ is the bias for the output, $\bf{C}_t$ is the cell state of the LSTM unit, and $\sigma$ denotes the sigmoid activation function and tanh is the hyperbolic tangent~\cite{hochreiter97:NC:long}. Initially, the hidden state vector without a prior state, is initialized randomly. The final dense vector $\bf{H}_t$ aggregates temporal pattern variations from the inputs $\textbf X_{\text{micro}}$'s of $n$ timestamps.

\vspace{1em}

{\noindent \bf \em Decoder.}
The Decoder predicts specific weather parameter values for consecutive time points over the given horizon after a lead time, if any. Including an LSTM nuit, the Decoder, initialized by the dense vector $\bf{H}_{t}$, also takes the starting value of the sequence, $\bf{Y}_0$, as its input to generate the output vector $\bf{O}_{t+1}$, denoted as:
\begin{equation}
    \bf{O}_{t+1} = \sigma({\bf{Y}_oW_{xo} + \bf{H}_t\bf{W}_{ho} + \bf{b}_o}),
    \label{eqn:Micro Output Vector}
\end{equation}
where the weight and bias variables are similar to those given in Eq. (\ref{eqn:Micro Hidden State}).
This output vector is passed to a fully connected network to obtain the forecast value of $\bf{Y}_{t+1}$, expressed by:
\begin{equation}
    \bf{Y}_{t+1} = \sigma({\bf{O}_{t+1}\bf{W}_{out} + b_{out}}),
    \label{eqn:Micro Pred Vector}
\end{equation}
where $\bf{W}_{out}$ are the weights of the fully connected network, and $\bf{b}_{out}$ is the bias.

During training, multivariate data sequences from a station are inputted in batches, with each training pass learned from appropriate sequences of $n$ values (e.g., 12 values for 1 hour of data at 5-minute intervals). For temperature prediction, as an instance, the modelet is inputted with the last hour's worth of most relevant parameter data in batches to predict the temperature 5 minutes immediately after, when the lead time is nil. During inference, the model generates a sequence of $n$ values by the decoder, one at a time iteratively. For improved accuracy, the encoder inputs suitable data (including predicted values) for encoding before $n$ iterations end.

\subsection {MiMa Model}

\begin{figure}
    \includegraphics[width=1.03\columnwidth, keepaspectratio]{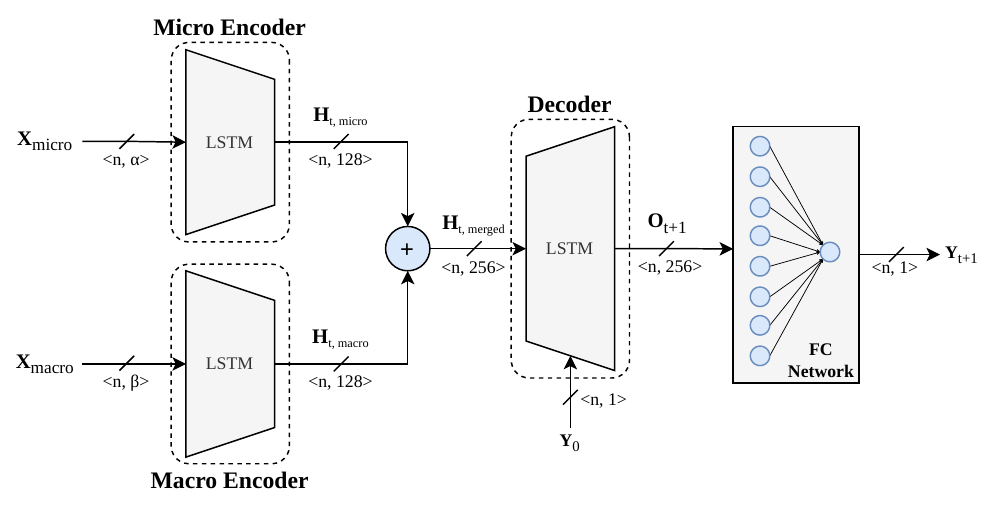}
    \vspace{-2em}
    \caption{\small Structure of MiMa model, with its Micro Encoder and its Decoder identical to those depicted in Fig.~\ref{fig:MiED} and with the hidden states of the Micro and the Macro Encoders concatenated as the Decoder's input.}
    \label{fig:MiMaED}
\end{figure}

Given that the number of weather parameters observed by Mesonet stations is limited and primarily indicates current near-surface readings without forward-looking information, forecasting based solely on the micro dataset is insufficient. The WRF-HRRR computed outputs (the Macro dataset) include atmospheric indicators at higher altitudes (e.g., 700/925 hPa geopotential height, low cloud cover, 3000m storm-relative helicity, etc.), as listed in Table \ref{parameterdescription}. They are useful for inferring future weather conditions near the surface. Thus, incorporating appropriately selected WRF-HRRR outputs into the model training process significantly enhances prediction accuracy, arriving at the MiMa model. Our proposed MiMa model takes the macro dataset as a complementary input to improve forecasting.

The macro dataset is generated on an hourly basis \cite{hrrr}, whereas surface observational data are collected every five minutes by Kentucky Mesonet stations. The MiMa model is obtained by adding a Macro Encoder to the Micro model (depicted in Fig. \ref{fig:MiED}) to integrate these two data sources with different temporal scales. Comprising a single LSTM unit, the Macro Encoder takes as its input, the geo-aligned WRF-HRRR data which are most relevant to the weather parameter under prediction (e.g., air temperature, relative humidity, etc.; see Table \ref{parameters}). The structure of our MiMa model is shown in Fig. \ref{fig:MiMaED}.

The subset of WRF-HRRR data chosen for a predicted weather parameter is based on parameter relevance degrees, as detailed in Section V-A. Since the Macro dataset and the Micro dataset are on different time scales, inputs to the Macro Encoder must be temporally downscaled from one hour to $T$-minutes (being the Mesonet station data sampling interval). Each hour is divided into $60/T$ time frames, using the hourly output from the Macro dataset to represent the first time frame's value. Values for the remaining time frames are computed using a polynomial function fitted to the last $l$ output parameter data points, with $l=3$. The polynomial $a * x^2 + b * x + c$ is used, fitting the immediate last three hourly WRF-HRRR computed values to find the best coefficients $a$, $b$, and $c$ for extrapolating the future $60/T - 1$ values at $T$-minutes intervals. This process is applied to every computed parameter listed in Table V, with their respective time frame populated according to the polynomial function.

\begin{table}
    \centering
    \caption{\small Predicted weather parameters of interest}
    \renewcommand{\arraystretch}{1.5} 
    \scalebox{0.88}{
        \normalsize
        \begin{tabular}{l|r|r}
            \hline
            \textbf{\normalsize Predicted Parameter} & \textbf{\normalsize Height}  & \textbf{\normalsize Reading Range} \\ \hline
            TEMP (Air Temperature)  &  2.0 m &  -40 to 60$^{\circ}$C \\
            HUMI (Relative Humidity)  &  2.0 m &  0 to 100 \% \\
            WSPD (Wind Speed)  &  10.0 m &  0 to 100 m/s \\
            PRES (Atmospheric Pressure) &  1.0 m &  600 to 1060 mb \\ \hline
        \end{tabular}
    }
    \label{parameters}
    \vspace{1em}
\end{table}

The most relevant parameters from the Micro dataset (or the Macro dataset) for each predicted parameter are listed in Table \ref{microparameter} (or Table \ref{WRFparameter}). In addition to the Micro encoder input $\bf{X}_{micro}$, the Macro encoder input $\bf{X}_{macro}$ is given by
\[
\mathbf{X}_{\text{macro}} = \begin{pmatrix}
    P_{1}^{1} & P_{2}^{1} & P_{3}^{1} & \cdots & P_{\beta}^{1} \\[0.3em]
    P_{1}^{2} & P_{2}^{2} & P_{3}^{2} & \cdots & P_{\beta}^{2} \\[0.3em]
    \vdots & \vdots & \vdots & \ddots & \vdots \\[0.3em]
    P_{1}^{n} & P_{2}^{n} & P_{3}^{n} & \cdots & P_{\beta}^{n}
\end{pmatrix},
\]
where the $i^{th}$ row, for $1 \leq i \leq n$, denotes those $\beta$ most relevant parameters from the WRF-HRRR dataset, for the lookback window of $n$ steps and the prediction horizon of $T \times n$ minutes when the time step equals $T$ minutes. Comprising an LSTM unit, the Macro Encoder takes its input $\textbf X_{\text{macro}}$ along with the hidden state vector from the previous time frame, to update its hidden state vector. It outputs a dense vector, $\textbf H_{t,\text{macro}}$, which is concatenated with the dense vector outputted from the Micro Encoder, $\textbf H_{t,\text{micro}}$, (as shown in Fig. \ref{fig:MiMaED}) to produce the vector of $\bf{H}_{t,\text{merged}}$:
\begin{equation}
    \bf{H}_{t,merged} = \text{concat}(\bf{H}_{t,micro}, \bf{H}_{t,macro}).
\end{equation}

The decoder in the MiMa model functions similarly to that in the Micro model, and its output is expressed by Eq.~(\ref{eqn:Micro Output Vector}), with ${\bf H}_{t}$ replaced by ${\bf H}_{t,merged}$.  
It is initialized by the concatenated dense vector $\bf{H}_{t,merged}$ to start forecasting for consecutive time points sequentially. During both training and prediction phases, the MiMa model utilizes the WRF-HRRR numerical data from the geo-grid where the observational station resides (known as spatial alignment) over the same time duration (known as temporal alignments). Note that while our MiMa model is encoder-decoder structured, its encoders are made to consider prediction outcomes adaptively, based on prediction errors observed at model validation immediately after training. This way allows the MiMa modelets to encode input data frames over the look-back window more frequently to lower their prediction error during inference, at the expense of longer inference times (in a few seconds, rather than tens of $\mu$s with just the decoder involved in predictions).

Note that our proposed MiMa model, built on the LSTM-based encoder-decoder architecture, exhibits very high accuracy in predicting weather variables over the short time horizons of our interest (up to a few hours, as demonstrated in the next section). Its high accuracy results mainly from inputting geo-aligned Micro and Macro data at the same time. When the prediction horizons are long (say, tens of hours), different model structures, like transformers with attention \cite{attention_need}, may be called for. A model built on the transformer with attention, however, typically requires a large amount of data for model training to have quality models with high prediction accuracy. For our weather forecasting, we employ small datasets (over two years), making it unsuitable to adopt any transformer-based model with attention.

	\section {Experiments and Results}

This section provides evaluation specifics, performance results and discussion, ensemble predictions, ablation studies, and extreme weather forecasting. Evaluation specifics include dataset details, parameter-relevant degree calculation, WRF-HRRR data preprocessing, and experiment setup details. Performance results are shown (1) for different prediction methods under prediction horizons of 1 hour and 24 hours and (2) for MiMa modelets with the prediction lead times of 1 hour and 4 hours for the 3-hour horizon.

\subsection{Evaluation Specifics}
\label{subsec:setting}

\noindent{\bf \em Dataset Details.} \ Two types of datasets are inputted to the Meteo Modelets we developed for performance evaluation, including the near-surface observational data gathered by Kentucky Mesonet~\cite{KAmesonet} and the WRF-HRRR~\cite{earth} atmospheric numerical values, called respectively Micro and Macro datasets because the former (or latter) data are available in 5-minute (hourly) temporal granularity. The Micro data comprise a set of weather parameters gathered by Mesonet stations for monitoring real-time meteorological phenomena, as shown in Fig.~\ref{fig:KentuckyMesonet}, where stations are signified by yellow circles.  
The monitored weather parameters include the readings of air temperature, relative humidity, wind speed, and atmospheric pressure,  among others, at various heights (see Table~\ref{parameters}), recorded once every 5 minutes, as opposed to the WRF-HRRR computed atmospheric values available hourly. Eleven Kentucky Mesonet stations are selected for evaluating MiMa models, with their geographical locations denoted in Fig.~\ref{fig:KentuckyMesonet}  by red line segments. For example, BMTN $(39.919, -82.906, 4031)$ is located at latitude: 36.919$^{\circ}$, longitude: -82.906$^{\circ}$, and elevation: 4031 feet, and it is in the Black Mountain. 
Those mesonet stations are scattered in the Eastern Kentucky area (with complex terrain), and their elevations range from 559 feet to 4031 feet.  

Four meteorological parameters of interest considered at each Mesonet station for model performance evaluation are listed in Table~\ref{parameters}, with their respective measuring heights and reading ranges included. The parameters of WSPD and PRES are measured respectively at $10 m$ and $1.0 m$, whereas the remaining two are measured at $2.0 m$. In the case of the individually trained MiMa modelets, there are 44 ($= 4 \times 11$) MiMa modelets involved. Each modelet is trained by inputting both the ground observational data gathered during, and WRF-HRRR atmospheric data computed for, Years 2018 and 2019, while tested via the data of Year 2020. The WRF-HRRR macro data employed are those corresponding to the $3$-km by $3$-km geo-grids of the eleven Kentucky Mesonet stations. There are 148 WRF-HRRR parameters computed for each geo-grid per hour, but only a few of them are relevant enough to a given parameter under prediction (say, TEMP) for consideration in its dedicated MiMa modelet. Similarly, each mesonet station gathers some 22 weather parameters periodically (mostly once per five minutes), with a few of them strongly relevant to the predicted parameter. The following describes a systematic way for identifying the relevant degrees of all parameters with respect to a predicted parameter under one dataset so that such identified relevant degrees permit each MiMa modelet to include a suitable set of strongly relevant parameters for superior prediction performance after training.

\begin{table}
    \centering
    \caption{\small Relevant parameters under the Micro dataset}
    \renewcommand{\arraystretch}{1.5} 
    \scalebox{0.8}{
    \normalsize
    \begin{tabular}{c|p{8.5cm}} 
        \hline
        \textbf{\normalsize Predicted} & \textbf{\normalsize Relevant parameters for prediction use$^{\textbf \S}$}                    \\ \hline
        \multirow{2}{*}{\normalsize TEMP} & TEMP, THMP, TDPT, ST02, SM02, SRAD, PRES, WDSD, HUMI, WSSD \\
        \multirow{2}{*}{\normalsize HUMI} & HUMI, SRAD, WSSD, WDSD, WSMX, WSPD, WSMN, THMP, PRES, TEMP \\
        \multirow{2}{*}{\normalsize WSPD} & WSPD, WSMX, WSMN, WSSD, PRES, ST02, HUMI, SM02             \\
        {\normalsize PRES} & PRES, TDPT, TEMP, THMP, ST02, WSMN, WSPD                    \\ \hline
    \end{tabular}
    }
    \begin{tablenotes}
        \footnotesize
        \item $^{\S}$ Description of relevant parameters is as follows: THMP -- 2-meter moisture sensor temperature (C); TDPT -- 2-meter dewpoint temperature (C); ST02 -- 2-cm soil temperature (C); SM02 -- 2-cm soil moisture (\%); SRAD -- 2-meter downwelling shortwave radiation ($W/m^2$); WSSD -- 10-meter wind speed standard deviation (within 5-minute window); WSMX -- 10-meter wind speed maximum ($m/s$) in the 5-minute interval; WSMN -- 10-meter wind speed minimum ($m/s$) in the 5-minute interval. 
    \end{tablenotes}
\label{microparameter}
\end{table}

\vspace{1em}

\noindent{\bf \em Parameter Relevance.} Weather parameter prediction by ML belongs to high-dimensional multivariate data analytics, with its performance dictated by involved dimensional features (i.e., parameters in the weather prediction context).  
Although selecting a proper set of parameters is challenging, it is an essential step in data preprocessing for removing irrelevant and redundant data to reduce time complexity and improve learning accuracy for the ML model at hand.  
In essence, this step involves identifying relevant degrees of all parameters with respect to a given parameter, so that the model takes into account only those parameters with high enough relevant degrees when predicting its target parameter. 
Typically, each parameter under prediction involves a different subset of parameters with high enough relevant degrees under a given dataset (be the near-surface gathered or the WRF-HRRR computed one), as listed in Tables~\ref{microparameter} and \ref{WRFparameter}.  
The subset of relevant parameters chosen for consideration in predicting a given parameter, however, is found to be identical at all geo-grids across a regional area.  
Hence, only one subset of relevant parameters is chosen systematically for each predicted parameter under the given dataset, irrespective of geo-grids where mesonet stations reside.  
A description of relevant WRF-HRRR parameters considered by prediction models is given in Table~\ref{parameterdescription}.

\vspace{3em}
\begin{table}
	\centering
	\caption{\small Relevant parameters under the WRF-HRRR dataset. The description of each parameter ID is shown in Table~\ref{parameterdescription}}
	\vspace{-0.5em}
    \renewcommand{\arraystretch}{1.2} 
    \small
	\scalebox{0.75}{
    \normalsize
	\begin{tabular}{l|l}
		\hline
		\textbf{\normalsize Predicted} &\textbf{\normalsize Considered WRF-HRRR param. ID for prediction} \hspace{8em}\\ \hline 
		\normalsize TEMP      & 66, 67, 59, 111, 32, 94, 28, 69, 109, 68    \\ \hline 
		\normalsize HUMI      & 70, 4, 108, 110, 112, 91, 90, 61, 113, 111   \\ \hline
		\normalsize WSPD      &  8, 73, 89, 115, 98, 23, 18, 121, 125 \\ \hline
		\normalsize PRES      & 57, 39, 40, 94, 28, 125, 128  \\ \hline
	\end{tabular}
	}
	\label{WRFparameter}
\end{table}

\begin{table}
	\centering
	\caption{\small Description of relevant WRF-HRRR parameters included in the MiMa modelets at hand}
    \renewcommand{\arraystretch}{1.1} 
	\scalebox{0.85}{
    \normalsize
	\begin{tabular}{l|l}
		\hline
		\textbf{\normalsize ID} \hspace{1em} & \hspace{6em} \textbf{\normalsize Parameter Description}                              \\ \hline
		4   & Surface Visibility (m)                             \\
		8   & Surface Wind speed (gust)  (m/s)                   \\
		18  & 700hpa Geopotential Height (gpm)                   \\
		28  & 925hpa Temperature (K)                             \\
		39  & MSLP (pa)                                          \\
		40  & 1000hpa Geopotential Height (gpm)                  \\
		57  & Surface Pressure (Pa)                              \\
		59  & Surface Temperature (K)                            \\
		61  & Ground Moisture (\%)                               \\
		66  & 2-meter Temperature (K)                            \\
		67  & 2-meter Potential temperature (K)                  \\
		68  & 2-metre specific humidity                          \\
		69  & 2-metre dewpoint temperature (K)                   \\
		70  & 2-metre relative humidity (\%)                     \\
		73  & 10-metre wind speed (m/s)                          \\
		89  & Surface Frictional velocity (m/s)                  \\
		90  & Surface Sensible heat net flux (w/m2)              \\
		91  & Surface Latent heat net flux (w/m2)                \\
		94  & Isobaric layer Surface lifted index (K)            \\
		98  & Low cloud cover (\%)                               \\
		108 & Surface Downward short-wave radiation flux (w/m2)  \\
		109 & Surface Downward long-wave radiation flux (w/m2)   \\
		110 & Surface Upward short-wave radiation flux (w/m2)    \\
		111 & Surface Upward long-wave radiation flux (w/m2)     \\
		112 & surface Visible Beam Downward Solar Flux (w/m2)    \\
		113 & Surface Visible Diffuse Downward Solar Flux (w/m2) \\
		115 & 3000m Storm relative helicity (J/kg)               \\
		121 & Vertical u-component shear (/s)                    \\
		125 & Isothermal Zero Pressure (Pa)                      \\
		128 & Highest Tropospheric Freezing Pressure (Pa)        \\ \hline
	\end{tabular}
	}
	\label{parameterdescription}
	\vspace{-2em}
\end{table}

\vspace{-3em}

Generally, two systematic approaches for choosing appropriate subsets of relevant parameters are wrapper and filter solutions.  
The former exhaustively searches all possible subsets of the parameters (i.e., dimensional features) to optimize the solution for a specific ML model.  
Such an exhaustive search-based method is known to be NP-hard, involving infeasibly high time complexity when the problem size is large (like the WRF-HRRR dataset with 148 parameters).  
In contrast, the latter usually applies a statistical method to determine a suboptimal solution with a feasible time, avoiding exhaustive search.  
An early filter solution based on the statistical method is realized by computing correlation coefficients among parameters~\cite{hall99:FLAIRS:feature}.  
A later statistical solution relied on mutual information present among parameters for determining their relevant degrees~\cite{frenay13:NN:mutual}, and it was shown to better capture relevant degrees of parameters with non-linear relationships in general when compared to its correlation coefficient-based counterpart~\cite{hall99:FLAIRS:feature}.  Since finding an optimal subset of major relevant parameters exhaustively by a wrapper solution is NP-hard, we employed the filter-based approach~\cite{hall99:FLAIRS:feature, frenay13:NN:mutual} to determine a proper parameter subset for each predicted parameter under a given dataset by calculating the relevant degrees (ranging from 1.0 to 0.0) of all parameters with respect to the predicted parameter.  
With similar time complexity, both filter-based solutions (corresponding to~\cite{hall99:FLAIRS:feature} and to \cite{frenay13:NN:mutual}) were adopted to compute the relevant degrees of parameters under the gathered weather dataset and the WRF-HRRR dataset, and their results were found to be identical possibly because parameters in our both datasets may not have a strong non-linear relationship to make the mutual information-based solution~\cite{frenay13:NN:mutual} outshine its correlation coefficient-based counterpart~\cite{hall99:FLAIRS:feature}, as shown previously.

A parameter with a higher degree of relevance tends to improve the accuracy of the predicted parameter when taken into account in the prediction model.  Intuitively, a proper subset of relevant parameters for inclusion in a prediction model should contain those parameters with relevant degrees exceeding a threshold $\Theta$ (say, $\Theta$ = 0.3), because including those lightly relevant parameters indiscriminately not only raises model time complexity but also may hurt accuracy.  To constrain model complexity without compromising its accuracy, a proper relevant parameter subset is limited to $\gamma$ (say, $\gamma = 10$) so that only those $\gamma$ most relevant parameters are included if there are more than $\gamma$ parameters with their relevant degrees $\geq \Theta$. Note that high model accuracy results provided that $\gamma$ is chosen reasonably and both dataset types adopt the same $\Theta$ and $\gamma$ values, as can be found in Tables~\ref{microparameter} and \ref{WRFparameter},   where TEMP (or PRES) involves 10 (or 7) most relevant parameters.  
A very small $\gamma$ value (say, 2) can yield unsatisfactorily low accuracy whereas an unnecessarily large $\gamma$ value (say, $\geq$ 20) incurs excessive time complexity without improving accuracy.  The subsequent results presented are under the choice of $\gamma = 10$.  Relevant WRF-HRRR parameters considered by prediction models are listed and described in Table~\ref{parameterdescription}.

\vspace{1em}

\noindent{\bf \em WRF-HRRR Data Preprocessing}.
The WRF-HRRR dataset~\cite{wrf} was recorded each hour, in a compressed format to contain a computed weather situation snapshot with 148 parameters for each of 1,905,141 ($= 1059 \times 1799$) geo-grids that cover the whole United States. 
An hourly weather snapshot held in the {\em grib2} format has the size of some 120 MB, and it has to be decompressed (into some 2.4 GB) before extracting the relevant parameters of interest at all geo-grids where mesonet stations reside for inclusion in weather prediction models.  
In our implementation, we employ the pygrib (a Python package) to extract the HRRR data. In particular, values of a parameter for all grids constitute one layer so there are 148 layers in total. 
Taking Surface Pressure as an example, it will be at Layer 57 (of 148 layers).
We read out details in Layer 57 from HRRR data, which also include the latitude and longitude of each grid.
Among $1059\times1799$ grids, we compute the distance between each grid and the Mesonet station and identify the one having the nearest distance to this station. 
The Surface Pressure at this grid is considered to be associated with the respective Mesonet station. 
Note that the pressure value corresponding to a station can also be the average of computed WRF-HRRR pressure results of 4 or 8 neighboring grids, but this averaged pressure value is found to be usually close to that of the nearest grid because weather parameter values are usually similar over a small region.

Considered as WRF-HRRR data preprocessing, uncompressing one weather snapshot followed by extracting relevant parameters takes about 2 minutes on a Dell T7910 workstation (with dual Xeon E5-2680v4 CPUs and 64GB memory).  
Hence, such data preprocessing on 24 snapshots (of one day) takes more than 48 minutes.  
For our model evaluation, total WRF-HRRR data preprocessing includes extracting 90 days of data per year (over one season) for three years, with two of them for model training and one for model testing, taking more than 216 hours ($= 2 \times 24 \times 90 \times 3$ minutes), apparently an extremely time-consuming task that calls for high parallelism to shorten its execution.  
The data preprocessing time rises when data from more years and/or from more days per year are employed to train models (for accuracy improvement).

To spatially align WRF-HRRR data and observational data for a station location, our data preprocessing searches each decompressed {\em grib2} file over its involved $1.9+$ million gridpoints for the one nearest to where the station lies.  
Time and space complexities involved in the search for the nearest gridpoint (according to the distance) are $O(n)$ and $O(1)$, respectively, where $n$ equals the number of gridpoints in a file.  
After such a gridpoint is identified for every station of interest in a region, all relevant parameters recorded under this gridpoint are extracted.  
Since the data extraction process is time-consuming and our workstation has 28 cores, 27 program instances can be launched concurrently, each handling 10 days' WRF-HRRR data files (one per hour) for high parallelism. 
Overall data preprocessing conducted in this way takes about eight hours ($= 48$ minutes $\times 10$) after the compressed WRF-HRRR data files had been downloaded from~\cite{earth} (for local access during preprocessing).

\vspace{1em}

\noindent{\bf \em Spatial Alignment of Micro and Macro Datasets.} Before training the MiMa modelets, we address the issue of spatial alignment between the two datasets of differing resolutions: the micro dataset, consisting of individual station point data, and the WRF-HRRR dataset, which provides gridded satellite data covering a larger geographical area. Since the WRF dataset represents broader spatial coverage while the micro dataset focuses on specific station locations, it is crucial to ensure proper alignment between the two to facilitate accurate comparisons and analyses.

To achieve this, we identify the latitude and longitude from the WRF dataset that is closest to each station's coordinates in the micro dataset. This is done by computing the distance between the station's latitude and longitude and the corresponding grid points in the WRF dataset. We then select the grid point with the minimum distance to the station as the representative location for extracting the WRF data. By doing so, we effectively convert the gridded satellite data into point data, ensuring that both the micro and macro datasets are spatially aligned. This allows for consistent comparison and integration of data from both sources, despite their differences in spatial resolution.

\vspace{1em}

\noindent{\bf \em Important Relationships.} In our work, we explore the important relationships between data assimilation, temporal downscaling, and weather prediction since each of these represents an important aspect of our work. Data assimilation involves integrating near-surface observational data (ground data) with atmospheric numerical data from the Weather Research and Forecasting (WRF) model. This fusion of data sources enhances the forecasting ability of our model by providing a multi-dimensional input that captures both ground-based and atmospheric information. Through this assimilation process, we align observational data with WRF’s outputs, resulting in a more comprehensive dataset that supports location-specific predictions and enriches the overall quality of the forecast input.

Temporal downscaling is applied to the WRF data to transform its hourly computational modeling into finer-grained 5-minute intervals. This process is essential for accurately capturing short-term weather fluctuations and aligning WRF’s coarser resolution with the 5-minute sampling rate of mesonet station data. By refining the temporal granularity of the WRF data, we prepare a synchronized dataset that meets the temporal requirements of our predictive model, ensuring consistency and precision in subsequent forecasts.

The assimilated and downscaled data then feed into our weather prediction model, where the combination of these two processes—data assimilation and temporal downscaling—enables the model to deliver high-precision weather predictions at fine temporal intervals. This integration ensures that real-time, location-specific weather conditions are accurately captured and forecasted, addressing both the spatial and temporal demands of our forecasting framework.

\vspace{1em}

\noindent{\bf \em Experiment Setup Details}. The proposed MiMa model takes data from the third season in 2018 and 2019 for its training to predict the weather conditions for 80\% of the 2020 data in the same season. The other 20\% of the 2020 data is used for validation data during the training to enable early stopping. Our early stopping setting has a patience of five consecutive epochs with no decrease in the mean absolute error. Four MiMa modelets are established to predict four weather parameters of interest (i.e., air temperature, relative humidity, wind speed, and atmospheric pressure) for each mesonet station location.   

As depicted in Fig. \ref{fig:MiMaED}, the Micro and Macro Encoders of a MiMa modelet utilize LSTM networks with 256 hidden states, while the decoder leverages LSTM networks with 512 hidden states. The LSTM networks utilize the hyperbolic tangent (tanh) function as the activation function and the sigmoid function for the recurrent activation~\cite{hochreiter97:NC:long}. The tanh activation function aids in maintaining the stability of the cell state by forcing the outputs to a range between -1 and 1. In contrast, the sigmoid function ensures that the recurrent connections are regulated between 0 and 1, facilitating a smooth gradient flow. Combined with the substantial hidden state sizes, this regularization technique allows the MiMa model to capture and learn complex temporal dependencies present in the input sequences. During training, a dropout rate of 50\% is applied to both encoders and the decoder of a MiMa modelet to mitigate the risk of overfitting, thereby enhancing the model's generalization capabilities. The high dropout rate ensures that approximately half of the neurons are randomly deactivated during each training iteration, promoting robustness and preventing reliance on specific neurons. In addition, the hidden state sizes, dropout rate, and activation functions are meticulously selected to balance the trade-off between model complexity and generalization performance, ensuring high weather forecasting accuracy. Each MiMa modelet is trained using a mini-batch size of 64 across 60 epochs. An early stopping mechanism halts training if no loss improvement is observed for 10 consecutive epochs. The LSTM block of each encoder involves 256 units, leading to a total of 512 units in the Decoder LSTM block. The Adam optimizer is employed with a learning rate of 0.001.

To compare the MiMa modelets, five other models are included in our evaluation: 
(1) the Micro model, 
(2) SARIMA, which is an autoregressive model supporting the direct modeling of the seasonal component of series ~\cite{hyndman18:book:forecasting},
(3) SNN, which is a simple neural network \cite{dueben18:GMD:challenges} that takes the Micro data to make predictions,
(4) SVR, which is a regression model based on support vector machines \cite{svm56}, with the Micro data as input features, and
(5) DUQ (Deep Uncertainty Quantification), which integrates deep learning techniques with one GRU layer of 128 hidden nodes \cite{wang2019deep57} to quantify uncertainties in weather predictions, enhancing the reliability of weather forecasts.
The Micro model is structured as shown in Fig.~\ref{fig:MiED}, taking just the micro dataset (i.e., gathered near-surface) for training an LSTM with 256 hidden states.
The computed atmospheric results of the WRF-HRRR model serve as the coarser prediction counterpart (in the hourly granularity). While these comparative models are evaluated alongside MiMa modelets, they consistently underperform as demonstrated in both our preliminary study \cite{original_mima55} and the results presented in Tables \ref{model_errors} and \ref{extra_model_errors} below.

Model prediction accuracy is measured according to two metrics that gauge the prediction error against those observed by mesonet stations:  
Root Mean Squared Error (RMSE) and Mean Absolute Error (MAE) [32].
Specifically, ${\rm RMSE} = (\frac{1}{n}\sum_{i=1}^{n}(Y_i-\hat{Y}_i)^2)^{0.5}$, ${\rm MAE} = \frac{1}{n}\sum_{i=1}^{n}|(Y_i-\hat{Y}_i)|$, 
where ${\bf Y}$ and $\hat{\bf Y}$ denote the observed and the predicted value vectors, respectively, and $n$ is the number of data values. These two metrics aim to depict the error amounts.
RMSE is sensitive to outliers (with extreme errors emphasized as their amounts are squared), and MAE simply averages all error amounts so that it better reflects prediction accuracy in the absence of extreme errors (like weather parameter forecasting).

\begin{table*}
\centering
\caption{\small RMSE and MAE values for all 11 mesonet stations along with their respective errors $^{\S}$}
\renewcommand{\arraystretch}{1.3} 
\scalebox{0.72}{
\begin{tabular}{|c|l|cc|cc|cc|cc|}
\hline
\multirow{2}{*}{\textbf{\normalsize Station}}                                                    & \multicolumn{1}{c|}{\multirow{2}{*}{\textbf{\normalsize Model}}} & \multicolumn{2}{c|}{\textbf{\normalsize TEMP}}         & \multicolumn{2}{c|}{\textbf{\normalsize HUMI}}            & \multicolumn{2}{c|}{\textbf{\normalsize WSPD}}             & \multicolumn{2}{c|}{\textbf{\normalsize PRES}}            \\ \cline{3-10} 
                                                                            & \multicolumn{1}{c|}{}                       & \textbf{RMSE}                   & \textbf{MAE}                    & \textbf{RMSE}                   & \textbf{MAE}                    & \textbf{RMSE}                     & \textbf{MAE}                     & \textbf{RMSE}                   & \textbf{MAE}                    \\ \hline
\multirow{7}{*}{\begin{tabular}[c]{@{}c@{}}\normalsize LSML \vspace{0.5em}\\ \normalsize @748 ft.\end{tabular}}  
& MiMa & \textbf{0.14 (0.67\%)} & \textbf{0.10 (0.12\%)} & \textbf{0.78 (0.97\%)} & \textbf{0.57 (0.71\%)} & \textbf{0.36 (37.11\%)} & 0.28 (28.87\%) & 0.40 (0.04\%) & 0.34 (0.03\%) \\
& Micro & 0.16 (0.77\%) & 0.13 (0.16\%) & 1.08 (1.35\%) & 0.85 (1.06\%) & 0.36 (37.11\%) & \underline{\textbf{0.27 (27.84\%)}} & 0.76 (0.08\%) & 0.63 (0.06\%) \\
& SARIMA & 0.28 (1.34\%) & 0.17 (0.21\%) & 2.18 (2.72\%) & 1.28 (1.60\%) & 0.51 (52.58\%) & 0.35 (36.08\%) & \underline{\textbf{0.11 (0.01\%)}} & \underline{\textbf{0.09 (0.01\%)}} \\
& WRF-HRRR & 1.04 (4.99\%) & 0.82 (1.02\%) & 6.70 (8.36\%) & 5.46 (6.81\%) & 4.73 (487.63\%) & 4.18 (430.93\%) & 3.86 (0.39\%) & 3.85 (0.39\%) \\
& SNN & 5.19 (18.86\%) & 4.37 (15.89\%) & 19.52 (25.75\%) & 15.25 (20.11\%) & 3.81 (19.17\%) & 3.11 (15.65\%) & 279.03 (50.14\%) & 256.67 (46.12\%) \\
& SVR & 6.24 (22.68\%) & 5.43 (19.72\%) & 23.35 (30.81\%) & 19.94 (26.31\%) & 3.01 (15.14\%) & 2.56 (12.87\%) & 239.63 (43.06\%) & 213.46 (38.36\%) \\
& DUQ & 2.50 (11.66\%) & 1.99 (9.27\%) & 12.52 (19.46\%) & 10.33 (16.06\%) & 0.88 (8.55\%) & 0.60 (5.89\%) & 3.00 (28.49\%) & 2.45 (23.52\%) \\ \hline
\multirow{7}{*}{\begin{tabular}[c]{@{}c@{}}\normalsize CCLA \vspace{0.5em}\\ \normalsize @764 ft.\end{tabular}}  
& MiMa & 0.28 (1.31\%) & 0.24 (1.12\%) & \textbf{1.49 (1.77\%)} & \textbf{0.66 (0.79\%)} & \textbf{0.38 (19.90\%)} & \textbf{0.16 (8.38\%)} & \textbf{0.06 (0.01\%)} & 0.06 (0.01\%) \\
& Micro & \underline{\textbf{0.24 (1.12\%)}} & \underline{\textbf{0.22 (1.03\%)}} & 1.86 (2.21\%) & 1.43 (1.70\%) & 0.38 (19.90\%) & 0.24 (12.57\%) & 0.16 (0.02\%) & \underline{\textbf{0.04 (0.00\%)}} \\
& SARIMA & 0.31 (1.45\%) & 0.27 (1.26\%) & 1.57 (1.87\%) & 1.20 (1.43\%) & 0.43 (22.51\%) & 0.32 (16.75\%) & 0.07 (0.01\%) & 0.05 (0.01\%) \\
& WRF-HRRR & 1.88 (8.79\%) & 1.61 (7.53\%) & 14.67 (17.46\%) & 13.45 (16.01\%) & 2.32 (121.47\%) & 1.78 (93.19\%) & 0.34 (0.03\%) & 0.28 (0.03\%) \\
& SNN & 2.34 (9.01\%) & 1.53 (5.88\%) & 8.78 (12.16\%) & 5.63 (7.80\%) & 0.73 (5.97\%) & 0.47 (3.86\%) & 1.05 (5.51\%) & 0.73 (3.85\%) \\
& SVR & 5.32 (20.49\%) & 4.30 (16.54\%) & 9.81 (13.58\%) & 6.64 (9.19\%) & 1.45 (11.82\%) & 1.21 (9.84\%) & 3.08 (16.23\%) & 2.34 (12.32\%) \\
& DUQ & 2.74 (15.01\%) & 2.33 (12.79\%) & 11.32 (19.11\%) & 8.94 (15.10\%) & 1.09 (7.56\%) & 0.78 (5.40\%) & 1.24 (12.47\%) & 1.13 (11.41\%) \\ \hline
\multirow{7}{*}{\begin{tabular}[c]{@{}c@{}}\normalsize LGRN \vspace{0.5em}\\ \normalsize @766 ft.\end{tabular}}  
& MiMa & \textbf{0.16 (0.77\%)} & \textbf{0.09 (0.43\%)} & \textbf{0.86 (1.09\%)} & \textbf{0.61 (0.77\%)} & \textbf{0.30 (15.63\%)} & \textbf{0.23 (11.98\%)} & \textbf{0.09 (0.01\%)} & \textbf{0.07 (0.01\%)} \\
& Micro & 0.19 (0.91\%) & 0.15 (0.72\%) & 1.01 (1.28\%) & 0.74 (0.94\%) & 0.31 (16.15\%) & 0.22 (11.46\%) & 0.09 (0.01\%) & 0.08 (0.01\%) \\
& SARIMA & 0.29 (1.39\%) & 0.18 (0.86\%) & 2.02 (2.57\%) & 1.23 (1.56\%) & 0.58 (30.21\%) & 0.41 (21.35\%) & 0.10 (0.01\%) & 0.09 (0.01\%) \\
& WRF-HRRR & 1.78 (8.52\%) & 1.49 (7.14\%) & 8.02 (10.19\%) & 6.36 (8.08\%) & 4.14 (215.63\%) & 3.84 (200.00\%) & 6.75 (0.68\%) & 6.75 (0.68\%) \\
& SNN & 2.65 (10.08\%) & 1.93 (7.34\%) & 8.83 (12.30\%) & 6.23 (8.68\%) & 0.69 (6.19\%) & 0.47 (4.27\%) & 1.19 (5.81\%) & 0.83 (4.07\%) \\
& SVR & 4.05 (15.44\%) & 3.25 (12.37\%) & 10.35 (14.41\%) & 7.41 (10.32\%) & 0.91 (8.21\%) & 0.71 (6.34\%) & 2.10 (10.30\%) & 1.57 (7.71\%) \\
& DUQ & 1.59 (9.02\%) & 1.11 (6.32\%) & 6.63 (11.58\%) & 3.89 (6.80\%) & 1.27 (10.51\%) & 1.02 (8.49\%) & 0.69 (6.98\%) & 0.49 (4.98\%) \\ \hline
\multirow{7}{*}{\begin{tabular}[c]{@{}c@{}}\normalsize FCHV \vspace{0.5em}\\ \normalsize @770 ft.\end{tabular}}  
& MiMa & \textbf{0.15 (0.70\%)} & \textbf{0.10 (0.47\%)} & \textbf{0.84 (1.10\%)} & \textbf{0.58 (0.76\%)} & \textbf{0.32 (15.46\%)} & \textbf{0.24 (11.59\%)} & 0.25 (0.03\%) & 0.22 (0.02\%) \\
& Micro & 0.24 (1.12\%) & 0.21 (0.98\%) & 0.88 (1.16\%) & 0.59 (0.78\%) & 0.33 (15.94\%) & 0.25 (12.08\%) & 0.57 (0.06\%) & 0.46 (0.05\%) \\
& SARIMA & 0.24 (1.12\%) & 0.14 (0.65\%) & 1.85 (2.43\%) & 1.22 (1.60\%) & 0.49 (23.67\%) & 0.34 (16.43\%) & \underline{\textbf{0.09 (0.01\%)}} & \underline{\textbf{0.09 (0.01\%)}} \\
& WRF-HRRR & 1.15 (5.36\%) & 0.91 (4.24\%) & 7.46 (9.80\%) & 6.13 (8.06\%) & 3.95 (190.82\%) & 3.65 (176.33\%) & 0.93 (0.09\%) & 0.67 (0.07\%) \\
& SNN & 3.40 (17.75\%) & 2.79 (14.58\%) & 12.99 (20.70\%) & 10.32 (16.46\%) & 1.07 (9.05\%) & 0.83 (6.96\%) & 1.90 (18.76\%) & 1.66 (16.36\%) \\
& SVR & 6.04 (23.70\%) & 5.05 (19.81\%) & 21.50 (29.33\%) & 18.07 (24.65\%) & 2.51 (18.71\%) & 2.28 (17.04\%) & 241.51 (48.56\%) & 219.69 (44.17\%) \\
& DUQ & 2.74 (14.31\%) & 2.31 (12.05\%) & 9.92 (15.81\%) & 8.21 (13.09\%) & 0.85 (7.19\%) & 0.65 (5.49\%) & 1.21 (11.99\%) & 1.06 (10.51\%) \\ \hline
\multirow{7}{*}{\begin{tabular}[c]{@{}c@{}}\normalsize CROP \vspace{0.5em}\\ \normalsize @858 ft.\end{tabular}}  
& MiMa & \textbf{0.18 (0.87\%)} & \textbf{0.12 (0.58\%)} & \textbf{0.91 (1.16\%)} & \textbf{0.64 (0.82\%)} & \textbf{0.32 (15.17\%)} & \textbf{0.25 (11.85\%)} & \textbf{0.08 (0.01\%)} & 0.07 (0.01\%) \\
& Micro & 0.21 (1.02\%) & 0.17 (0.83\%) & 0.91 (1.16\%) & 0.65 (0.83\%) & 0.33 (15.64\%) & 0.26 (12.32\%) & 0.08 (0.01\%) & \underline{\textbf{0.06 (0.01\%)}} \\
& SARIMA & 0.24 (1.17\%) & 0.16 (0.78\%) & 1.60 (2.04\%) & 1.02 (1.30\%) & 0.42 (19.91\%) & 0.34 (16.11\%) & 0.10 (0.01\%) & 0.08 (0.01\%) \\
& WRF-HRRR & 0.87 (4.23\%) & 0.64 (3.11\%) & 5.12 (6.53\%) & 3.74 (4.77\%) & 4.82 (228.44\%) & 4.24 (200.95\%) & 0.35 (0.04\%) & 0.29 (0.03\%) \\
& SNN & 2.26 (8.14\%) & 1.55 (5.58\%) & 7.89 (10.73\%) & 5.01 (6.82\%) & 0.90 (7.55\%) & 0.61 (5.15\%) & 1.08 (5.25\%) & 0.74 (3.62\%) \\
& SVR & 3.99 (14.35\%) & 3.16 (11.35\%) & 11.50 (15.64\%) & 8.67 (11.79\%) & 1.29 (10.86\%) & 0.96 (8.07\%) & 2.43 (11.81\%) & 1.78 (8.66\%) \\
& DUQ & 8.07 (38.02\%) & 6.85 (32.25\%) & 9.90 (14.50\%) & 8.19 (11.99\%) & 1.27 (11.82\%) & 0.89 (8.29\%) & 1.81 (17.92\%) & 1.61 (16.00\%) \\ \hline
\multirow{7}{*}{\begin{tabular}[c]{@{}c@{}}\normalsize ELST \vspace{0.5em}\\ \normalsize @860 ft.\end{tabular}}  
& MiMa & \textbf{0.22 (1.06\%)} & \textbf{0.13 (0.62\%)} & \textbf{0.93 (1.05\%)} & \textbf{0.64 (0.72\%)} & \textbf{0.35 (27.78\%)} & \textbf{0.26 (20.63\%)} & \textbf{0.08 (0.01\%)} & \textbf{0.07 (0.01\%)} \\
& Micro & 0.23 (1.10\%) & 0.15 (0.72\%) & 1.03 (1.16\%) & 0.77 (0.87\%) & 0.36 (28.57\%) & 0.27 (21.43\%) & 0.33 (0.03\%) & 0.17 (0.02\%) \\
& SARIMA & 0.23 (1.10\%) & 0.16 (0.77\%) & 1.55 (1.75\%) & 0.96 (1.08\%) & 0.45 (35.71\%) & 0.32 (25.40\%) & 0.09 (0.01\%) & 0.08 (0.01\%) \\
& WRF-HRRR & 1.39 (6.67\%) & 0.86 (4.12\%) & 10.41 (11.76\%) & 7.70 (8.70\%) & 4.76 (377.78\%) & 4.02 (319.05\%) & 0.53 (0.05\%) & 0.46 (0.05\%) \\
& SNN & 2.86 (9.33\%) & 1.95 (6.37\%) & 9.34 (11.79\%) & 5.80 (7.32\%) & 0.78 (7.28\%) & 0.53 (5.00\%) & 0.91 (5.69\%) & 0.67 (4.18\%) \\
& SVR & 4.20 (13.69\%) & 3.30 (10.77\%) & 12.10 (15.26\%) & 9.01 (11.36\%) & 1.12 (10.47\%) & 0.91 (8.49\%) & 1.75 (10.90\%) & 1.22 (7.60\%) \\
& DUQ & 1.87 (9.40\%) & 1.34 (6.72\%) & 8.72 (14.87\%) & 5.75 (9.81\%) & 0.94 (8.76\%) & 0.69 (6.40\%) & 3.32 (29.43\%) & 2.78 (24.66\%) \\ \hline
\multirow{7}{*}{\begin{tabular}[c]{@{}c@{}}\normalsize HUEY \vspace{0.5em}\\ \normalsize @896 ft.\end{tabular}}  
& MiMa & \textbf{0.27 (1.28\%)} & \textbf{0.21 (0.99\%)} & \textbf{1.56 (2.20\%)} & \textbf{0.75 (1.06\%)} & \textbf{0.29 (107.41\%)} & \textbf{0.22 (81.48\%)} & \textbf{0.06 (0.01\%)} & \textbf{0.04 (0.00\%)} \\
& Micro & 0.30 (1.42\%) & 0.58 (2.74\%) & 1.63 (2.30\%) & 2.64 (3.73\%) & 0.31 (114.81\%) & 0.25 (92.59\%) & 0.08 (0.01\%) & 0.52 (0.05\%) \\
& SARIMA & 0.31 (1.47\%) & 0.27 (1.28\%) & 1.77 (2.50\%) & 1.54 (2.17\%) & 0.59 (218.52\%) & 0.46 (170.37\%) & 0.07 (0.01\%) & 0.06 (0.01\%) \\
& WRF-HRRR & 1.19 (5.63\%) & 0.93 (4.40\%) & 8.17 (11.53\%) & 6.65 (9.38\%) & 2.16 (800.00\%) & 1.80 (666.67\%) & 0.54 (0.05\%) & 0.50 (0.05\%) \\
& SNN & 2.90 (10.72\%) & 2.23 (8.23\%) & 8.16 (11.34\%) & 5.53 (7.70\%) & 0.64 (6.34\%) & 0.46 (4.59\%) & 1.19 (5.38\%) & 0.89 (4.02\%) \\
& SVR & 5.85 (21.59\%) & 4.69 (17.32\%) & 9.98 (13.88\%) & 7.37 (10.25\%) & 1.49 (14.79\%) & 1.28 (12.68\%) & 3.56 (16.06\%) & 2.81 (12.70\%) \\
& DUQ & 4.62 (23.52\%) & 3.98 (20.27\%) & 11.35 (17.26\%) & 9.44 (14.36\%) & 1.25 (13.51\%) & 1.08 (11.72\%) & 1.16 (9.75\%) & 1.00 (8.41\%) \\ \hline
\multirow{7}{*}{\begin{tabular}[c]{@{}c@{}}\normalsize LXGN \vspace{0.5em}\\ \normalsize @1044 ft.\end{tabular}}  
& MiMa & \textbf{0.16 (0.72\%)} & \textbf{0.09 (0.41\%)} & 1.03 (1.41\%) & \textbf{0.52 (0.71\%)} & \textbf{0.40 (17.32\%)} & \textbf{0.19 (8.23\%)} & \textbf{0.06 (0.01\%)} & \textbf{0.05 (0.01\%)} \\
& Micro & 0.45 (2.04\%) & 0.10 (0.45\%) & \underline{\textbf{1.02 (1.40\%)}} & 1.28 (1.75\%) & 0.40 (17.32\%) & 0.24 (10.39\%) & 0.06 (0.01\%) & 0.21 (0.02\%) \\
& SARIMA & 0.18 (0.81\%) & 0.11 (0.50\%) & 0.79 (1.08\%) & 0.56 (0.77\%) & 0.47 (20.35\%) & 0.36 (15.58\%) & 0.06 (0.01\%) & 0.06 (0.01\%) \\
& WRF-HRRR & 1.12 (5.07\%) & 0.93 (4.21\%) & 4.24 (5.80\%) & 3.46 (4.73\%) & 2.15 (93.07\%) & 1.73 (74.89\%) & 1.10 (0.11\%) & 1.07 (0.11\%) \\
& SNN & 4.94 (28.33\%) & 4.53 (25.95\%) & 20.94 (31.87\%) & 19.77 (30.09\%) & 0.87 (7.65\%) & 0.65 (5.60\%) & 0.81 (7.50\%) & 0.61 (5.64\%) \\
& SVR & 5.13 (20.33\%) & 4.17 (16.52\%) & 10.02 (12.50\%) & 7.13 (8.89\%) & 1.54 (13.67\%) & 1.30 (11.41\%) & 3.46 (17.23\%) & 2.69 (13.38\%) \\
& DUQ & 3.55 (20.36\%) & 3.13 (17.94\%) & 8.98 (13.67\%) & 7.08 (10.78\%) & 0.98 (8.64\%) & 0.78 (6.85\%) & 1.17 (10.91\%) & 1.04 (9.67\%) \\ \hline
\multirow{7}{*}{\begin{tabular}[c]{@{}c@{}}\normalsize FARM \vspace{0.5em}\\ \normalsize @559 ft.\end{tabular}}  
& MiMa & \textbf{0.19 (0.86\%)} & \textbf{0.10 (0.45\%)} & \textbf{1.49 (1.78\%)} & \textbf{0.61 (0.73\%)} & \textbf{0.37 (19.47\%)} & \textbf{0.16 (8.42\%)} & \textbf{0.06 (0.01\%)} & \textbf{0.06 (0.01\%)} \\
& Micro & 0.25 (1.13\%) & 0.12 (0.54\%) & 1.65 (1.97\%) & 0.93 (1.11\%) & 0.38 (20.00\%) & 0.21 (11.05\%) & 0.08 (0.01\%) & 0.06 (0.01\%) \\
& SARIMA & 0.22 (1.00\%) & 0.19 (0.86\%) & 1.55 (1.85\%) & 0.86 (1.03\%) & 0.40 (21.05\%) & 0.31 (16.32\%) & 0.07 (0.01\%) & 0.06 (0.01\%) \\
& WRF-HRRR & 1.00 (4.53\%) & 0.80 (3.63\%) & 5.47 (6.54\%) & 4.19 (5.01\%) & 1.47 (77.37\%) & 1.19 (62.63\%) & 0.77 (0.08\%) & 0.72 (0.07\%) \\
& SNN & 7.69 (42.85\%) & 7.13 (39.72\%) & 59.04 (88.49\%) & 58.34 (87.44\%) & 1.03 (8.98\%) & 0.81 (7.09\%) & 0.70 (6.79\%) & 0.47 (4.60\%) \\
& SVR & 5.08 (20.22\%) & 4.19 (16.69\%) & 9.41 (13.14\%) & 6.58 (9.19\%) & 1.80 (12.53\%) & 1.47 (10.26\%) & 2.57 (14.41\%) & 1.92 (10.77\%) \\
& DUQ & 6.67 (37.19\%) & 5.45 (30.36\%) & 10.58 (15.85\%) & 8.71 (13.06\%) & 0.90 (7.89\%) & 0.68 (5.94\%) & 0.59 (5.78\%) & 0.42 (4.03\%) \\ \hline
\multirow{7}{*}{\begin{tabular}[c]{@{}c@{}}\normalsize DANV \vspace{0.5em}\\ \normalsize @981 ft.\end{tabular}}  
& MiMa & 0.16 (0.75\%) & 0.14 (0.66\%) & \textbf{0.89 (1.09\%)} & \textbf{0.62 (0.76\%)} & \textbf{0.34 (22.97\%)} & \textbf{0.27 (18.24\%)} & \textbf{0.08 (0.01\%)} & \textbf{0.06 (0.01\%)} \\
& Micro & \underline{\textbf{0.15 (0.70\%)}} & \underline{\textbf{0.12 (0.56\%)}} & 0.91 (1.11\%) & 0.68 (0.83\%) & 0.36 (24.32\%) & 0.28 (18.92\%) & 0.11 (0.01\%) & 0.09 (0.01\%) \\
& SARIMA & 0.24 (1.13\%) & 0.16 (0.75\%) & 1.60 (1.95\%) & 1.02 (1.25\%) & 0.42 (28.38\%) & 0.34 (22.97\%) & 0.10 (0.01\%) & 0.08 (0.01\%) \\
& WRF-HRRR & 0.87 (4.08\%) & 0.64 (3.00\%) & 5.12 (6.26\%) & 3.74 (4.57\%) & 4.82 (325.68\%) & 4.24 (286.49\%) & 0.35 (0.04\%) & 0.29 (0.03\%) \\
& SNN & 2.45 (8.88\%) & 1.78 (6.45\%) & 9.10 (11.65\%) & 6.06 (7.77\%) & 0.87 (6.74\%) & 0.58 (4.49\%) & 0.98 (5.08\%) & 0.68 (3.50\%) \\
& SVR & 3.47 (12.59\%) & 2.64 (9.57\%) & 10.12 (12.95\%) & 7.04 (9.01\%) & 1.06 (8.17\%) & 0.78 (6.02\%) & 1.89 (9.75\%) & 1.33 (6.89\%) \\
& DUQ & 6.76 (31.61\%) & 5.69 (26.59\%) & 17.76 (26.39\%) & 15.15 (22.50\%) & 0.91 (7.96\%) & 0.66 (5.78\%) & 1.45 (13.28\%) & 1.25 (11.51\%) \\ \hline
\multirow{7}{*}{\begin{tabular}[c]{@{}c@{}}\normalsize BMTN \vspace{0.5em}\\ \normalsize @4031 ft.\end{tabular}}  
& MiMa & \textbf{0.20 (1.18\%)} & \textbf{0.11 (0.65\%)} & \textbf{1.49 (1.59\%)} & \textbf{0.75 (0.80\%)} & \textbf{0.32 (18.39\%)} & \textbf{0.15 (8.62\%)} & \textbf{0.08 (0.01\%)} & \textbf{0.10 (0.01\%)} \\
& Micro & 0.22 (1.30\%) & 0.19 (1.12\%) & 1.52 (1.62\%) & 1.25 (1.33\%) & 0.32 (18.39\%) & 0.15 (8.62\%) & 0.08 (0.01\%) & 0.10 (0.01\%) \\
& SARIMA & 0.20 (1.18\%) & 0.15 (0.89\%) & 1.65 (1.76\%) & 0.87 (0.93\%) & 0.62 (35.63\%) & 0.47 (27.01\%) & 0.11 (0.01\%) & 0.22 (0.02\%) \\
& WRF-HRRR & 1.70 (10.04\%) & 1.55 (9.15\%) & 14.14 (15.06\%) & 11.72 (12.48\%) & 1.77 (101.72\%) & 1.51 (86.78\%) & 24.04 (2.73\%) & 24.04 (2.73\%) \\
& SNN & 8.43 (61.29\%) & 8.22 (59.75\%) & 6.95 (13.33\%) & 5.56 (10.66\%) & 0.71 (8.73\%) & 0.57 (7.06\%) & 3.37 (30.41\%) & 3.28 (29.55\%) \\
& SVR & 4.62 (22.61\%) & 3.72 (18.23\%) & 10.87 (13.47\%) & 8.01 (9.92\%) & 1.18 (16.15\%) & 0.98 (13.37\%) & 3.78 (19.16\%) & 3.02 (15.27\%) \\
& DUQ & 3.91 (28.44\%) & 3.30 (24.00\%) & 7.67 (14.72\%) & 6.32 (12.14\%) & 0.82 (10.04\%) & 0.67 (8.21\%) & 1.40 (12.58\%) & 1.20 (10.84\%) \\
 \hline
\end{tabular}
}
\label{model_errors}
\begin{tablenotes}
    \item $^{\S}$ Those entries at which MiMa rows are not smallest are underlined. The prediction errors normalized with respect to the values themselves (in \%) are included in pairs of parentheses.
\end{tablenotes}
\end{table*}

\vspace{0.3em}
\subsection{Performance Results and Discussion}

\noindent {\bf \em Under Different Prediction Methods.} We have conducted multiple experiments to forecast four weather parameters of interest using the MiMa modelets at all time points over 16 days chosen arbitrarily in the third season of 2020. 
There are 4608 ($= 12\times 24 \times16$) predicted time points for each parameter per chosen Kentucky Mesonet station, given that each hour incurs 12 prediction points with 5 minutes apart. 
Accuracy metrics under MiMa modeling, averaged over all prediction time points, are listed at the first row of each station in Table~\ref{model_errors}.
As can be seen, temperature prediction at BMTN (with the highest elevation among all eleven stations) has an RMSE of 0.20 while the prediction at FARM (with the lowest elevation among all stations) has an RMSE of 0.19. Those RMSE amounts translate respectively to 1.18\% and 0.86\% (given in pairs of parentheses) when normalized against their observed temperature readings, which are smaller at altitude-highest BMTN.  
Overall, the MiMa modelets predict temperature accurately for all stations, with their TEMP's RMSE values $\leq$ 0.27 (or 1.28\%), irrespective of their altitude.  
Among the 44 forecasting instances (due to four parameters at eleven locations), the MiMa modelets demonstrated superior prediction accuracy, exhibiting the smallest RMSE values in all but five cases among all forecasters included in Table \ref{model_errors}. Note that the normalized prediction errors of WSPD (shown in pairs of parentheses) tend to be large, signifying that the wind speeds are usually very low, making small error RMSE values become large after normalization.

The RMSE values of two other models as well as WRF-HRRR on forecasting the four weather parameters are also included in Table~\ref{model_errors} for comparison.  
The MiMa models exhibit the best accuracy (in terms of RMSE) consistently for all four parameters at eleven stations, except for five cases (i.e., TEMP at CCLA and DANV, HUMI at LXGN, PRES at LSML, and FCHV). Furthermore, the MiMa model outperforms its closest counterpart, the Micro model, in all but three cases (i.e., TEMP at CCLA and DANV, HUMI at LXGN), signifying the advantage of employing both datasets for the model input, as opposed to utilizing just gathered observational data, like the Micro model.

\begin{figure*}[htbp]
    \centering
    \begin{minipage}[b]{0.90\columnwidth}
        \captionsetup{skip=0.1pt} 
        \centering
        \includegraphics[width=1.1\textwidth, height=0.2\textheight]{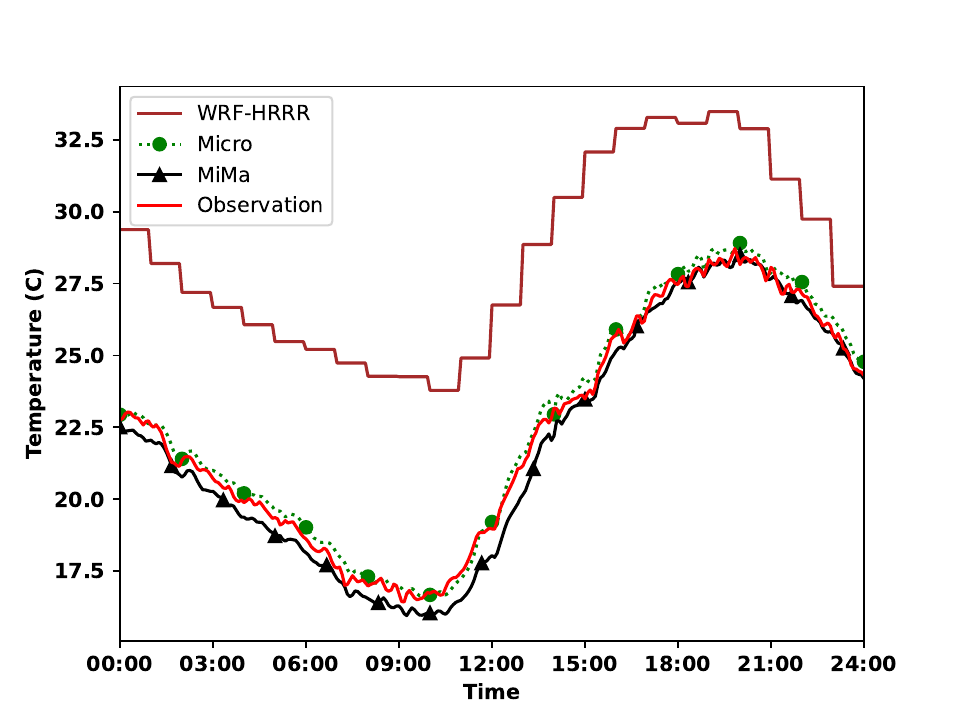}
        \caption*{\parbox{1.1\textwidth}{\centering (a) Temperature}}
    \end{minipage}
    \hspace{0.05\columnwidth}
    \begin{minipage}[b]{0.90\columnwidth}
        \captionsetup{skip=0.1pt} 
        \centering
        \includegraphics[width=1.1\textwidth, height=0.2\textheight]{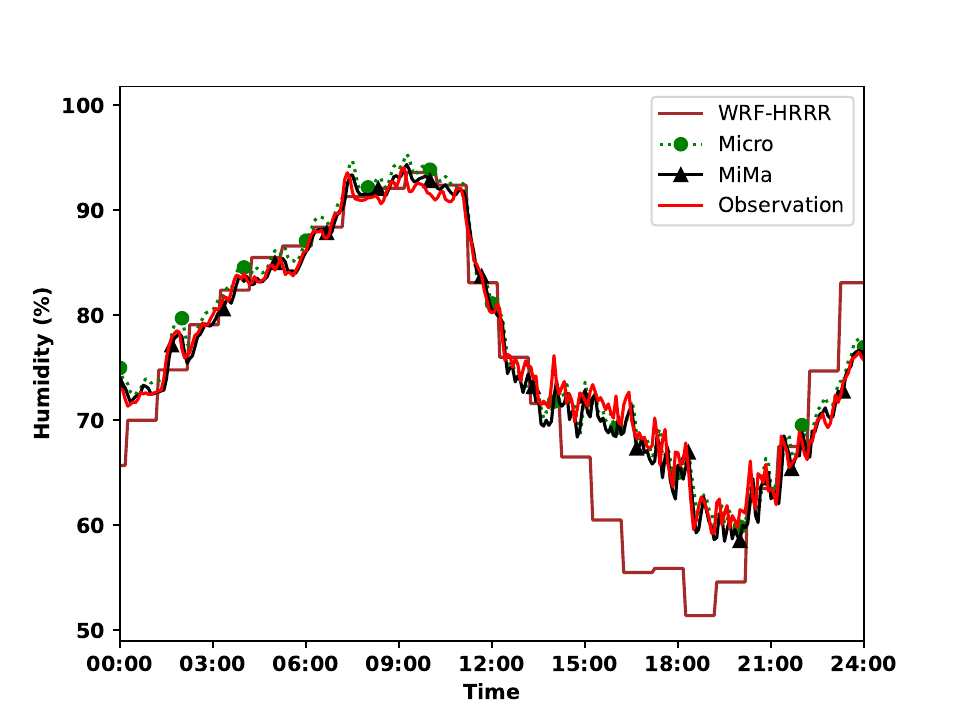}
        \caption*{\parbox{1.1\textwidth}{\centering (b) Humidity}}
    \end{minipage}

    \vspace{-0.2em}
    
    \begin{minipage}[b]{0.90\columnwidth}
        \captionsetup{skip=0.1pt} 
        \centering
        \includegraphics[width=1.1\textwidth, height=0.2\textheight]{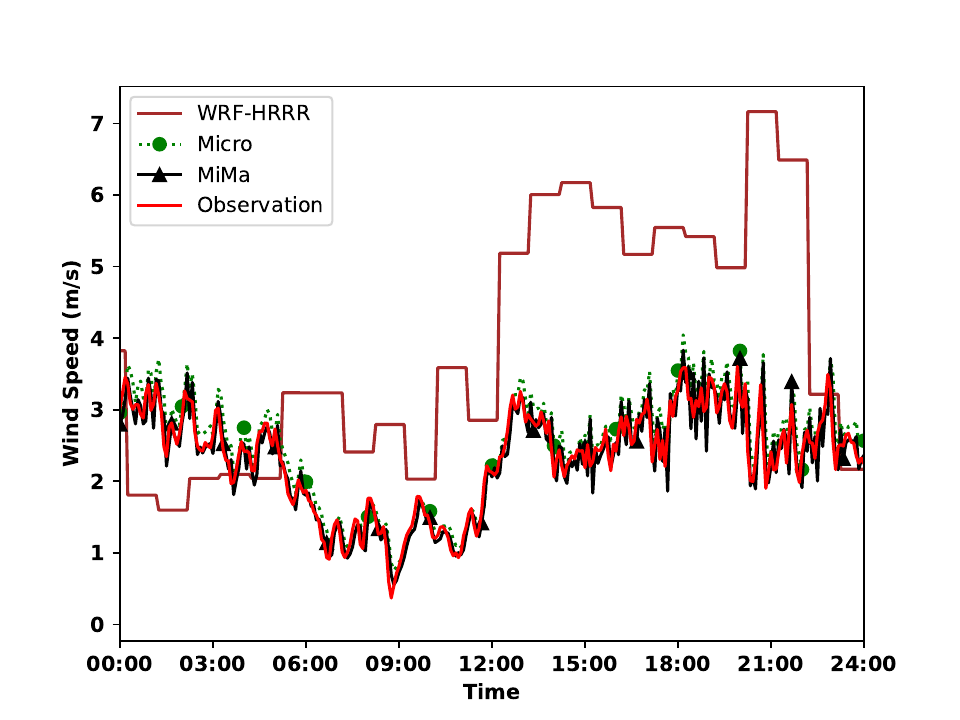}
        \caption*{\parbox{1.2\textwidth}{\centering (c) Wind Speed}}
    \end{minipage}
    \hspace{0.05\columnwidth}
    \begin{minipage}[b]{0.90\columnwidth}
        \captionsetup{skip=0.1pt} 
        \centering
        \includegraphics[width=1.1\textwidth, height=0.2\textheight]{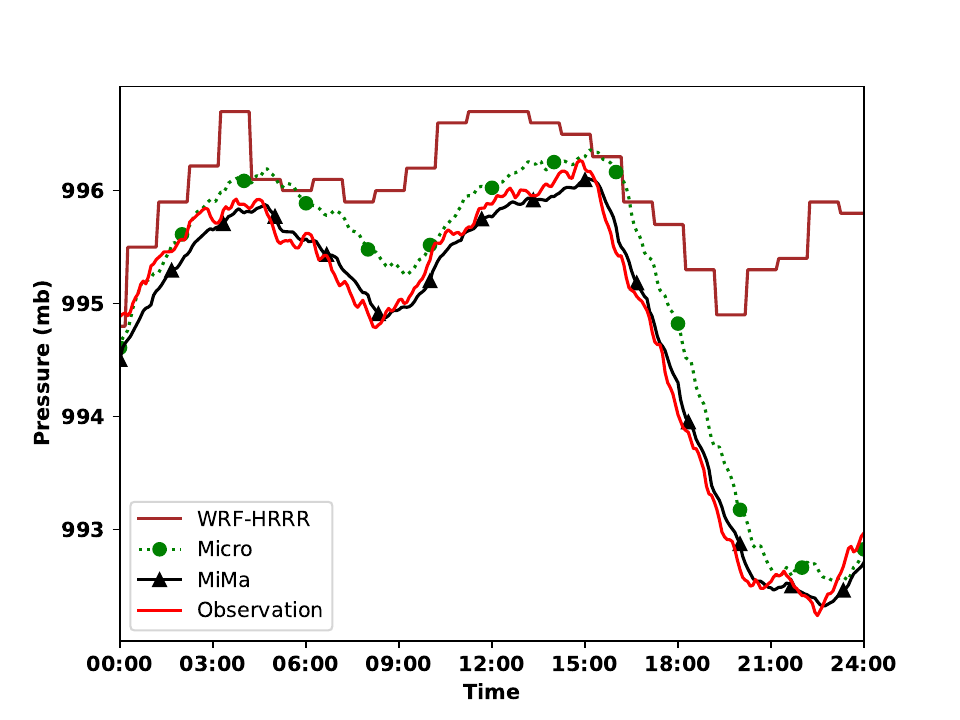}
        \caption*{\parbox{1.1\textwidth}{\centering (d) Pressure}}
    \end{minipage}
    \caption{\small Prediction results for 24 hours for those four prediction parameters with the Micro, MiMa, and WRF models compared to the observed data.}
    \label{24_hour_predictions}
\end{figure*}

When taking the RMSE values of all four predicted parameters at each station into aggregation consideration, the MiMa model outperforms its Micro counterpart noticeably, able to deliver significantly better prediction on aggregated weather parameters at every station.  
From the individual parameter prediction's perspective, the MiMa model achieves better pressure prediction over all six stations in terms of the RMSE metric, when compared with the Micro model.
For predicting TEMP, HUMI, and WSPD, SARIMA underperforms the Micro model except for TEMP at BMTN and LXGN, and for HUMI at CCLA, FARM, and LXGN.
When forecasting PRES, SARIMA may perform better than the Micro model.
Additionally, the WRF-HRRR model is seen to perform the worst among all models, implying that its widely available computed results are far from satisfaction and that the superior MiMa model is indispensable for precise forecasting regionally, with the aid of near-surface gathered data. 

From the MAE metric standpoint, it is found in Table~\ref{model_errors} that our MiMa model consistently outperforms its two better counterparts (the Micro model and the SARIMA model) in all but seven cases, with four of them for the PRES prediction. 
With the MAE values of all four predicted parameters at each station taken into aggregation consideration, 
the MiMa model outperforms its Micro counterpart by larger margins than those under the RMSE metric. 
This may be due to the fact that MAE is less sensitive to the prediction error amount than RMSE. Among four types of weather parameter predictions, the MiMa model enjoys the largest gap against the Micro model for PRES prediction under the MAE metric over all stations aggregately, followed by HUMI prediction.

\begin{table*}
\centering
\caption{\small RMSE values of MiMa modelets at each 15-minute interval over a 3-hour horizon under 1-hour (4-hour) lead time}
\renewcommand{\arraystretch}{1.2} 
\scalebox{0.68}{
\begin{tabular}{|c|l|cccccccccccc|}
\hline
\textbf{\normalsize Station}      & \multicolumn{1}{c|}{\textbf{\normalsize Param.}} & \textbf{\normalsize 15 min.} & \textbf{\normalsize30 min} & \textbf{\normalsize 45 min.} & \textbf{\normalsize 60 min.} & \textbf{\normalsize 75 min.} & \textbf{\normalsize 90 min.} & \textbf{\normalsize 105 min.} & \textbf{\normalsize 120 min.} & \textbf{\normalsize 135 min.} & \textbf{\normalsize 150 min.} & \textbf{\normalsize 165 min.} & \textbf{\normalsize 180 min.} \\ \hline
\multirow{4}{*}{\normalsize LSML} & TEMP                                    & 0.23 (0.34)   & 0.23 (0.34)  & 0.23 (0.36)   & 0.24 (0.39)   & 0.24 (0.39)   & 0.24 (0.41)   & 0.24 (0.43)    & 0.25 (0.42)    & 0.26 (0.44)    & 0.27 (0.44)    & 0.29 (0.45)    & 0.32 (0.46)    \\
                      & HUMI                                    & 0.88 (0.92)   & 0.88 (0.93)  & 0.89 (0.96)   & 0.89 (0.97)   & 0.90 (1.00)   & 0.90 (1.00)   & 0.90 (1.00)    & 0.91 (1.03)    & 0.91 (1.02)    & 0.91 (1.02)    & 0.91 (1.02)    & 0.92 (1.02)    \\
                      & WSPD                                    & 0.47 (0.58)   & 0.49 (0.61)  & 0.50 (0.61)   & 0.52 (0.61)   & 0.53 (0.61)   & 0.54 (0.61)   & 0.56 (0.61)    & 0.57 (0.63)    & 0.57 (0.65)    & 0.58 (0.66)    & 0.58 (0.69)    & 0.58 (0.70)    \\
                      & PRES                                    & 0.32 (0.39)   & 0.32 (0.39)  & 0.34 (0.39)   & 0.34 (0.41)   & 0.34 (0.41)   & 0.35 (0.44)   & 0.35 (0.46)    & 0.35 (0.49)    & 0.35 (0.49)    & 0.35 (0.50)    & 0.36 (0.52)    & 0.37 (0.52)    \\ \hline
\multirow{4}{*}{\normalsize CCLA} & TEMP                                    & 0.27 (0.29)   & 0.27 (0.31)  & 0.27 (0.33)   & 0.27 (0.35)   & 0.28 (0.37)   & 0.28 (0.38)   & 0.28 (0.39)    & 0.28 (0.39)    & 0.28 (0.39)    & 0.28 (0.40)    & 0.29 (0.40)    & 0.29 (0.40)    \\
                      & HUMI                                    & 1.65 (1.83)   & 1.67 (1.84)  & 1.68 (1.87)   & 1.70 (1.88)   & 1.75 (1.90)   & 1.79 (1.90)   & 1.82 (1.92)    & 1.82 (1.92)    & 1.83 (1.95)    & 1.83 (1.95)    & 1.84 (1.95)    & 1.84 (1.95)    \\
                      & WSPD                                    & 0.48 (0.55)   & 0.48 (0.55)  & 0.49 (0.55)   & 0.50 (0.55)   & 0.51 (0.56)   & 0.51 (0.58)   & 0.51 (0.60)    & 0.51 (0.61)    & 0.51 (0.60)    & 0.52 (0.61)    & 0.54 (0.61)    & 0.55 (0.61)    \\
                      & PRES                                    & 0.09 (0.17)   & 0.09 (0.18)  & 0.10 (0.19)   & 0.10 (0.22)   & 0.11 (0.24)   & 0.12 (0.24)   & 0.12 (0.26)    & 0.13 (0.28)    & 0.14 (0.28)    & 0.14 (0.29)    & 0.15 (0.30)    & 0.15 (0.32)    \\ \hline
\multirow{4}{*}{\normalsize LGRN} & TEMP                                    & 0.36 (0.61)   & 0.37 (0.61)  & 0.40 (0.64)   & 0.42 (0.65)   & 0.42 (0.67)   & 0.43 (0.68)   & 0.44 (0.70)    & 0.44 (0.71)    & 0.46 (0.71)    & 0.53 (0.71)    & 0.59 (0.71)    & 0.59 (0.72)    \\
                      & HUMI                                    & 0.77 (0.78)   & 0.77 (0.80)  & 0.77 (0.80)   & 0.77 (0.82)   & 0.77 (0.85)   & 0.77 (0.88)   & 0.77 (0.88)    & 0.77 (0.90)    & 0.78 (0.90)    & 0.78 (0.91)    & 0.78 (0.91)    & 0.78 (0.91)    \\
                      & WSPD                                    & 0.32 (0.37)   & 0.32 (0.37)  & 0.32 (0.37)   & 0.32 (0.39)   & 0.33 (0.40)   & 0.34 (0.43)   & 0.34 (0.46)    & 0.34 (0.48)    & 0.34 (0.48)    & 0.35 (0.48)    & 0.35 (0.48)    & 0.35 (0.49)    \\
                      & PRES                                    & 0.11 (0.13)   & 0.11 (0.13)  & 0.11 (0.13)   & 0.11 (0.13)   & 0.11 (0.15)   & 0.12 (0.16)   & 0.12 (0.16)    & 0.12 (0.18)    & 0.12 (0.19)    & 0.12 (0.19)    & 0.13 (0.20)    & 0.13 (0.23)    \\ \hline
\multirow{4}{*}{\normalsize FCHV} & TEMP                                    & 0.23 (0.41)   & 0.23 (0.42)  & 0.25 (0.43)   & 0.27 (0.43)   & 0.28 (0.44)   & 0.28 (0.44)   & 0.29 (0.44)    & 0.29 (0.43)    & 0.32 (0.44)    & 0.35 (0.44)    & 0.39 (0.45)    & 0.40 (0.47)    \\
                      & HUMI                                    & 0.92 (1.02)   & 0.95 (1.02)  & 0.96 (1.04)   & 0.96 (1.06)   & 0.97 (1.08)   & 0.97 (1.12)   & 0.99 (1.16)    & 1.00 (1.16)    & 1.00 (1.17)    & 1.01 (1.17)    & 1.01 (1.18)    & 1.01 (1.20)    \\
                      & WSPD                                    & 0.34 (0.44)   & 0.34 (0.45)  & 0.36 (0.46)   & 0.38 (0.46)   & 0.38 (0.46)   & 0.39 (0.46)   & 0.39 (0.46)    & 0.42 (0.46)    & 0.43 (0.46)    & 0.43 (0.47)    & 0.43 (0.47)    & 0.44 (0.47)    \\
                      & PRES                                    & 0.07 (0.12)   & 0.07 (0.13)  & 0.08 (0.16)   & 0.09 (0.19)   & 0.09 (0.19)   & 0.09 (0.21)   & 0.09 (0.21)    & 0.10 (0.24)    & 0.10 (0.25)    & 0.10 (0.27)    & 0.10 (0.27)    & 0.10 (0.29)    \\ \hline
\multirow{4}{*}{\normalsize CROP} & TEMP                                    & 0.31 (0.34)   & 0.31 (0.34)  & 0.32 (0.34)   & 0.32 (0.35)   & 0.32 (0.35)   & 0.32 (0.36)   & 0.33 (0.36)    & 0.33 (0.36)    & 0.33 (0.36)    & 0.33 (0.37)    & 0.33 (0.37)    & 0.33 (0.37)    \\
                      & HUMI                                    & 1.12 (1.26)   & 1.14 (1.28)  & 1.14 (1.29)   & 1.14 (1.29)   & 1.15 (1.31)   & 1.18 (1.31)   & 1.18 (1.32)    & 1.20 (1.35)    & 1.22 (1.38)    & 1.22 (1.38)    & 1.23 (1.38)    & 1.24 (1.39)    \\
                      & WSPD                                    & 0.28 (0.34)   & 0.29 (0.34)  & 0.29 (0.34)   & 0.30 (0.34)   & 0.31 (0.35)   & 0.31 (0.36)   & 0.31 (0.36)    & 0.32 (0.37)    & 0.33 (0.40)    & 0.33 (0.40)    & 0.33 (0.40)    & 0.33 (0.40)    \\
                      & PRES                                    & 0.14 (0.19)   & 0.14 (0.22)  & 0.14 (0.24)   & 0.14 (0.24)   & 0.15 (0.24)   & 0.15 (0.26)   & 0.15 (0.26)    & 0.17 (0.26)    & 0.17 (0.29)    & 0.17 (0.30)    & 0.17 (0.31)    & 0.18 (0.32)    \\ \hline
\multirow{4}{*}{\normalsize ELST} & TEMP                                    & 0.38 (0.65)   & 0.41 (0.66)  & 0.48 (0.66)   & 0.49 (0.67)   & 0.58 (0.67)   & 0.59 (0.68)   & 0.63 (0.68)    & 0.63 (0.68)    & 0.64 (0.68)    & 0.64 (0.69)    & 0.65 (0.70)    & 0.65 (0.70)    \\
                      & HUMI                                    & 0.65 (0.76)   & 0.66 (0.76)  & 0.66 (0.76)   & 0.68 (0.77)   & 0.68 (0.78)   & 0.69 (0.78)   & 0.70 (0.80)    & 0.71 (0.80)    & 0.72 (0.80)    & 0.72 (0.80)    & 0.74 (0.81)    & 0.75 (0.81)    \\
                      & WSPD                                    & 0.55 (0.55)   & 0.55 (0.56)  & 0.55 (0.56)   & 0.55 (0.56)   & 0.55 (0.56)   & 0.55 (0.56)   & 0.55 (0.56)    & 0.55 (0.56)    & 0.55 (0.56)    & 0.55 (0.56)    & 0.55 (0.57)    & 0.55 (0.57)    \\
                      & PRES                                    & 0.10 (0.12)   & 0.11 (0.12)  & 0.11 (0.14)   & 0.11 (0.14)   & 0.11 (0.14)   & 0.11 (0.14)   & 0.11 (0.15)    & 0.12 (0.15)    & 0.12 (0.15)    & 0.12 (0.16)    & 0.12 (0.18)    & 0.12 (0.18)    \\ \hline
\multirow{4}{*}{\normalsize HUEY} & TEMP                                    & 0.33 (0.43)   & 0.33 (0.46)  & 0.33 (0.46)   & 0.33 (0.46)   & 0.33 (0.47)   & 0.34 (0.48)   & 0.35 (0.50)    & 0.36 (0.52)    & 0.38 (0.53)    & 0.39 (0.54)    & 0.41 (0.55)    & 0.42 (0.55)    \\
                      & HUMI                                    & 0.97 (1.16)   & 0.97 (1.18)  & 1.03 (1.23)   & 1.03 (1.24)   & 1.03 (1.26)   & 1.05 (1.29)   & 1.06 (1.31)    & 1.07 (1.33)    & 1.09 (1.33)    & 1.09 (1.36)    & 1.11 (1.39)    & 1.13 (1.43)    \\
                      & WSPD                                    & 0.34 (0.34)   & 0.34 (0.34)  & 0.34 (0.35)   & 0.34 (0.35)   & 0.34 (0.35)   & 0.34 (0.35)   & 0.34 (0.35)    & 0.34 (0.35)    & 0.34 (0.35)    & 0.34 (0.35)    & 0.34 (0.35)    & 0.34 (0.35)    \\
                      & PRES                                    & 0.18 (0.18)   & 0.18 (0.18)  & 0.18 (0.19)   & 0.18 (0.20)   & 0.18 (0.21)   & 0.18 (0.21)   & 0.18 (0.21)    & 0.18 (0.21)    & 0.18 (0.22)    & 0.18 (0.22)    & 0.18 (0.22)    & 0.18 (0.23)    \\ \hline
\multirow{4}{*}{\normalsize LXGN} & TEMP                                    & 0.22 (0.24)   & 0.22 (0.24)  & 0.22 (0.24)   & 0.22 (0.25)   & 0.22 (0.25)   & 0.23 (0.26)   & 0.23 (0.27)    & 0.23 (0.30)    & 0.23 (0.31)    & 0.23 (0.31)    & 0.23 (0.33)    & 0.23 (0.35)    \\
                      & HUMI                                    & 1.00 (1.12)   & 1.01 (1.13)  & 1.05 (1.15)   & 1.05 (1.16)   & 1.06 (1.18)   & 1.07 (1.18)   & 1.07 (1.20)    & 1.08 (1.21)    & 1.08 (1.22)    & 1.09 (1.23)    & 1.09 (1.25)    & 1.10 (1.29)    \\
                      & WSPD                                    & 0.52 (0.60)   & 0.53 (0.62)  & 0.53 (0.63)   & 0.53 (0.64)   & 0.54 (0.64)   & 0.58 (0.66)   & 0.58 (0.67)    & 0.59 (0.67)    & 0.56 (0.67)    & 0.56 (0.67)    & 0.58 (0.68)    & 0.59 (0.68)    \\
                      & PRES                                    & 0.15 (0.19)   & 0.15 (0.19)  & 0.15 (0.19)   & 0.15 (0.19)   & 0.16 (0.21)   & 0.16 (0.21)   & 0.16 (0.21)    & 0.16 (0.21)    & 0.16 (0.23)    & 0.16 (0.23)    & 0.16 (0.23)    & 0.16 (0.23)    \\ \hline
\multirow{4}{*}{\normalsize FARM} & TEMP                                    & 0.18 (0.29)   & 0.19 (0.29)  & 0.20 (0.32)   & 0.20 (0.33)   & 0.21 (0.34)   & 0.21 (0.34)   & 0.22 (0.34)    & 0.23 (0.34)    & 0.24 (0.34)    & 0.26 (0.34)    & 0.26 (0.34)    & 0.27 (0.34)    \\
                      & HUMI                                    & 1.98 (2.31)   & 2.00 (2.34)  & 2.02 (2.34)   & 2.03 (2.37)   & 2.06 (2.37)   & 2.07 (2.37)   & 2.08 (2.37)    & 2.10 (2.38)    & 2.13 (2.38)    & 2.20 (2.39)    & 2.26 (2.41)    & 2.28 (2.41)    \\
                      & WSPD                                    & 0.36 (0.39)   & 0.36 (0.39)  & 0.36 (0.41)   & 0.37 (0.41)   & 0.37 (0.41)   & 0.37 (0.41)   & 0.37 (0.41)    & 0.37 (0.41)    & 0.37 (0.42)    & 0.37 (0.42)    & 0.37 (0.42)    & 0.38 (0.42)    \\
                      & PRES                                    & 0.08 (0.13)   & 0.08 (0.13)  & 0.08 (0.14)   & 0.08 (0.15)   & 0.08 (0.17)   & 0.08 (0.17)   & 0.09 (0.17)    & 0.10 (0.17)    & 0.10 (0.18)    & 0.10 (0.18)    & 0.11 (0.19)    & 0.13 (0.19)    \\ \hline
\multirow{4}{*}{\normalsize DANV} & TEMP                                    & 0.21 (0.24)   & 0.22 (0.24)  & 0.22 (0.25)   & 0.22 (0.25)   & 0.22 (0.25)   & 0.23 (0.25)   & 0.23 (0.26)    & 0.23 (0.26)    & 0.23 (0.26)    & 0.23 (0.26)    & 0.24 (0.26)    & 0.24 (0.26)    \\
                      & HUMI                                    & 0.90 (1.05)   & 0.92 (1.05)  & 0.94 (1.05)   & 0.95 (1.05)   & 0.96 (1.05)   & 0.99 (1.06)   & 0.99 (1.08)    & 1.01 (1.10)    & 1.03 (1.11)    & 1.03 (1.11)    & 1.04 (1.13)    & 1.04 (1.13)    \\
                      & WSPD                                    & 0.52 (0.53)   & 0.52 (0.53)  & 0.52 (0.53)   & 0.52 (0.53)   & 0.52 (0.53)   & 0.52 (0.53)   & 0.52 (0.53)    & 0.52 (0.53)    & 0.52 (0.53)    & 0.52 (0.53)    & 0.52 (0.53)    & 0.52 (0.54)    \\
                      & PRES                                    & 0.12 (0.30)   & 0.13 (0.32)  & 0.14 (0.34)   & 0.15 (0.36)   & 0.16 (0.36)   & 0.16 (0.38)   & 0.18 (0.39)    & 0.20 (0.40)    & 0.22 (0.41)    & 0.26 (0.42)    & 0.28 (0.42)    & 0.30 (0.44)    \\ \hline
\multirow{4}{*}{\normalsize BMTN} & TEMP                                    & 0.25 (0.29)   & 0.26 (0.29)  & 0.27 (0.32)   & 0.27 (0.32)   & 0.28 (0.32)   & 0.28 (0.32)   & 0.28 (0.32)    & 0.28 (0.34)    & 0.28 (0.33)    & 0.29 (0.34)    & 0.29 (0.35)    & 0.29 (0.36)    \\
                      & HUMI                                    & 1.73 (1.85)   & 1.73 (1.85)  & 1.73 (1.90)   & 1.73 (1.92)   & 1.73 (1.93)   & 1.73 (1.96)   & 1.73 (1.98)    & 1.74 (1.99)    & 1.74 (2.02)    & 1.76 (2.04)    & 1.82 (2.05)    & 1.83 (2.09)    \\
                      & WSPD                                    & 0.45 (0.47)   & 0.46 (0.47)  & 0.46 (0.48)   & 0.46 (0.48)   & 0.46 (0.48)   & 0.46 (0.49)   & 0.46 (0.49)    & 0.46 (0.50)    & 0.46 (0.50)    & 0.47 (0.50)    & 0.47 (0.50)    & 0.47 (0.51)    \\
                      & PRES                                    & 0.07 (0.42)   & 0.09 (0.46)  & 0.08 (0.53)   & 0.09 (0.56)   & 0.09 (0.59)   & 0.09 (0.60)   & 0.10 (0.62)    & 0.16 (0.67)    & 0.21 (0.71)    & 0.27 (0.72)    & 0.34 (0.78)    & 0.39 (0.84)    \\ \hline
\end{tabular}
}
\label{pointwise_error_lt}
\end{table*}

As the weather parameters have different units and reading ranges (see Table ~\ref{parameters}), the prediction errors normalized with respect to the values themselves (in \%) are included in pairs of parentheses in Table ~\ref{model_errors}, providing scale-independent measures for comparing prediction quality across parameter types. Under the scale-independent normalized error amounts, the WSPD column usually contains large \% values, because the WSPD prediction tends to be less accurate and the WSPD itself typically is low, whereas the PRES column all has very small \% values due mainly to pressure readings in the large range of 600 - 1060 MB (see Table \ref{parameters}). From the RMSE metric results listed in Table ~\ref{model_errors}, our MiMa model consistently outperforms its best counterpart (the Micro or SARIMA model) in all 44 cases, except five (i.e., three for the Micro model and two for the SARIMA model as underlined in Table \ref{model_errors}). When all four predicted parameters at each station are taken into aggressive consideration, the MiMa model outperforms its Micro counterpart markedly under RMSE, e.g. 40.5\% at LSML and 23.4\% at ELST. Likewise, the MiMa model enjoys 93.5\% better PRES prediction over all eleven stations aggregately against the Micro model in terms of the RMSE metric, followed by 25.1\% on TEMP prediction. Similarly, large gaps exist between the MiMa model and its Micro counterpart under MAE. Note that RMSE and MAE comparative values for 17 additional Kentucky Mesonet stations, whose observed parameter data are complete, are listed in Table \ref{extra_model_errors} (see Appendix).

\vspace{1em}

\noindent {\bf \em Comparative Prediction Outcomes for 24 Hours.} To illustrate the prediction details of the four meteorological parameters continuously over time under different models, we randomly select one day in the third season of 202 to forecast its weather parameter values at Station BMTN, starting from 00:00 am to 11:59 pm. Comparative prediction results obtained from the MiMa model, the Micro model, and the WRF-HRRR numerical computation together with near-ground observational readings, are depicted in Figs. \ref{24_hour_predictions}(a), \ref{24_hour_predictions}(b), \ref{24_hour_predictions}(c), and \ref{24_hour_predictions}(d), respectively for temperature, humidity, wind speed, and pressure. It is clear from the figures that the curves of our MiMa model are always closest to those of observational readings for all four weather parameters examined. This demonstrates that our model continuously provides the best prediction results throughout the entire duration (of 24 hours) for all parameters, in comparison to its counterparts. Note that the WRF-HRRR computed parameter values, despite readily available for all geo-grids of the whole US, are substantially far away from the near-surface observations for most of the examined duration, signifying their apparent inadequacies in practical applications.

\vspace{1em}

\vspace{0.3em}
\noindent {\bf \em Performance of MiMa Modelets under 1-hour and 4-hour Lead Times.} The developed MiMa modelets can provide fine predictions temporally for any horizon flexibly with solid accuracy, provided that they are trained with data over an adequate look-back time window (typically to equal the prediction horizon). We have obtained the experimental outcomes of MiMa modelets when forecasting four weather variables of interest in the fine temporal granularity (of 15 minutes) over a 3-hour horizon to involve 12 consecutive prediction points. Such experimental results indicate short-term parameter forecasting effectiveness, particularly relevant to nowcasting that supports the real-world socioeconomic needs of many sectors that rely on weather-oriented decision-making~\cite{xingjian15::convolutional, mass12:BAMS:nowcasting, ravuri21:Nature:skillful}. The prediction errors in RMSE of MiMa modelets under 1-hour and 4-hour lead times with a 15-minute interval over the following 3 hours for all stations are listed in Table \ref{pointwise_error_lt}. They rise gradually at fairly slow paces as prediction progresses from the first time point (at 15 min.) to the 12\textsuperscript{th} time point (at 180 min.) over the prediction horizon for all four weather parameters. The table also reveals that larger lead times yield higher prediction errors, as expected. Overall, the proposed MiMa modelets offer very precise short-term weather parameter forecasting for small to medium lead times, with the prediction of humidity being less accurate in general. 

\begin{figure}
    \centering
    \includegraphics[width=\columnwidth]{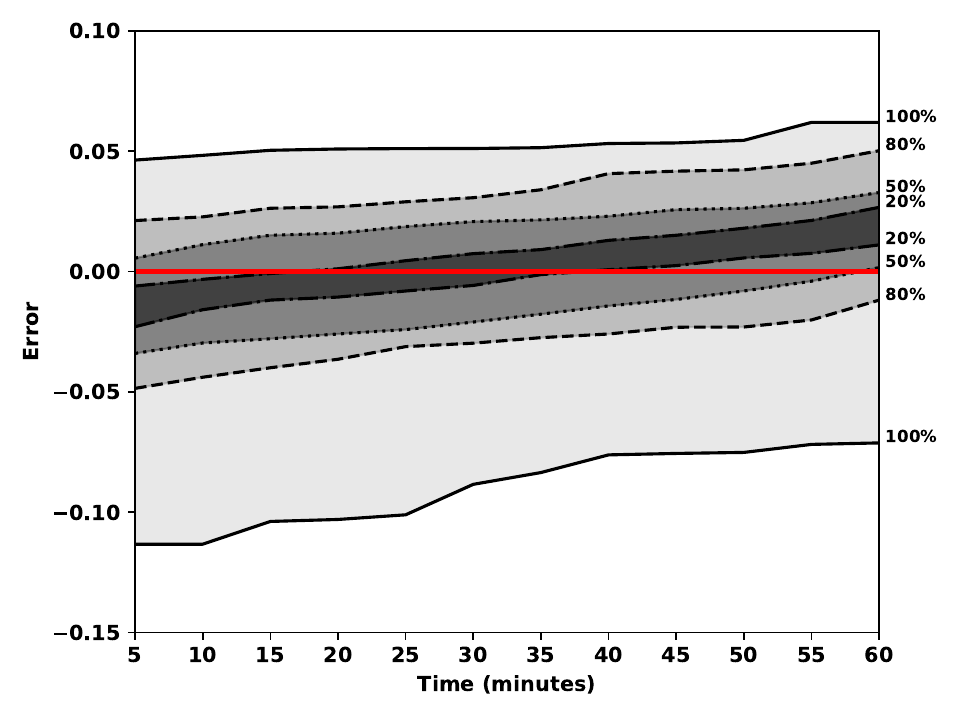}
    \caption{\small Ensemble temperature prediction plot of MiMa modelet for Station FARM.}
    \label{fig:ensemble}
\end{figure}

\subsection{Ensemble Predictions}

Weather parameter value predictions, like any data time series forecasting, come with uncertainty for predicting values over a given time duration, say one hour. Probabilistic weather forecasts are often adopted to quantify uncertainty by post-processing results obtained under various predictors or for different time sans, realizing ensemble predictions \cite{51rasp2018neural}, \cite{52scher2021ensemble}. While the ensemble approach to general forecasting is reviewed in \cite{54wu2021ensemble}, a data-driven method based on neural networks is considered specifically for processing the outcomes of the weather predictor to learn their distribution that enables probabilistic forecasts, with low computational complexity \cite{51rasp2018neural}. Meanwhile, transforming a deterministic ML-based weather forecaster into an ensemble model by extending the forecaster to incorporate probabilistic information is studied \cite{52scher2021ensemble}. Lately, ML algorithms have been employed to learn the statistical properties of prediction outcomes under a given deterministic forecast \cite{53brecht2023computing}, like our proposed MiMa modeling.

Our ensemble predictions are realized by post-processing the outcomes of MiMa modelets to get probabilistic weather forecasts. Instead of what has been the case so far by presenting forecasting results in averaged values, probabilistic forecasting for a MiMa modelet over a duration can be derived by analyzing the modelet's predictions outcomes over the duration. Taking the MiMa modelet for predicting the air temperature of Station FARM over the 1-hour horizon at the 5-minute interval (for 12 prediction values per hour) with nil lead time as an example. Its ensemble prediction results for one hour versus prediction errors are illustrated in Fig. \ref{fig:ensemble}, with 4 probabilistic confidence levels (100\%, 80\%, 50\%, and 20\%) marked. They are obtained by examining the modeled prediction values for a number of random days, each involving 24 sets of 12 predicted values, one for a prediction time point. Those predicted values at each time point of an hour are compared with their corresponding observational temperature readings, to get their prediction errors. The modelets' prediction confidence level at a time point is obtained from the distribution of all errors for the time point.

The pair of solid curves in Fig. \ref{fig:ensemble} denote the ranges that errors always (100\%) fall for the 12 prediction time points (of the 1-hour horizon). They signify the largest errors above and below the observational readings, whose errors are always zero as marked by the bold red line on the x-axis. At the time point of 5 minutes (or 60 minutes), the MiMa modelet is sure to have its prediction errors of no more than +0.049$^{\circ}$C and -0.12$^{\circ}$C (or +0.06$^{\circ}$C and -0.075$^{\circ}$C) at Station FARM. Likewise, the two dotted curves indicate that the MiMa modelet is confident with 50\% to have its temperature prediction errors of no more than +0.01$^{\circ}$C and -0.035$^{\circ}$C (or +0.045$^{\circ}$C and -0.0$^{\circ}$C) at the time point of 5 minutes (or 60 minutes).

A similar plot can be generated for each MiMa modelet, so can its Micro and SARIMA counterparts, enabling the comparison of their prediction error ranges under varying confidence levels at every time point. Since MiMa modelets usually exhibit the most accurate predictions on an average according to their comparative results shown in Table \ref{model_errors} and Fig. \ref{24_hour_predictions}, they are to yield narrower prediction error ranges (i.e., smallest prediction value variations) for a given confidence level at every time point.

\begin{table*}[t]
	\centering
	\caption{\small Mean RMSE and MAE values of ablation study variants for the hourly forecasting horizon over prediction time points in 16 chosen days, with the best ones shown in bold}
	\vspace{-0.5em}
    \renewcommand{\arraystretch}{1.0} 
	
	\begin{tabular}{l|l|cc|cc|cc|cc}
	    \hline
		\multirow{2}{*}{\textbf{\normalsize Station}} \hspace{2em} & \multirow{2}{*}{\textbf{\normalsize Model}} \hspace{2em} & \multicolumn{2}{c|}{\textbf{TEMP}}   & \multicolumn{2}{c|}{\textbf{HUMI}}  & \multicolumn{2}{c|}{\textbf{WSPD}}   & \multicolumn{2}{c}{\textbf{PRES}}    \\ \cline{3-10} 
												&       								& \textbf{RMSE} & \textbf{MAE} & \textbf{RMSE} & \textbf{MAE} & \textbf{RMSE} & \textbf{MAE}  & \textbf{RMSE} & \textbf{MAE}  \\
		\hline
        \multicolumn{1}{l|}{\multirow{3}{*}{\normalsize CCLA}} & MiMa  & 0.28 & 0.24 &{\bf 1.49} & {\bf 0.66} &{\bf 0.38} &{\bf 0.16} &{\bf 0.06} & {\bf 0.03} \\
		\multicolumn{1}{l|}{}                      & Micro & {\bf 0.24} & {\bf 0.22} & 1.86 & 1.43 & 0.38 & 0.24 & 0.16 & 0.04 \\
		\multicolumn{1}{l|}{}                      & Macro & 0.83 & 0.54 & 4.84 & 3.70 & 0.52 & 0.41 & 0.24 & 0.18 \\ \hline
        \multicolumn{1}{l|}{\multirow{3}{*}{\normalsize HUEY}} & MiMa  & {\bf 0.27} &{\bf 0.21} &{\bf 1.56} &{\bf 0.75} & {\bf 0.29} &{\bf 0.22} & {\bf 0.06} & {\bf 0.04} \\
		\multicolumn{1}{l|}{}                      & Micro & 0.30 & 0.58 & 1.63 & 2.64 & 0.31 & 0.25 & 0.08 & 0.52 \\
		\multicolumn{1}{l|}{}                      & Macro & 0.63 & 0.46 & 3.81 & 2.98 & 0.67 & 0.65 & 0.28 & 0.22 \\ \hline
        \multicolumn{1}{l|}{\multirow{3}{*}{\normalsize LXGN}} & MiMa  & {\bf 0.16} &{\bf 0.09} & 1.03 &{\bf 0.52} &{\bf 0.40} & {\bf 0.19} &{\bf 0.06} &{\bf 0.05} \\
		\multicolumn{1}{l|}{}                      & Micro & 0.45 & 0.10 &{\bf 1.02} & 1.28 & 0.40 & 0.24 & 0.06 & 0.21 \\
		\multicolumn{1}{l|}{}                      & Macro & 0.47 & 0.38 & 3.02 & 2.36 & 0.55 & 0.44 & 0.24 & 0.19 \\ \hline
        \multicolumn{1}{l|}{\multirow{3}{*}{\normalsize FARM}} & MiMa  & {\bf 0.19} & {\bf 0.10} &{\bf 1.49} &{\bf 0.61} &{\bf 0.37} &{\bf 0.16} & {\bf 0.06} & {\bf 0.06} \\
		\multicolumn{1}{l|}{}                      & Micro & 0.25 & 0.12 & 1.65 & 0.93 & 0.38 & 0.21 & 0.08 & 0.06 \\
		\multicolumn{1}{l|}{}                      & Macro & 1.00 & 0.81 & 3.67 & 2.88 & 0.58 & 0.44 & 0.25 & 0.19 \\ \hline
		\multicolumn{1}{l|}{\multirow{3}{*}{\normalsize BMTN}} & MiMa  & {\bf 0.20} & {\bf 0.11} & {\bf 1.49} & {\bf 0.75} &{\bf 0.32} &{\bf 0.15} &{\bf 0.08} & {\bf 0.10} \\
		\multicolumn{1}{l|}{}                      & Micro & 0.22 & 0.19 & 1.52 & 1.25 & 0.32 & 0.15 & 0.08 & 0.23 \\
		\multicolumn{1}{l|}{}                      & Macro & 0.59 & 0.41 & 3.68 & 2.52 & 0.66 & 0.53 & 0.34 & 0.28 \\ \hline
	\end{tabular}
 \label{ablation}
\end{table*}

\subsection{Ablation Study}
\label{subsec:ablation}

We have conducted the ablation study to demonstrate the necessity and importance of both Micro and Macro datasets in our MiMa model, by ablating 
(1) the macro components to arrive at the Micro model, which is a counterpart included in the earlier comparative evaluation and 
(2) the micro components to yield the Macro model. For this study, the Macro model is trained and evaluated in the same process as described previously for the MiMa model, but without involving the micro dataset.  
Table~\ref{ablation} lists the RMSE and the MAE values of two MiMa variants for the hourly forecasting horizon at five mesonet stations, averaged over prediction time points in 16 days chosen arbitrarily in the third season of 2020.
It unveils that MiMa outperforms its Micro variant in predicting all four weather parameters at five different mesonet locations under the RMSE accuracy metric, except TEMP at CCLA and HUMI at LXGN (with small margins of 0.04 and 0.01). 
Moreover, MiMa surpasses its Macro variant consistently by substantial gaps for all 20 cases in terms of RMSE.  
Under the MAE metric, the MiMa model outperforms its Micro variant in all cases except TEMP at CCLA (both by the negligible margin of 0.02), and it always outperforms its Macro variant by huge margins for all cases. 
Hence, this study indicates the necessity for the MiMa model to incorporate both ground observational data and atmospheric numerical outputs for superior weather parameter forecasting.

\begin{table}
	\centering
	\caption{\small Mean RMSE and MAE values for predicting extreme weather situations over 12 consecutive time points with 5 minutes apart at Station BMTN, where shrunk data time series are derived from 3rd season of 2018, 2019, and 2020}
	\vspace{-0.5em}
    \renewcommand{\arraystretch}{1.2} 
    \scalebox{0.85}{
	
	\begin{tabular}{l|rr|rr|rr}
		\hline
		& \multicolumn{2}{c|}{\textbf{MiMa}} & \multicolumn{2}{c|}{\textbf{Micro}} & \multicolumn{2}{c}{\textbf{WRF-HRRR}} \\ \cline{2-7} 
		& \textbf{RMSE} & \textbf{MAE} & \textbf{RMSE} & \textbf{MAE} & \textbf{RMSE} & \textbf{MAE} \\
		\hline
		Frigidity  &{\bf 0.63} &{\bf 0.50} & 0.71 & 0.53 & 1.97 & 1.76 \\
		Torridity  &{\bf 0.41} &{\bf 0.31} & 0.45 & 0.37 & 1.28 & 1.06 \\
		Storm      &{\bf 0.36} &{\bf 0.27} & 0.39 & 0.27 & 6.59 & 5.95 \\
		Aridity    &{\bf 8.38} &{\bf 5.49} & 9.44 & 6.01 & 8.41 & 6.35 \\
		Steaminess &{\bf 0.00} &{\bf 0.00} & 0.00 & 0.00 & 12.26 & 8.73 \\
		EHP        &{\bf 0.17} &{\bf 0.14} & 0.25 & 0.19 & 24.81 & 24.81 \\
		\hline
	\end{tabular}}
	
	\label{consecutive}
\end{table}

\subsection{Extreme Weather Forecasting}

It is critical and interesting to evaluate the effectiveness of MiMa modelets in handling extreme weather predictions, which can be more challenging than typical weather forecasting.  
To this end, each time series of real-world weather parameter readings is shrunk to retain only those extreme readings, according to a chosen threshold. 
Here, the threshold is set to be 5\%, signifying that the top 5\% or the bottom 5\% of original data will remain in shrunk time series for evaluation use.  
This article includes only the evaluation results of Station BMTN for one season in 2018, 2019, and 2020, with the 2018 and 2019 shrunk time series for model training.  
Six extreme weather situations are considered, i.e., frigidity, torridity, storm, aridity, steaminess, and EHP, which are assumed to be associated with the lowest temperature, highest temperature, highest wind speed, lowest humidity, highest humidity, and highest pressure, respectively.
Specifically, the shrunk time series composed of the lowest (or highest) 5\% original temperature-relevant data in the third season of 2018 and 2019 are employed to train modelets for predicting frigidity (or torridity), with those of 2020 employed to compute prediction accuracy metrics. 
The temperature-relevant parameter data can be found in Tables~\ref{microparameter} and \ref{WRFparameter}.  
Likewise, the shrunk time series with the lowest (or highest) 5\% original humidity-relevant data are obtained to train modelets for predicting aridity (or steaminess).    
In addition, modelets for predicting storm and EHP are trained respectively by the shrunk data time series of the top 5\% wind speed-relevant data and the top 5\% pressure-relevant data.
For each extreme weather phenomena, we have $2592$ ($=2\times90\times24\times12\times5\%$) training samples, and $1296$ ($=90\times24\times12\times5\%$) testing samples.

\begin{table*}
\centering
\caption{RMSE and MAE values for Re-MiMa at 8 stations not involved in modelet training $^\diamond$}
\renewcommand{\arraystretch}{1.4} 
\scalebox{0.95}{
\begin{tabular}{|c|cc|cc|cc|cc|}
\hline
\multirow{2}{*}{\textbf{\normalsize Station}}                                    & \multicolumn{2}{c|}{\textbf{TEMP}} & \multicolumn{2}{c|}{\textbf{HUMI}} & \multicolumn{2}{c|}{\textbf{WSPD}} & \multicolumn{2}{c|}{\textbf{PRES}} \\ \cline{2-9} 
                                                            & \textbf{RMSE}               & \textbf{MAE}                & \textbf{RMSE}              & \textbf{MAE}              & \textbf{RMSE}               & \textbf{MAE }              & \textbf{RMSE}              & \textbf{MAE}              \\ \hline
\begin{tabular}[c]{@{}c@{}}\normalsize LSML \\  @748 ft.\end{tabular}  & 0.22 (1.05\%)      & 0.16 (0.77\%)      & 0.85 (1.06\%)     & 0.62 (0.77\%)    & 0.25 (25.77\%)     & 0.20 (20.62\%)    & 0.06 (0.01\%)     & 0.06 (0.01\%)    \\ \hline
\begin{tabular}[c]{@{}c@{}}\normalsize CCLA \\  @764 ft.\end{tabular}  & 0.13 (0.61\%)      & 0.12 (0.56\%)      & 0.89 (1.06\%)     & 0.65 (0.77\%)    & 0.22 (11.52\%)     & 0.16 (8.38\%)     & 0.06 (0.01\%)     & 0.05 (0.01\%)    \\ \hline
\begin{tabular}[c]{@{}c@{}}\normalsize LGRN \\  @766 ft.\end{tabular}  & 0.19 (0.91\%)      & 0.15 (0.72\%)      & 0.95 (1.20\%)     & 0.71 (0.90\%)    & 0.23 (11.98\%)     & 0.16 (8.33\%)     & 0.07 (0.01\%)     & 0.06 (0.01\%)    \\ \hline
\begin{tabular}[c]{@{}c@{}}\normalsize FCHV \\  @770 ft.\end{tabular}  & 0.10 (0.47\%)      & 0.08 (0.37\%)      & 0.92 (1.21\%)     & 0.63 (0.83\%)    & 0.20 (9.66\%)      & 0.15 (7.25\%)     & 0.05 (0.01\%)     & 0.04 (0.00\%)    \\ \hline
\begin{tabular}[c]{@{}c@{}}\normalsize CROP \\  @858 ft.\end{tabular}  & 0.12 (0.58\%)      & 0.09 (0.44\%)      & 0.82 (1.05\%)     & 0.61 (0.78\%)    & 0.21 (9.95\%)      & 0.17 (8.06\%)     & 0.07 (0.01\%)     & 0.06 (0.01\%)    \\ \hline
\begin{tabular}[c]{@{}c@{}}\normalsize ELST \\  @860 ft.\end{tabular}  & 0.07 (0.34\%)      & 0.05 (0.24\%)      & 0.68 (0.77\%)     & 0.43 (0.49\%)    & 0.43 (34.13\%)     & 0.40 (31.75\%)    & 0.19 (0.02\%)     & 0.14 (0.01\%)    \\ \hline
\begin{tabular}[c]{@{}c@{}}\normalsize HUEY \\  @896 ft.\end{tabular}  & 0.11 (0.52\%)      & 0.10 (0.47\%)      & 0.97 (1.37\%)     & 0.73 (1.03\%)    & 0.27 (103.85\%)    & 0.26 (100.00\%)   & 0.17 (0.02\%)     & 0.11 (0.01\%)    \\ \hline
\begin{tabular}[c]{@{}c@{}}\normalsize LXGN \\  @1044 ft.\end{tabular} & 0.11 (0.50\%)      & 0.08 (0.36\%)      & 0.61 (0.83\%)     & 0.45 (0.61\%)    & 0.24 (10.53\%)     & 0.18 (7.89\%)     & 0.30 (0.03\%)     & 0.25 (0.03\%)    \\ \hline
\end{tabular}
}
\label{Re-MiMa_errors}
\begin{tablenotes}
    \item $^\diamond$ Those stations are treated as ungauged stations for evaluation purposes. The prediction errors normalized with respect to the values themselves (in \%) are included in pairs of parentheses.
\end{tablenotes}
\end{table*}

MiMa modelets are trained in the same way by shrunk data time series as by original data time series described earlier, for predicting parameter values at 12 subsequent time points 5 minutes apart.  
The forecasting accuracy results of MiMa modelet and their counterparts, with respect to the shrunk data time series derived from the third season of 2020, are given in Table~\ref{consecutive}, where the best ones are bold.  
As can be observed from the table results, MiMa modelets accurately predict all extreme weather situations except one (aridity), with their MAE values staying within 0.50.  

Forecasting extreme low humidity suffers from relatively large inaccuracy, yielding MAE (or RMSE) equal to 5.49 (or 8.38).  When compared with their counterparts (i.e., Micro and WRF-HRRR), MiMa modelets predict all extreme weather situations more precisely, by solid margins.  For example, the MiMa modelet enjoys 8.5\% (or 11.2\%) better aridity prediction under the MAE (or RMSE) metric, than its best counterpart, the Micro model.  
It achieves 26.3\% (or 32.0\%) better EHP prediction than the Micro model in terms of MAE (or RMSE).

    \section{Regional MiMa Modeling}

To enhance the utility of MiMa modelets, we leverage the transfer learning capabilities of modelets by incorporating elevation data into the training process to let MiMa modelets be trained on data from multiple stations. Specifically, we inject elevation-specific knowledge to modelets by adding the elevation as an input parameter to both the Micro and the Macro encoders depicted in Fig. \ref{fig:MiMaED}. This allows the elevation data to be tagged with corresponding Micro and Macro data during encoding, thus correlating Micro and Macro data with the elevation data, giving rise to the development of Re-MiMa (regional MiMa) modeling, whose structure is identical to Fig. \ref{fig:MiMaED}, except for replacing its $X_{micro}$ (or $X_{macro}$) input to the Micro (or Macro) Encoder with $X_{micro_{Re}}$ (or $X_{macro_{Re}})$, which is specified below with the size of $\langle n\times \zeta, \alpha+1\rangle$ (or $\langle n\times \zeta, \beta+1 \rangle$). In general, Re-MiMa modelets are trained using data from a few (say, 3 or 4) representative stations in the region of interest, along with their elevations. The representative stations should cover the elevation range of the region, including those with two extremes and one (or two) in-between elevations. The input data frame of the Micro Encoder comprises parameter readings observed at those $\zeta$ representative stations, together with their corresponding elevations, as
\[
\mathbf{\textit{X}}_{micro_{Re}} = \begin{bmatrix}
\text{SF}(R^1),
\text{SF}(R^2), 
\hdots,
\text{SF}(R^i),
\hdots,
\text{SF}(R^n)
\end{bmatrix}^{T},
\]
where $\text{SF}(R^i)$ indicates the shuffling operation of $R^i$, which signifies those $\zeta$ streams of $\alpha$ most relevant parameters observed at those $\zeta$ representative stations involved in training, plus stations' elevations, at the $i^{th}$ timestamp, for $1 \leq i \leq n$, under the look-back window time window of $n$. The group of $\zeta$ streams of $R^i$ is given by
\[
R^i = \begin{pmatrix}
    P_{1, 1}^{i} & P_{1, 2}^{i} & P_{1, 3}^{i} & \cdots & P_{1, \alpha}^{i} & E_{sta-1} \\[0.3em]
    P_{2, 1}^{i} & P_{2, 2}^{i} & P_{2, 3}^{i} & \cdots & P_{2, \alpha}^{2} & E_{sta-2} \\[0.3em]
    \vdots & \vdots & \vdots & \ddots & \vdots &\vdots \\[0.3em]
    P_{\zeta, 1}^{i} & P_{\zeta, 2}^{i} & P_{\zeta, 3}^{i} & \cdots & P_{\zeta, \alpha}^{i} & E_{sta-\zeta}
\end{pmatrix},
\]
where $E_{sta}$ is the elevation of the station involved in modelet training. Those $\zeta$ streams of ($\alpha + 1$) values in Group $R^i$ for the $i^{th}$ timestamp are shuffled randomly, to yield different orders from one group to the next, expressed by $\text{SF}(R^i)$. This shuffling strategy aims to prevent model training from being biased towards the data observed at any single station, ensuring that data from those $\zeta$ representative stations in a given region are employed for training fairly without criticality. The Micro Encoder has its input size of ($n \times \zeta$) by ($ \alpha + 1$).

Similarly, the input data frame of the Macro Encoder comprises the values of the $\beta$ most relevant parameters obtained by WRF-HRRR computation of the $\zeta$ geo-grids aligned with those representative stations, plus their corresponding elevations, as expressed by
\[
\mathbf{\textit{X}}_{macro_{Re}} = \begin{bmatrix}
\text{SF}(R^1),
\text{SF}(R^2), 
\hdots,
\text{SF}(R^i),
\hdots,
\text{SF}(R^n)
\end{bmatrix}^{T},
\]
where $R^i$ is the same as what is outlined above, except that each row contains $\beta$ most relevant WRF-HRRR parameters. Its size equals ($n \times \zeta$) by ($\beta + 1$).

Training Re-MiMa modelets to cover the whole elevation range of a given region, we can perform inference on stations not included in the training set, making it possible to expand the model's applicability and usability. For performance evaluation, those 8 Kentucky Mesonet stations (out of 11 stations marked in Fig. \ref{fig:KentuckyMesonet}) not involved in modelet training serve to benchmark Re-MiMa modelets, with the results listed in Table \ref{Re-MiMa_errors}. Our evaluation adopts a proximity-based approach, where observational data borrowed from the training station closest in elevation to the target station (which has no observational data but whose corresponding WRF-HRRR data and its elevation information) are used for prediction. As an example, for predicting the CCLA station, which is situated at 764 feet, the observational data from the FARM station (which is involved in modelet training and located at 559 feet), plus its elevation data serve as input to the Micro Encoders of Re-MiMa modelets. Leveraging elevation similarity within the region for high accuracy, this approach lets Re-MiMa modeling generalize well to forecast new (ungauged) locations by borrowing the elevation-closest observational data available in the region, thereby enhancing the robustness and reliability of predicting ungauged locations.

Table \ref{Re-MiMa_errors} lists the mean errors when predicting weather parameters at all time points under the 5-minute granularity over 16 days chosen arbitrarily in the third season of 2020. They demonstrate that Re-MiMa modeling achieves superior prediction performance for stations that are not included in model training. This ability of Re-MiMa modelets stems from transfer learning, enabling accurate forecasts for any location in a given region. For example, the RMSE values of TEMP (or PRES) for all 8 stations are limited to 1.05\% (or 0.03\%). Note that the RMSE value of WSPD at ELST is the largest, equal to 0.43, which translates to $\sim$104\%, implying that the station experiences light to no wind most of the time. Interestingly, the Re-MiMa modelets often outperform their original MiMa counterparts. When comparing Table \ref{model_errors} (for the results of MiMa modelets) and Table \ref{Re-MiMa_errors}, it reveals that Re-MiMa modelets provide better forecasts for 22 cases (out of 32), in terms of the RMSE metric. This is likely due to two key reasons: (1) data from $\zeta$ representative stations are employed for training, instead of data from one single station and (2) training data from those $\zeta$ stations are shuffled randomly, able to effectively mitigate the risk of the model overfitting to the unique characteristics of one particular station. With data from multiple stations shuffled randomly upon training, it not only prevents the modelets from learning station-specific patterns that might not generalize well but also encourages them to recognize broader trends and relationships present across different stations. Re-MiMa modelets are valuable in practice due to their high prediction accuracy at any location (be ungauged or gauged) in a given region based on observational data from just a few representative stations therein.

	\section{Conclusion}

This paper presents a novel machine learning model that integrates ground measurements (the micro dataset) and atmospheric numerical outputs (the macro dataset) for the first time. This model, referred to as the MiMa model, aims to deliver precise, location-specific weather parameter predictions over short-term time horizons in fine resolutions (e.g., 5 or 15 minutes). Utilizing the transformer structure with two encoders and one decoder that all comprise Long Short-Term Memory (LSTM) units, our model effectively captures temporal variations in weather conditions and incorporates two key data sources to forecast relevant weather parameters for each Mesonet station location via a single model instance per parameter, termed a MiMa modelet. Furthermore, transfer learning is leveraged to generalize MiMa modelets for accurately predicting weather variables at any location in a given region, utilizing the observational data of just a few representative stations (often 3 or 4), plus stations' elevations, to train modelets. It arrives at Re-MiMa modelets, with one for each parameter type throughout the whole region.

Experimental results from various Kentucky Mesonet station locations demonstrate that our modelets usually achieve the best meteorological forecasting (for 39 cases out of 44, under the RMSE metric) with fine temporal granularity among all examined models. Furthermore, Re-MiMa modelets are observed to perform as well as, or even better than (in 22 cases out of 32), their location-specific MiMa counterparts, making it possible to reduce the modelet count without compromising forecasting accuracy. The developed Re-MiMa modelets effectively meet the long-standing challenge of precise forecasts at ungauged locations. Providing accurate regional forecasts over short time horizons in the fine temporal resolution (e.g., at 5 or 15 minutes), MiMa and Re-MiMa modelets address the real-world socioeconomic needs of various sectors that rely on real-time, weather-oriented decision-making. They are ready for widespread deployment in any region where near-surface observational data is available for superior forecasting accuracy.

\vspace{-1em}

\section*{Acknowledgement}
\vspace{-0.3em}
This work was supported in part by NSF under Grants 1763620, 1948374, and 2019511.

\flushend
\bibliographystyle{IEEEtran}
\bibliography{main}

\vspace{-2em}
\begin{IEEEbiography}
[{\includegraphics[width=1in,height=1.25in,clip,keepaspectratio]{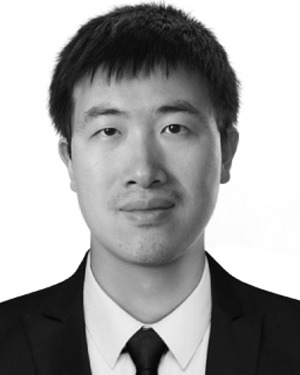}}]
  {Yihe Zhang} 
  (M'24) received M.S. degree from SYSU-CMU Joint Institute of Engineering, Sun-Yat Sen University and Carnegie Mellon University, in 2016. He is currently working toward a Ph.D. degree in the School of Computing and Informatics, University of Louisiana at Lafayette. His research interests include machine learning applications, knowledge graphs, and data-driven security. He received the Best Paper Award and the Distinguished Paper Award from DSN 2023.
\end{IEEEbiography}

\vspace{-4em}

\begin{IEEEbiography}  [{\includegraphics[width=1in,height=1.25in,clip,keepaspectratio]{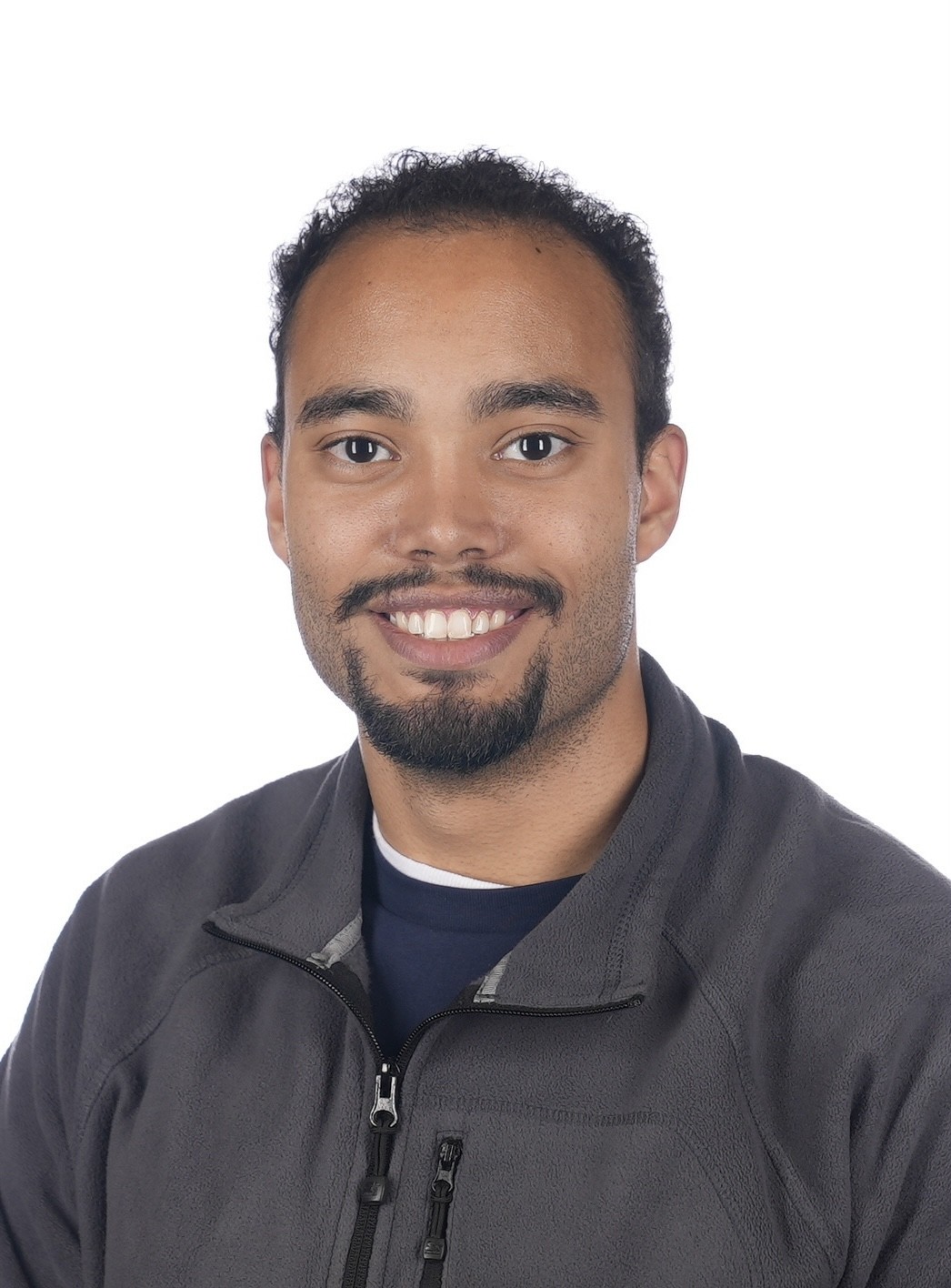}}]
  {Bryce Turney} 
  (S'22) received his B.S. degree in Electrical Engineering from the Department of Electrical and Computer Engineering at the University of Louisiana at Lafayette. He is currently pursuing a Ph.D. degree in Computer Engineering from the Center for Advanced Computer Studies (CACS), at the University of Louisiana at Lafayette. His current areas of interest include machine learning, parallel computing, embedded systems, and autonomous systems.
\end{IEEEbiography}

\vspace{-4em}

\begin{IEEEbiography}
  [{\includegraphics[width=1in,height=1.25in,clip,keepaspectratio]{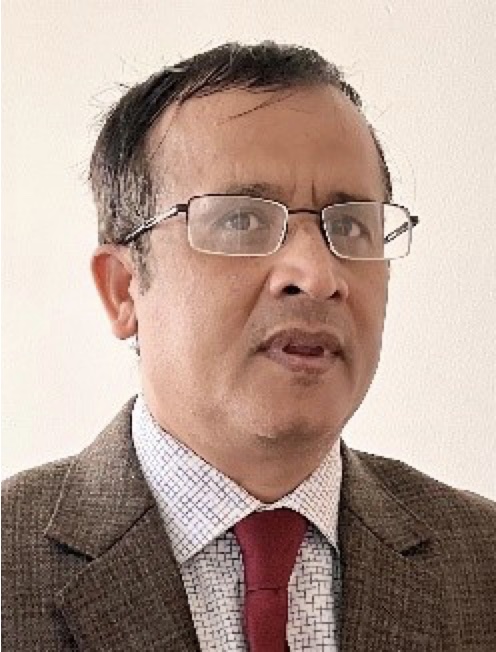}}]
  {Purushottam Sigdel} 
  (M'21) completed his Ph.D. at the Center for Advanced Computer Studies, University of Louisiana at Lafayette, in 2020. Presently, he serves as a Silicon Architecture Engineer for the Data Center Processor Architecture division at Intel, San Jose, CA.  His areas of interest encompass high-performance computer systems, network-on-chip, system-on-chip, parallel and distributed processing, application-aware architectural optimization, and privacy-preserved machine learning.
\end{IEEEbiography}

\vspace{-4em}

\begin{IEEEbiography}
  [{\includegraphics[width=1in,height=1.25in,clip,keepaspectratio]{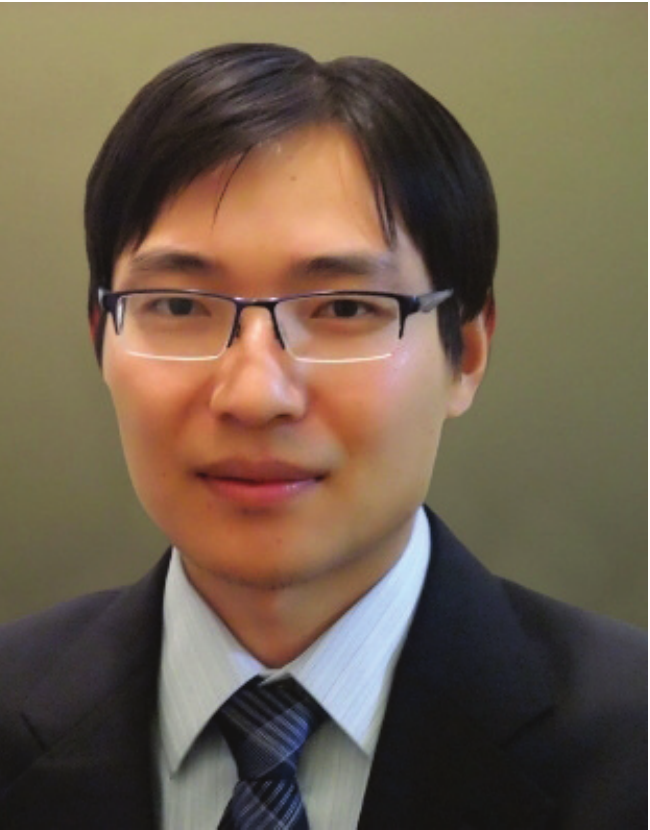}}]
  {Xu Yuan} 
  (M'16-SM'22) received the Ph.D. degree from the Bradley Department of Electrical and Computer Engineering, Virginia Tech, Blacksburg, in 2016. He is an Associate Professor in the Department of Computer and Information Sciences at the University of Delaware, Newark. His research focuses on artificial intelligence (AI), cybersecurity, networking, and cyber-physical systems. He received the Best Paper Award and the Distinguished Paper Award from DSN 2023.
\end{IEEEbiography}

\vspace{-4em}

\begin{IEEEbiography}
  [{\includegraphics[width=1in,height=1.2in,clip,keepaspectratio]{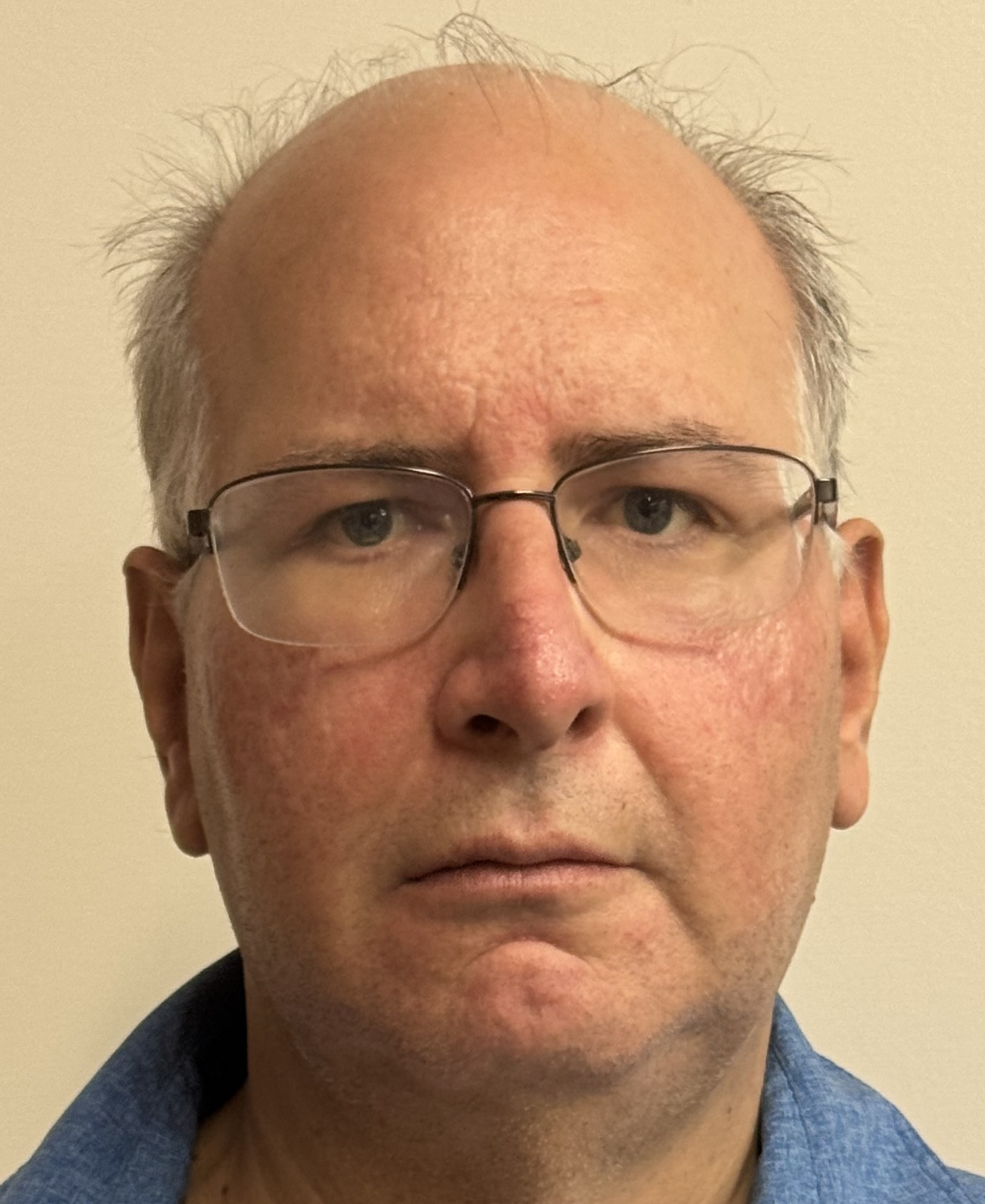}}]
  {Eric Rappin} 
  received his Ph.D. in Atmospheric Sciences from the University of Wisconsin-Madison in 2004.  He currently is the lead researcher at the Kentucky Climate Center at Western Kentucky University.  His research focus is on how land cover and land use transitions impact the way the land surface interacts with the atmosphere through conservation of energy and water.
\end{IEEEbiography}

\vspace{-4em}

\begin{IEEEbiography}
    [{\includegraphics[width=1in,height=1.2in,clip,keepaspectratio]{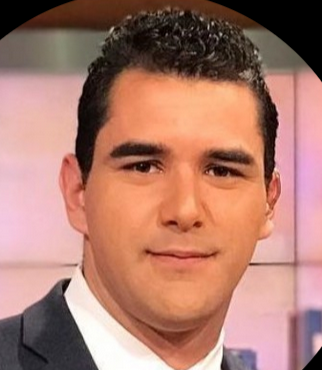}}]
    {Adrian Lago}
    received his Ph.D. in Atmospheric Sciences from Florida International University in 2022. He currently works as a meteorologist at the National Oceanic and Atmospheric Administration (NOAA). His research interests lie in weather forecasting, radar analysis, and remote sensing.
\end{IEEEbiography}

\vspace{-4em}

\begin{IEEEbiography}
  [{\includegraphics[width=1in,height=1.25in,clip,keepaspectratio]{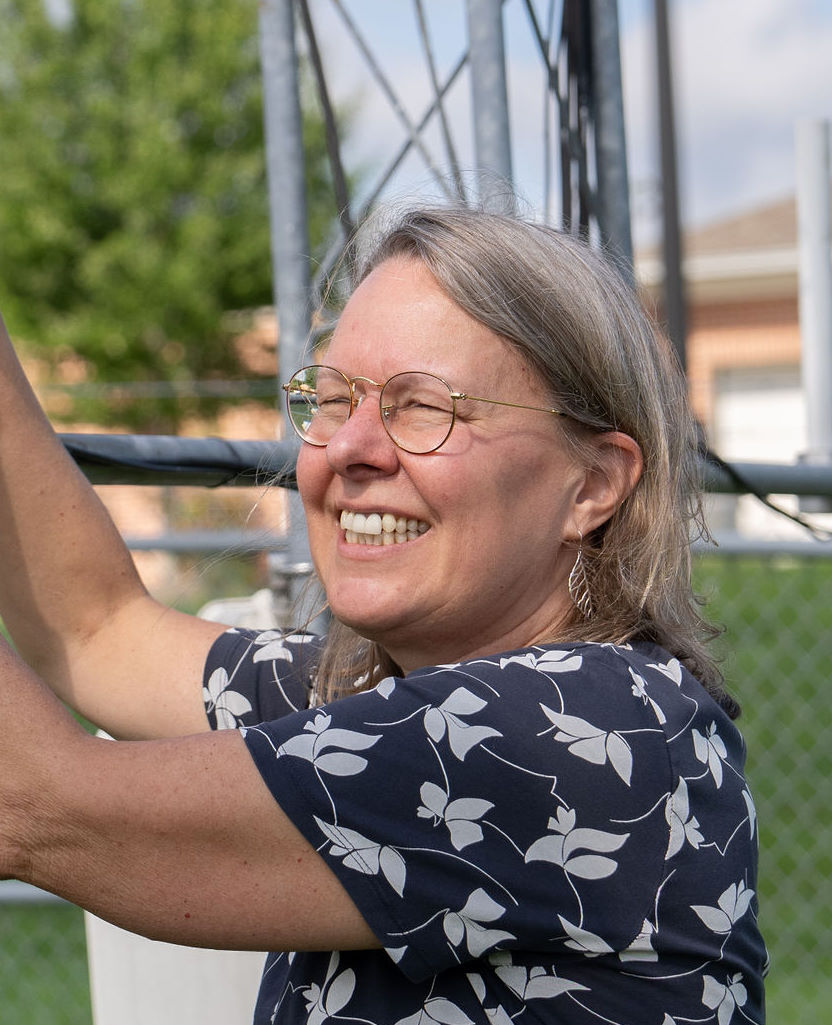}}]
  {Sytske Kimball} 
  received her Ph.D. in Meteorology from Pennsylvania State University in 2000.  She currently is professor of meteorology and serves as department chair of the Department of Earth Sciences at the University of South Alabama. Her research interests include landfalling hurricanes, local weather and weather phenomena including low-level temperature inversions and sea breeze-driven thunderstorm formation.
\end{IEEEbiography}

\vspace{-3em}

\begin{IEEEbiography}
  [{\includegraphics[width=1in,height=1.25in,clip,keepaspectratio]{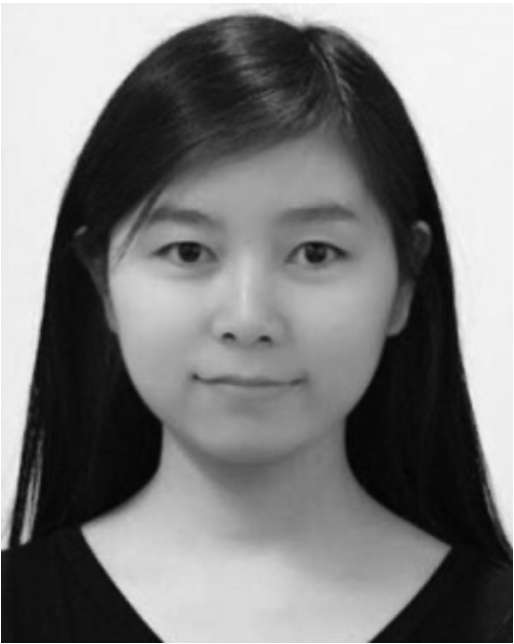}}]
  {Li Chen} 
  (M'18-SM'24) received her Ph.D. degree from the Department of Electrical and Computer Engineering, University of Toronto, in 2018.  She is currently an associate professor with the School of Computing and Informatics, University of Louisiana at Lafayette, USA. Her research interests include big data analytics systems, machine learning systems, cloud computing, datacenter networking, and resource allocation.
\end{IEEEbiography}

\vspace{-3em}

\begin{IEEEbiography}
  [{\includegraphics[width=1in,height=1.25in,clip,keepaspectratio]{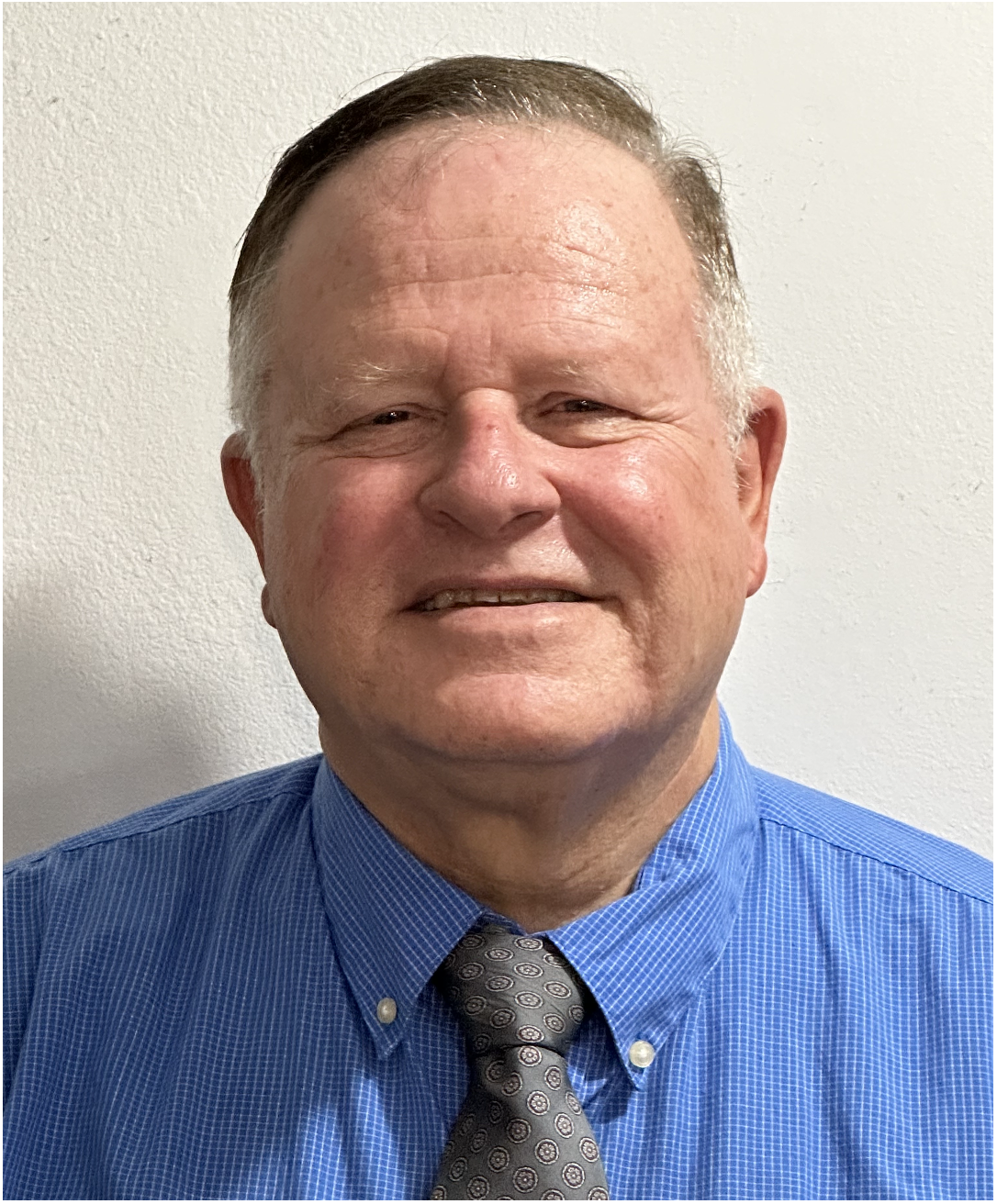}}]
  {Paul Darby} 
  (M'76-SM'15) received his Ph.D. in Computer Engineering from the Center for Advanced Computer Studies (CACS), at the University of Louisiana at Lafayette. Currently he is a faculty member professor of practice with the Department of Electrical and Computer Engineering at the University of Louisiana at Lafayette. His research interests include mobile computing and systems, swarm robotics, and satellite experimental smartphone Grid (ESG-Grid) technologies and engineering.
\end{IEEEbiography}

\vspace{-3em}

\begin{IEEEbiography}
  [{\includegraphics[width=1in,height=1.25in,clip,keepaspectratio]{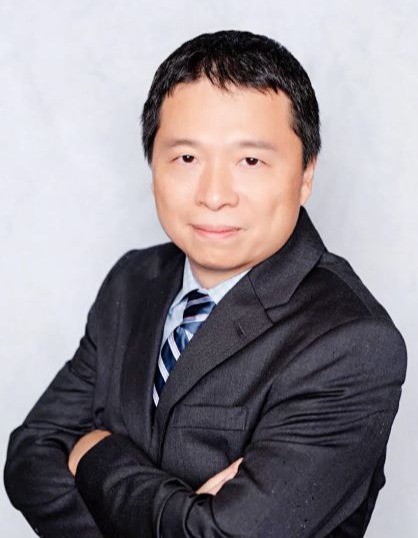}}]
  {Lu Peng} 
  (M'05-SM'12) obtained a Ph.D. degree in Computer Engineering from the University of Florida in 2005.  He is the Yahoo! Founder Chair in Science and Engineering in the Computer Science Department at Tulane University.  His current research focuses on computer architecture, reliability, and big data analytics. He received the Best Paper Award from both IGSC 2019 and ICCD 2001 (in Processor Architecture Track).
\end{IEEEbiography}

\vspace{-3em}

\begin{IEEEbiography}
  [{\includegraphics[width=1in,height=1.25in,clip,keepaspectratio]{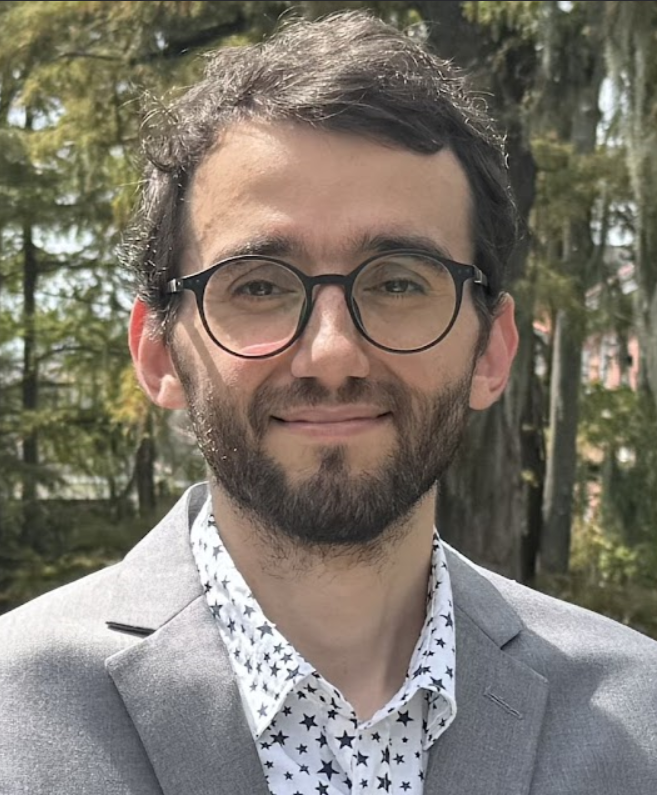}}]
  {Sercan Aygun} 
  (M'22-SM'24) received his Ph.D. in electronics engineering from Istanbul Technical University in 2022. Currently, he is an Assistant Professor at the University of Louisiana, Lafayette. His research focuses on emerging computing technologies. He has received multiple accolades, including Best Scientific Research Award (ACM SIGBED ESWEEK 2022), Best Paper (GLSVLSI’23), Best Poster (GLSVLSI’24), TESID’s Best Scientific Application Ph.D. Award, and first place in TUBA’s Science and Engineering PhD Thesis Awards.
\end{IEEEbiography}

\vspace{-3em}

\begin{IEEEbiography}
  [{\includegraphics[width=1in,height=1.25in,clip,keepaspectratio]{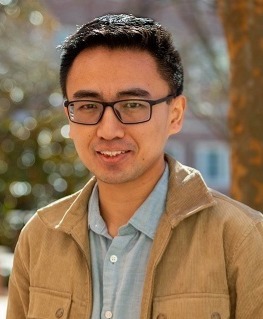}}]
  {Yazhou Tu} 
  (M’19) received his Ph.D. in Computer Science from the University of Louisiana at Lafayette. He is currently an Assistant Professor at the Computer Science and Software Engineering Department, at Auburn University. His research interests include cyber-physical security, side channels, sensing security, and embedded systems. He received the Distinguished Paper Award from IEEE S\&P 2024.
\end{IEEEbiography}

\vspace{-3em}

\begin{IEEEbiography}
  [{\includegraphics[width=1in,height=1.25in,clip,keepaspectratio]{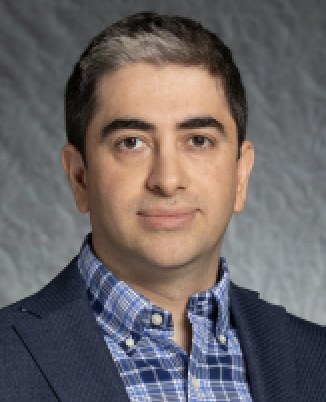}}]
  {M. Hassan Najafi} 
  (M'18-SM'23) received his Ph.D. degree in Electrical and Electronics Engineering from the University of Minnesota-Twin Cities in 2018. He is an Assistant Professor at the School of Computing and Informatics, University of Louisiana at Lafayette.  His research includes stochastic and approximate computing, unary processing, in-memory computing, and hyperdimensional computing. He received the Best Paper Award from ICCD 2017 and GLSVLSI 2023.
\end{IEEEbiography}

\vspace{-3em}

\begin{IEEEbiography}
  [{\includegraphics[width=1in,height=1.25in,clip,keepaspectratio]{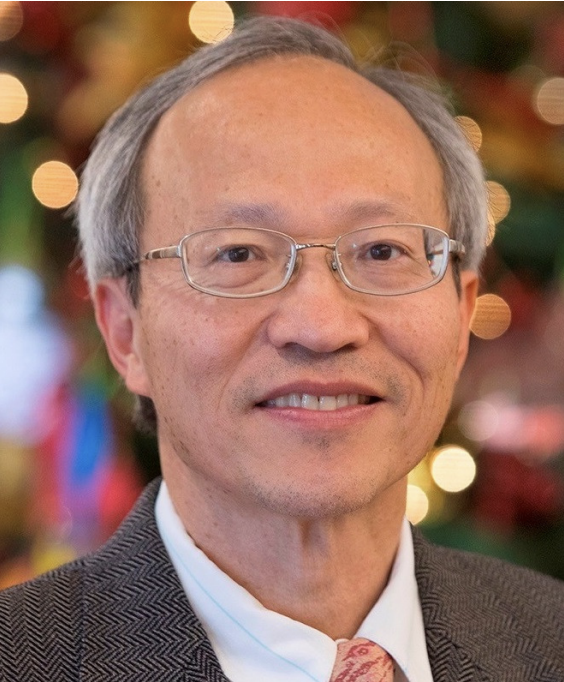}}]
  {Nian-Feng Tzeng} 
  (M'86-SM'92-F'10-LF'22) received his Ph.D. degree in Computer Science from the University of Illinois, Urbana-Champaign in 1986.  He has been with the Center for Advanced Computer Studies, School of Computing and Informatics, the University of Louisiana at Lafayette, since 1987.  His current research interest includes high-performance computer systems, computer networks, and parallel and distributed processing.
\end{IEEEbiography}

\clearpage

\onecolumn
\appendices

\section*{Appendix}

\begin{table*}[h]
\caption{\footnotesize RMSE and MAE values for 17 additional mesonet stations in the Kentucky region along with their respective errors $^{\S\S}$}
\renewcommand{\arraystretch}{1.3} 
\centering
\scalebox{0.75}{
\begin{tabular}{|c|l|cc|cc|cc|cc|}
\hline
\multirow{2}{*}{\textbf{\normalsize Station}}                                                    & \multicolumn{1}{c|}{\multirow{2}{*}{\textbf{\normalsize Model}}} & \multicolumn{2}{c|}{\textbf{\normalsize TEMP}}         & \multicolumn{2}{c|}{\textbf{\normalsize HUMI}}            & \multicolumn{2}{c|}{\textbf{\normalsize WSPD}}             & \multicolumn{2}{c|}{\textbf{\normalsize PRES}}            \\ \cline{3-10} 
                                                                            & \multicolumn{1}{c|}{}                       & \textbf{RMSE}                   & \textbf{MAE}                    & \textbf{RMSE}                   & \textbf{MAE}                    & \textbf{RMSE}                     & \textbf{MAE}                     & \textbf{RMSE}                   & \textbf{MAE}                    \\ \hline
\multirow{7}{*}{\begin{tabular}[c]{@{}c@{}}\normalsize HCKM \vspace{0.5em}\\ \normalsize @345 ft.\end{tabular}}  
& MiMa & \textbf{0.12 (0.65\%)} & \textbf{0.09 (0.11\%)} & \textbf{0.72 (0.91\%)} & \textbf{0.55 (0.68\%)} & \textbf{0.34 (35.00\%)} & \textbf{0.23 (24.20\%)} & \textbf{0.10 (0.01\%)} & \textbf{0.08 (0.01\%)} \\
& Micro & 0.15 (0.74\%) & 0.12 (0.15\%) & 1.05 (1.32\%) & 0.82 (1.02\%) & 0.34 (35.00\%) & 0.26 (26.00\%) & 0.28 (0.04\%) & 0.32 (0.02\%) \\
& SARIMA & 0.26 (1.30\%) & 0.16 (0.20\%) & 2.10 (2.65\%) & 1.25 (1.55\%) & 0.50 (51.50\%) & 0.34 (34.00\%) & 0.74 (0.07\%) & 0.60 (0.05\%) \\
& WRF-HRRR & 1.00 (4.90\%) & 0.80 (1.00\%) & 6.50 (8.20\%) & 5.30 (6.60\%) & 4.50 (460.00\%) & 4.00 (410.00\%) & 3.70 (0.38\%) & 3.75 (0.38\%) \\
& SNN & 2.84 (11.28\%) & 2.11 (8.38\%) & 7.56 (11.26\%) & 4.96 (7.39\%) & 0.84 (5.69\%) & 0.58 (3.96\%) & 1.00 (5.57\%) & 0.63 (3.53\%) \\
& SVR & 3.53 (14.03\%) & 2.62 (10.44\%) & 8.91 (13.27\%) & 6.66 (9.92\%) & 0.97 (6.62\%) & 0.73 (4.96\%) & 1.60 (8.90\%) & 1.09 (6.05\%) \\
& DUQ & 3.09 (20.58\%) & 2.69 (17.93\%) & 11.59 (20.05\%) & 8.75 (15.14\%) & 0.90 (8.08\%) & 0.70 (6.24\%) & 0.64 (5.86\%) & 0.44 (4.08\%) \\
\hline

\multirow{7}{*}{\begin{tabular}[c]{@{}c@{}}\normalsize PVRT \vspace{0.5em}\\ \normalsize @404 ft.\end{tabular}}  
& MiMa & \textbf{0.13 (0.66\%)} & \textbf{0.09 (0.11\%)} & \textbf{0.75 (0.94\%)} & \textbf{0.56 (0.69\%)} & \textbf{0.35 (36.00\%)} & 0.27 (27.50\%) & \textbf{0.08 (0.00\%)} & \textbf{0.05 (0.00\%)} \\
& Micro & 0.16 (0.75\%) & 0.12 (0.15\%) & 1.07 (1.33\%) & 0.83 (1.04\%) & 0.35 (36.00\%) & \underline{\textbf{0.27 (27.00\%)}} & 0.39 (0.03\%) & 0.33 (0.03\%) \\
& SARIMA & 0.27 (1.32\%) & 0.17 (0.20\%) & 2.15 (2.68\%) & 1.27 (1.58\%) & 0.50 (51.50\%) & 0.34 (34.50\%) & 0.75 (0.07\%) & 0.62 (0.06\%) \\
& WRF-HRRR & 1.02 (4.95\%) & 0.81 (1.01\%) & 6.60 (8.25\%) & 5.40 (6.70\%) & 4.60 (475.00\%) & 4.05 (420.00\%) & 3.80 (0.38\%) & 3.78 (0.38\%) \\
& SNN & 2.24 (8.41\%) & 1.38 (5.20\%) & 7.87 (10.96\%) & 4.40 (6.13\%) & 0.82 (6.54\%) & 0.55 (4.36\%) & 0.93 (4.99\%) & 0.59 (3.17\%) \\
& SVR & 3.42 (12.88\%) & 2.55 (9.60\%) & 10.58 (14.74\%) & 7.79 (10.85\%) & 1.12 (8.94\%) & 0.89 (7.09\%) & 1.84 (9.85\%) & 1.27 (6.79\%) \\
& DUQ & 1.90 (10.82\%) & 1.35 (7.68\%) & 8.99 (15.31\%) & 6.28 (10.70\%) & 1.03 (9.53\%) & 0.77 (7.18\%) & 2.84 (27.33\%) & 2.57 (24.68\%) \\
\hline

\multirow{7}{*}{\begin{tabular}[c]{@{}c@{}}\normalsize WDBY \vspace{0.5em}\\ \normalsize @421 ft.\end{tabular}}  
& MiMa & \textbf{0.11 (0.60\%)} & 0.10 (0.14\%) & \textbf{0.70 (0.85\%)} & \textbf{0.53 (0.65\%)} & \textbf{0.33 (34.50\%)} & 0.25 (25.40\%) & 0.37 (0.03\%) & 0.31 (0.02\%) \\
& Micro & 0.15 (0.72\%) & \underline{\textbf{0.08 (0.12\%)}} & 1.02 (1.30\%) & 0.80 (1.00\%) & 0.34 (34.50\%) & \underline{\textbf{0.25 (25.30\%)}} & 0.73 (0.07\%) & 0.59 (0.05\%) \\
& SARIMA & 0.26 (1.30\%) & 0.16 (0.19\%) & 2.05 (2.60\%) & 1.22 (1.53\%) & 0.49 (50.80\%) & 0.33 (33.20\%) & \underline{\textbf{0.10 (0.01\%)}} & \underline{\textbf{0.09 (0.01\%)}} \\
& WRF-HRRR & 1.00 (4.90\%) & 0.79 (1.00\%) & 6.45 (8.15\%) & 5.25 (6.60\%) & 4.50 (460.00\%) & 4.00 (410.00\%) & 3.70 (0.37\%) & 3.75 (0.37\%) \\
& SNN & 2.61 (9.76\%) & 1.70 (6.35\%) & 8.05 (11.95\%) & 4.56 (6.76\%) & 0.98 (8.50\%) & 0.67 (5.80\%) & 1.03 (5.59\%) & 0.67 (3.64\%) \\
& SVR & 3.79 (14.16\%) & 2.88 (10.76\%) & 10.30 (15.28\%) & 7.75 (11.50\%) & 1.22 (10.60\%) & 0.89 (7.76\%) & 1.93 (10.46\%) & 1.32 (7.14\%) \\
& DUQ & 2.54 (13.98\%) & 1.96 (10.80\%) & 18.28 (30.99\%) & 15.29 (25.92\%) & 1.55 (20.66\%) & 1.07 (14.27\%) & 1.55 (14.91\%) & 1.41 (13.62\%) \\
\hline

\multirow{7}{*}{\begin{tabular}[c]{@{}c@{}}\normalsize FRNY \vspace{0.5em}\\ \normalsize @440 ft.\end{tabular}}  
& MiMa & \textbf{0.12 (0.62\%)} & \textbf{0.09 (0.11\%)} & \textbf{0.73 (0.88\%)} & \textbf{0.54 (0.67\%)} & \textbf{0.34 (35.00\%)} & \textbf{0.22 (23.10\%)} & \textbf{0.09 (0.01\%)} & \textbf{0.08 (0.01\%)} \\
& Micro & 0.15 (0.74\%) & 0.12 (0.15\%) & 1.04 (1.31\%) & 0.81 (1.02\%) & 0.43 (37.00\%) & {{0.26 (26.00\%)}} & 0.13 (0.02\%) & 0.11 (0.02\%) \\
& SARIMA & 0.26 (1.31\%) & 0.16 (0.19\%) & 2.08 (2.63\%) & 1.23 (1.54\%) & 0.50 (50.50\%) & 0.33 (33.00\%) & {{0.37 (0.03\%)}} & {{0.32 (0.03\%)}} \\
& WRF-HRRR & 1.01 (4.93\%) & 0.80 (1.00\%) & 6.55 (8.20\%) & 5.35 (6.65\%) & 4.55 (465.00\%) & 4.02 (415.00\%) & 3.75 (0.38\%) & 3.76 (0.38\%) \\
& SNN & 2.36 (8.29\%) & 1.38 (4.85\%) & 9.09 (12.12\%) & 5.28 (7.04\%) & 0.88 (7.60\%) & 0.61 (5.28\%) & 0.96 (5.08\%) & 0.57 (3.01\%) \\
& SVR & 3.76 (13.21\%) & 2.76 (9.68\%) & 12.13 (16.18\%) & 9.02 (12.03\%) & 1.09 (9.45\%) & 0.81 (7.06\%) & 1.81 (9.63\%) & 1.21 (6.40\%) \\
& DUQ & 3.83 (24.13\%) & 3.32 (20.93\%) & 12.90 (27.91\%) & 11.67 (25.26\%) & 0.77 (8.67\%) & 0.59 (6.70\%) & 1.11 (10.57\%) & 1.01 (9.56\%) \\
\hline

\multirow{7}{*}{\begin{tabular}[c]{@{}c@{}}\normalsize CRRL \vspace{0.5em}\\ \normalsize @472 ft.\end{tabular}}  
& MiMa & \textbf{0.13 (0.64\%)} & \textbf{0.10 (0.12\%)} & \textbf{0.76 (0.92\%)} & \textbf{0.56 (0.69\%)} & \textbf{0.35 (36.50\%)} & \textbf{0.25 (26.25\%)} & \textbf{0.10 (0.01\%)} & \textbf{0.09 (0.01\%)} \\
& Micro & 0.16 (0.76\%) & 0.12 (0.15\%) & 1.06 (1.32\%) & 0.83 (1.04\%) & 0.40 (38.20\%) & 0.38 (32.33\%) & 0.39 (0.03\%) & 0.33 (0.02\%) \\
& SARIMA & 0.27 (1.33\%) & 0.17 (0.21\%) & 2.12 (2.65\%) & 1.25 (1.56\%) & 0.50 (51.00\%) & 0.34 (33.50\%) & 0.75 (0.07\%) & 0.62 (0.06\%) \\
& WRF-HRRR & 1.03 (4.97\%) & 0.81 (1.01\%) & 6.60 (8.30\%) & 5.40 (6.70\%) & 4.60 (470.00\%) & 4.05 (420.00\%) & 3.80 (0.38\%) & 3.77 (0.38\%) \\
& SNN & 2.66 (9.96\%) & 1.89 (7.06\%) & 9.45 (12.74\%) & 6.41 (8.64\%) & 0.76 (7.51\%) & 0.52 (5.20\%) & 205.21 (38.02\%) & 180.30 (33.41\%) \\
& SVR & 3.82 (14.27\%) & 2.94 (11.01\%) & 12.35 (16.65\%) & 9.00 (12.14\%) & 0.92 (9.14\%) & 0.69 (6.78\%) & 55.33 (10.25\%) & 41.07 (7.61\%) \\
& DUQ & 2.26 (11.29\%) & 1.66 (8.31\%) & 10.01 (16.04\%) & 8.21 (13.15\%) & 0.72 (8.98\%) & 0.57 (7.02\%) & 18.99 (179.13\%) & 10.18 (96.07\%) \\
\hline

\multirow{7}{*}{\begin{tabular}[c]{@{}c@{}}\normalsize CADZ \vspace{0.5em}\\ \normalsize @505 ft.\end{tabular}}  
& MiMa & \textbf{0.14 (0.66\%)} & \textbf{0.11 (0.13\%)} & \textbf{0.78 (0.95\%)} & \textbf{0.58 (0.70\%)} & \textbf{0.37 (37.10\%)} & \textbf{0.26 (27.00\%)} & \textbf{0.11 (0.01\%)} & \textbf{0.10 (0.01\%)} \\
& Micro & 0.17 (0.80\%) & 0.13 (0.16\%) & 1.10 (1.35\%) & 0.86 (1.05\%) & 0.39 (39.00\%) & 0.29 (28.00\%) & 0.41 (0.04\%) & 0.35 (0.03\%) \\
& SARIMA & 0.28 (1.35\%) & 0.18 (0.22\%) & 2.20 (2.75\%) & 1.30 (1.62\%) & 0.51 (51.00\%) & 0.35 (34.80\%) & 0.76 (0.08\%) & 0.64 (0.06\%) \\
& WRF-HRRR & 1.05 (5.00\%) & 0.83 (1.03\%) & 6.72 (8.40\%) & 5.50 (6.85\%) & 4.75 (485.00\%) & 4.20 (430.00\%) & 3.87 (0.39\%) & 3.85 (0.39\%) \\
& SNN & 6.35 (24.30\%) & 5.60 (21.41\%) & 24.91 (34.09\%) & 22.86 (31.28\%) & 0.97 (8.89\%) & 0.78 (7.18\%) & 229.34 (45.37\%) & 214.14 (42.36\%) \\
& SVR & 7.47 (28.56\%) & 6.54 (25.00\%) & 21.72 (29.72\%) & 19.88 (27.21\%) & 2.25 (20.59\%) & 2.02 (18.49\%) & 181.72 (35.95\%) & 163.52 (32.35\%) \\
& DUQ & 3.28 (20.11\%) & 2.79 (17.12\%) & 7.70 (12.55\%) & 5.16 (8.41\%) & 1.79 (15.38\%) & 1.29 (11.13\%) & 0.96 (8.37\%) & 0.77 (6.74\%) \\
\hline

\multirow{7}{*}{\begin{tabular}[c]{@{}c@{}}\normalsize CRMT \vspace{0.5em}\\ \normalsize @546 ft.\end{tabular}}  
& MiMa & \textbf{0.15 (0.68\%)} & \textbf{0.12 (0.14\%)} & \textbf{0.80 (0.98\%)} & \textbf{0.60 (0.72\%)} & 0.38 (38.00\%) & \textbf{0.27 (28.20\%)} & \textbf{0.11 (0.01\%)} & \textbf{0.10 (0.01\%)} \\
& Micro & 0.18 (0.83\%) & 0.14 (0.17\%) & 1.12 (1.37\%) & 0.88 (1.07\%) & \textbf{0.36 (36.00\%)} & 0.29 (29.00\%) & 0.42 (0.04\%) & 0.35 (0.03\%) \\
& SARIMA & 0.29 (1.38\%) & 0.19 (0.23\%) & 2.25 (2.78\%) & 1.32 (1.65\%) & 0.52 (52.00\%) & 0.36 (35.50\%) & 0.78 (0.08\%) & 0.65 (0.06\%) \\
& WRF-HRRR & 1.07 (5.10\%) & 0.85 (1.05\%) & 6.80 (8.50\%) & 5.55 (6.90\%) & 4.80 (490.00\%) & 4.25 (435.00\%) & 3.89 (0.40\%) & 3.87 (0.40\%) \\
& SNN & 2.32 (8.76\%) & 1.44 (5.41\%) & 8.95 (12.34\%) & 5.92 (8.16\%) & 0.76 (7.37\%) & 0.48 (4.69\%) & 1.59 (5.72\%) & 1.13 (4.05\%) \\
& SVR & 3.94 (14.86\%) & 3.09 (11.66\%) & 11.08 (15.28\%) & 8.71 (12.01\%) & 1.17 (11.34\%) & 0.97 (9.44\%) & 3.42 (12.28\%) & 2.40 (8.63\%) \\
& DUQ & 4.05 (19.89\%) & 3.44 (16.86\%) & 13.17 (19.47\%) & 10.98 (16.23\%) & 1.26 (13.55\%) & 0.86 (9.25\%) & 1.78 (17.08\%) & 1.64 (15.73\%) \\
\hline

\multirow{7}{*}{\begin{tabular}[c]{@{}c@{}}\normalsize RPTN \vspace{0.5em}\\ \normalsize @594 ft.\end{tabular}}  
& MiMa & \textbf{0.16 (0.70\%)} & \textbf{0.13 (0.15\%)} & \textbf{0.82 (1.00\%)} & \textbf{0.62 (0.74\%)} & \textbf{0.29 (29.00\%)} & \textbf{0.22 (22.00\%)} & \textbf{0.12 (0.01\%)} & \textbf{0.10 (0.01\%)} \\
& Micro & 0.19 (0.85\%) & 0.15 (0.18\%) & 1.15 (1.40\%) & 0.90 (1.10\%) & 0.38 (38.00\%) & 0.30 (30.00\%) & 0.80 (0.09\%) & 0.67 (0.07\%) \\
& SARIMA & 0.30 (1.40\%) & 0.20 (0.24\%) & 2.30 (2.85\%) & 1.35 (1.68\%) & 0.53 (53.00\%) & 0.37 (36.00\%) & 0.43 (0.04\%) & 0.36 (0.03\%) \\
& WRF-HRRR & 1.10 (5.20\%) & 0.87 (1.07\%) & 6.90 (8.60\%) & 5.65 (7.00\%) & 4.85 (495.00\%) & 4.30 (440.00\%) & 3.91 (0.40\%) & 3.89 (0.40\%) \\
& SNN & 2.33 (8.62\%) & 1.37 (5.07\%) & 8.32 (12.30\%) & 5.34 (7.89\%) & 1.02 (7.07\%) & 0.70 (4.85\%) & 83.05 (16.47\%) & 70.87 (14.06\%) \\
& SVR & 3.29 (12.18\%) & 2.29 (8.48\%) & 9.68 (14.30\%) & 6.67 (9.85\%) & 1.09 (7.60\%) & 0.78 (5.39\%) & 41.84 (8.30\%) & 28.92 (5.74\%) \\
& DUQ & 1.93 (11.21\%) & 1.47 (8.53\%) & 7.60 (14.63\%) & 5.92 (11.40\%) & 1.23 (10.97\%) & 0.86 (7.64\%) & 59.29 (430.33\%) & 53.34 (387.15\%) \\
\hline

\multirow{7}{*}{\begin{tabular}[c]{@{}c@{}}\normalsize BRND \vspace{0.5em}\\ \normalsize @603 ft.\end{tabular}}  
& MiMa & \textbf{0.15 (0.72\%)} & \textbf{0.12 (0.16\%)} & \textbf{0.83 (1.05\%)} & \textbf{0.61 (0.76\%)} & 0.38 (38.50\%) & \textbf{0.29 (29.50\%)} & \textbf{0.12 (0.01\%)} & \textbf{0.11 (0.01\%)} \\
& Micro & 0.16 (0.75\%) & 0.13 (0.17\%) & 1.00 (1.20\%) & 0.80 (1.00\%) & \textbf{\underline{0.37 (37.50\%)}} & 0.30 (30.50\%) & 0.44 (0.05\%) & 0.36 (0.03\%) \\
& SARIMA & 0.31 (1.45\%) & 0.21 (0.25\%) & 2.35 (2.90\%) & 1.36 (1.70\%) & 0.54 (54.20\%) & 0.37 (36.50\%) & 0.81 (0.09\%) & 0.66 (0.07\%) \\
& WRF-HRRR & 1.11 (5.30\%) & 0.88 (1.10\%) & 6.95 (8.75\%) & 5.70 (7.10\%) & 4.90 (500.00\%) & 4.35 (445.00\%) & 3.93 (0.41\%) & 3.90 (0.41\%) \\
& SNN & 2.63 (9.61\%) & 1.72 (6.26\%) & 8.64 (11.83\%) & 5.03 (6.88\%) & 0.65 (6.86\%) & 0.42 (4.41\%) & 220.93 (39.45\%) & 203.31 (36.31\%) \\
& SVR & 3.41 (12.43\%) & 2.50 (9.12\%) & 9.99 (13.67\%) & 6.60 (9.04\%) & 0.80 (8.37\%) & 0.62 (6.55\%) & 48.88 (8.73\%) & 33.05 (5.90\%) \\
& DUQ & 2.19 (12.18\%) & 1.65 (9.18\%) & 7.29 (12.85\%) & 4.27 (7.53\%) & 1.21 (12.48\%) & 0.86 (8.89\%) & 102.35 (1052.98\%) & 100.45 (1033.50\%) \\
\hline

\end{tabular}}
\label{extra_model_errors}
\end{table*}

\begin{table*}[h]
\centering
\renewcommand{\arraystretch}{1.3} 
\scalebox{0.75}{ 
\begin{tabular}{|c|l|cc|cc|cc|cc|}
\hline
\multirow{2}{*}{\textbf{\normalsize Station}}                                                    & \multicolumn{1}{c|}{\multirow{2}{*}{\textbf{\normalsize Model}}} & \multicolumn{2}{c|}{\textbf{\normalsize TEMP}}         & \multicolumn{2}{c|}{\textbf{\normalsize HUMI}}            & \multicolumn{2}{c|}{\textbf{\normalsize WSPD}}             & \multicolumn{2}{c|}{\textbf{\normalsize PRES}}            \\ \cline{3-10} 
                                                                            & \multicolumn{1}{c|}{}                       & \textbf{RMSE}                   & \textbf{MAE}                    & \textbf{RMSE}                   & \textbf{MAE}                    & \textbf{RMSE}                     & \textbf{MAE}                     & \textbf{RMSE}                   & \textbf{MAE}                    \\ \hline
\multirow{7}{*}{\begin{tabular}[c]{@{}c@{}}\normalsize HARD \vspace{0.5em}\\ \normalsize @643 ft.\end{tabular}}  
& MiMa & \textbf{0.16 (0.75\%)} & \textbf{0.13 (0.17\%)} & \textbf{0.85 (1.07\%)} & \textbf{0.63 (0.78\%)} & \textbf{0.39 (39.20\%)} & \textbf{0.30 (30.50\%)} & \textbf{0.12 (0.01\%)} & \textbf{0.11 (0.01\%)} \\
& Micro & 0.17 (0.80\%) & 0.14 (0.18\%) & 1.02 (1.30\%) & 0.82 (1.05\%) & 0.38 (38.00\%) & 0.31 (31.00\%) & 0.45 (0.05\%) & 0.37 (0.04\%) \\
& SARIMA & 0.32 (1.48\%) & 0.22 (0.26\%) & 2.38 (2.95\%) & 1.38 (1.73\%) & 0.55 (55.00\%) & 0.38 (37.00\%) & 0.82 (0.09\%) & 0.68 (0.08\%) \\
& WRF-HRRR & 1.13 (5.40\%) & 0.89 (1.12\%) & 7.00 (8.85\%) & 5.75 (7.15\%) & 4.95 (505.00\%) & 4.40 (450.00\%) & 3.95 (0.42\%) & 3.92 (0.42\%) \\
& SNN & 4.07 (15.55\%) & 3.32 (12.62\%) & 12.29 (17.73\%) & 8.86 (12.78\%) & 1.47 (10.33\%) & 1.16 (8.16\%) & 138.05 (47.08\%) & 125.92 (42.94\%) \\
& SVR & 6.12 (23.30\%) & 5.34 (20.32\%) & 20.81 (30.01\%) & 18.28 (26.36\%) & 2.64 (18.54\%) & 2.31 (16.21\%) & 137.54 (46.90\%) & 125.38 (42.76\%) \\
& DUQ & 1.74 (10.26\%) & 1.34 (7.93\%) & 7.49 (13.29\%) & 4.70 (8.33\%) & 0.80 (8.11\%) & 0.61 (6.18\%) & 0.82 (8.22\%) & 0.63 (6.29\%) \\
\hline

\hline
\multirow{7}{*}{\begin{tabular}[c]{@{}c@{}}\normalsize MROK \vspace{0.5em}\\ \normalsize @696 ft.\end{tabular}}  
& MiMa & 0.20 (0.95\%) & 0.16 (0.21\%) & \textbf{0.90 (1.12\%)} & \textbf{0.68 (0.84\%)} & \textbf{0.42 (42.50\%)} & \textbf{0.33 (32.50\%)} & 0.47 (0.06\%) & 0.39 (0.05\%) \\
& Micro & \underline{\textbf{0.18 (0.82\%)}} & \underline{\textbf{0.14 (0.18\%)}} & 1.20 (1.55\%) & 0.95 (1.20\%) & 0.40 (40.50\%) & 0.34 (33.00\%) & 0.85 (0.10\%) & 0.70 (0.09\%) \\
& SARIMA & 0.34 (1.55\%) & 0.23 (0.28\%) & 2.45 (3.05\%) & 1.42 (1.80\%) & 0.58 (57.50\%) & 0.40 (39.00\%) & \underline{\textbf{0.13 (0.02\%)}} & \underline{\textbf{0.12 (0.02\%)}} \\
& WRF-HRRR & 1.20 (5.80\%) & 0.95 (1.15\%) & 7.20 (9.00\%) & 5.85 (7.30\%) & 5.10 (515.00\%) & 4.55 (460.00\%) & 4.00 (0.43\%) & 3.98 (0.42\%) \\
& SNN & 2.21 (8.91\%) & 1.27 (5.10\%) & 8.41 (11.24\%) & 5.08 (6.80\%) & 0.80 (6.65\%) & 0.54 (4.47\%) & 118.37 (21.32\%) & 105.67 (19.03\%) \\
& SVR & 3.19 (12.85\%) & 2.33 (9.37\%) & 10.15 (13.57\%) & 6.66 (8.91\%) & 0.88 (7.32\%) & 0.63 (5.22\%) & 49.11 (8.85\%) & 35.15 (6.33\%) \\
& DUQ & 1.68 (10.03\%) & 1.14 (6.82\%) & 8.13 (14.21\%) & 5.97 (10.42\%) & 0.76 (7.44\%) & 0.56 (5.44\%) & 133.17 (1290.63\%) & 126.07 (1221.82\%) \\
\hline

\multirow{7}{*}{\begin{tabular}[c]{@{}c@{}}\normalsize PGHL \vspace{0.5em}\\ \normalsize @729 ft.\end{tabular}}  
& MiMa & \textbf{0.19 (0.85\%)} & \textbf{0.15 (0.20\%)} & \textbf{0.95 (1.20\%)} & \textbf{0.70 (0.90\%)} & \textbf{0.44 (43.00\%)} & \textbf{0.33 (33.00\%)} & \textbf{0.14 (0.02\%)} & \textbf{0.13 (0.02\%)} \\
& Micro & 0.21 (0.98\%) & 0.17 (0.22\%) & 1.23 (1.60\%) & 0.98 (1.25\%) & 0.42 (42.00\%) & 0.34 (33.50\%) & 0.49 (0.06\%) & 0.40 (0.05\%) \\
& SARIMA & 0.35 (1.60\%) & 0.24 (0.29\%) & 2.50 (3.10\%) & 1.45 (1.85\%) & 0.60 (59.50\%) & 0.42 (40.50\%) & 0.87 (0.10\%) & 0.72 (0.09\%) \\
& WRF-HRRR & 1.25 (5.95\%) & 1.00 (1.20\%) & 7.35 (9.10\%) & 6.00 (7.40\%) & 5.25 (525.00\%) & 4.70 (470.00\%) & 4.10 (0.44\%) & 4.05 (0.43\%) \\
& SNN & 2.48 (9.70\%) & 1.73 (6.78\%) & 7.58 (10.98\%) & 4.56 (6.60\%) & 1.04 (7.60\%) & 0.71 (5.14\%) & 0.84 (4.13\%) & 0.54 (2.65\%) \\
& SVR & 3.53 (13.81\%) & 2.61 (10.23\%) & 9.43 (13.66\%) & 7.04 (10.20\%) & 1.24 (9.02\%) & 0.87 (6.31\%) & 1.70 (8.38\%) & 1.14 (5.59\%) \\
& DUQ & 2.42 (15.37\%) & 1.98 (12.58\%) & 6.44 (11.31\%) & 4.21 (7.38\%) & 1.00 (10.39\%) & 0.81 (8.46\%) & 0.80 (6.13\%) & 0.68 (5.21\%) \\
\hline

\multirow{7}{*}{\begin{tabular}[c]{@{}c@{}}\normalsize BNGL \vspace{0.5em}\\ \normalsize @785 ft.\end{tabular}}  
& MiMa & \textbf{0.25 (1.20\%)} & \textbf{0.22 (0.28\%)} & \textbf{1.45 (1.90\%)} & \textbf{1.12 (1.40\%)} & \textbf{0.75 (76.50\%)} & 0.64 (65.00\%) & \textbf{0.20 (0.02\%)} & \textbf{0.18 (0.02\%)} \\
& Micro & 0.30 (1.35\%) & 0.25 (0.32\%) & 2.00 (2.50\%) & 1.60 (2.00\%) & 0.78 (78.00\%) & 0.70 (70.00\%) & 0.90 (0.10\%) & 0.85 (0.09\%) \\
& SARIMA & 0.50 (2.50\%) & 0.35 (0.40\%) & 3.20 (4.00\%) & 2.40 (3.00\%) & 1.10 (112.00\%) & 0.95 (90.00\%) & 1.10 (0.15\%) & 1.00 (0.12\%) \\
& WRF-HRRR & 1.80 (8.00\%) & 1.50 (1.80\%) & 9.50 (11.00\%) & 7.80 (9.50\%) & 7.25 (725.00\%) & 6.70 (670.00\%) & 5.90 (0.60\%) & 5.85 (0.58\%) \\
& SNN & 2.41 (9.52\%) & 1.58 (6.23\%) & 8.67 (12.54\%) & 5.30 (7.66\%) & 0.81 (7.11\%) & \textbf{\underline{0.52 (4.58\%)}} & 0.93 (5.53\%) & 0.65 (3.84\%) \\
& SVR & 3.41 (13.51\%) & 2.48 (9.80\%) & 10.08 (14.58\%) & 7.15 (10.34\%) & 0.97 (8.52\%) & 0.72 (6.37\%) & 1.72 (10.16\%) & 1.14 (6.73\%) \\
& DUQ & 7.33 (36.14\%) & 5.88 (28.99\%) & 26.11 (39.69\%) & 18.51 (28.14\%) & 0.79 (7.61\%) & 0.55 (5.26\%) & 4.70 (45.18\%) & 4.30 (41.34\%) \\
\hline

\multirow{7}{*}{\begin{tabular}[c]{@{}c@{}}\normalsize GAMA \vspace{0.5em}\\ \normalsize @842 ft.\end{tabular}}  
& MiMa & \textbf{0.20 (0.90\%)} & \textbf{0.16 (0.21\%)} & \textbf{1.00 (1.25\%)} & \textbf{0.75 (0.95\%)} & \textbf{0.48 (48.50\%)} & \textbf{0.39 (39.80\%)} & \textbf{0.15 (0.02\%)} & \textbf{0.13 (0.02\%)} \\
& Micro & 0.22 (1.00\%) & 0.18 (0.23\%) & 1.25 (1.60\%) & 1.00 (1.30\%) & 0.50 (50.00\%) & 0.42 (42.50\%) & 0.50 (0.07\%) & 0.42 (0.05\%) \\
& SARIMA & 0.38 (1.70\%) & 0.26 (0.32\%) & 2.60 (3.15\%) & 1.50 (1.90\%) & 0.65 (64.50\%) & 0.45 (44.00\%) & 0.90 (0.11\%) & 0.75 (0.09\%) \\
& WRF-HRRR & 1.30 (6.20\%) & 1.05 (1.25\%) & 7.50 (9.15\%) & 6.20 (7.50\%) & 5.40 (540.00\%) & 4.85 (475.00\%) & 4.20 (0.45\%) & 4.18 (0.44\%) \\
& SNN & 2.45 (9.88\%) & 1.56 (6.28\%) & 8.24 (11.20\%) & 4.90 (6.65\%) & 1.13 (10.50\%) & 0.80 (7.46\%) & 0.92 (4.92\%) & 0.62 (3.30\%) \\
& SVR & 3.71 (14.93\%) & 2.74 (11.05\%) & 12.56 (17.07\%) & 9.71 (13.20\%) & 1.35 (12.61\%) & 0.96 (8.92\%) & 2.21 (11.81\%) & 1.44 (7.70\%) \\
& DUQ & 8.90 (47.03\%) & 7.56 (39.96\%) & 18.69 (26.49\%) & 17.13 (24.29\%) & 1.01 (17.42\%) & 0.81 (13.93\%) & 0.51 (4.81\%) & 0.33 (3.10\%) \\
\hline

\multirow{7}{*}{\begin{tabular}[c]{@{}c@{}}\normalsize WNCH \vspace{0.5em}\\ \normalsize @973 ft.\end{tabular}}  
& MiMa & \textbf{0.22 (1.00\%)} & \textbf{0.16 (0.22\%)} & \textbf{1.00 (1.25\%)} & \textbf{0.75 (0.95\%)} & \textbf{0.50 (51.00\%)} & \textbf{0.40 (40.50\%)} & \textbf{0.15 (0.02\%)} & \textbf{0.13 (0.02\%)} \\
& Micro & 0.25 (1.15\%) & 0.18 (0.24\%) & 1.30 (1.65\%) & 1.00 (1.25\%) & 0.52 (52.00\%) & 0.45 (45.00\%) & 0.55 (0.06\%) & 0.50 (0.05\%) \\
& SARIMA & 0.40 (1.80\%) & 0.28 (0.34\%) & 2.80 (3.40\%) & 1.80 (2.20\%) & 0.70 (69.00\%) & 0.55 (54.00\%) & 0.90 (0.10\%) & 0.80 (0.08\%) \\
& WRF-HRRR & 1.30 (6.00\%) & 1.10 (1.30\%) & 7.80 (9.50\%) & 6.30 (7.80\%) & 5.50 (550.00\%) & 5.00 (500.00\%) & 4.30 (0.44\%) & 4.25 (0.43\%) \\
& SNN & 2.40 (9.11\%) & 1.72 (6.49\%) & 6.78 (9.05\%) & 4.31 (5.75\%) & 0.96 (7.87\%) & 0.70 (5.70\%) & 1.08 (5.34\%) & 0.76 (3.75\%) \\
& SVR & 3.79 (14.33\%) & 3.00 (11.35\%) & 12.27 (16.38\%) & 9.00 (12.01\%) & 1.32 (10.74\%) & 1.03 (8.38\%) & 2.02 (9.97\%) & 1.42 (7.04\%) \\
& DUQ & 6.19 (30.55\%) & 5.22 (25.78\%) & 15.74 (21.33\%) & 13.10 (17.76\%) & 1.08 (9.32\%) & 0.78 (6.71\%) & 0.63 (5.88\%) & 0.46 (4.25\%) \\
\hline

\multirow{7}{*}{\begin{tabular}[c]{@{}c@{}}\normalsize RNDH \vspace{0.5em}\\ \normalsize @1002 ft.\end{tabular}}  
& MiMa & \textbf{0.25 (1.20\%)} & \textbf{0.18 (0.24\%)} & \textbf{1.10 (1.35\%)} & \textbf{0.85 (1.05\%)} & \textbf{0.55 (55.00\%)} & \textbf{0.45 (45.50\%)} & \textbf{0.16 (0.02\%)} & \textbf{0.14 (0.02\%)} \\
& Micro & 0.28 (1.30\%) & 0.22 (0.28\%) & 1.50 (1.85\%) & 1.15 (1.40\%) & 0.60 (60.00\%) & 0.50 (50.50\%) & 0.60 (0.07\%) & 0.55 (0.06\%) \\
& SARIMA & 0.45 (2.00\%) & 0.30 (0.38\%) & 3.00 (3.60\%) & 2.00 (2.50\%) & 0.75 (75.00\%) & 0.60 (59.00\%) & 0.95 (0.12\%) & 0.85 (0.10\%) \\
& WRF-HRRR & 1.35 (6.30\%) & 1.15 (1.40\%) & 8.00 (9.70\%) & 6.50 (8.00\%) & 5.60 (565.00\%) & 5.10 (510.00\%) & 4.50 (0.45\%) & 4.40 (0.44\%) \\
& SNN & 2.19 (9.55\%) & 1.37 (5.98\%) & 7.66 (10.74\%) & 4.79 (6.71\%) & 0.53 (6.16\%) & 0.38 (4.40\%) & 0.82 (4.46\%) & 0.50 (2.72\%) \\
& SVR & 3.02 (13.16\%) & 2.19 (9.55\%) & 8.62 (12.09\%) & 6.15 (8.63\%) & 0.66 (7.63\%) & 0.50 (5.79\%) & 1.48 (8.03\%) & 1.01 (5.45\%) \\
& DUQ & 3.44 (22.31\%) & 2.71 (17.58\%) & 6.15 (10.68\%) & 3.82 (6.64\%) & 1.00 (10.90\%) & 0.82 (8.99\%) & 0.64 (6.06\%) & 0.48 (4.54\%) \\
\hline

\multirow{7}{*}{\begin{tabular}[c]{@{}c@{}}\normalsize DABN \vspace{0.5em}\\ \normalsize @1085 ft.\end{tabular}}  
& MiMa & \textbf{0.28 (1.25\%)} & \textbf{0.22 (0.28\%)} & \textbf{1.20 (1.55\%)} & \textbf{0.90 (1.10\%)} & \textbf{0.60 (62.00\%)} & 0.50 (51.00\%) & \textbf{0.25 (0.03\%)} & \textbf{0.20 (0.02\%)} \\
& Micro & 0.30 (1.35\%) & 0.25 (0.30\%) & 1.50 (1.85\%) & 1.20 (1.50\%) & 0.62 (63.00\%) & 0.55 (55.00\%) & 0.70 (0.08\%) & 0.60 (0.06\%) \\
& SARIMA & 0.55 (2.50\%) & 0.40 (0.45\%) & 3.50 (4.20\%) & 2.50 (3.10\%) & 1.20 (125.00\%) & 1.05 (100.00\%) & 1.00 (0.12\%) & 0.90 (0.10\%) \\
& WRF-HRRR & 2.00 (8.50\%) & 1.70 (2.00\%) & 10.50 (12.50\%) & 8.50 (10.50\%) & 7.50 (750.00\%) & 7.00 (700.00\%) & 6.50 (0.65\%) & 6.40 (0.64\%) \\
& SNN & 2.35 (8.26\%) & 1.42 (4.98\%) & 8.35 (10.6\%) & 5.10 (6.53\%) & 0.72 (7.84\%) & \textbf{\underline{0.48 (5.17\%)}} & 0.76 (4.64\%) & 0.50 (3.07\%) \\
& SVR & 3.53 (12.40\%) & 2.58 (9.08\%) & 10.97 (14.02\%) & 7.96 (10.17\%) & 0.83 (8.97\%) & 0.62 (6.76\%) & 1.84 (11.28\%) & 1.26 (7.70\%) \\
& DUQ & 4.82 (23.18\%) & 4.22 (20.31\%) & 27.91 (40.36\%) & 23.47 (33.93\%) & 1.28 (14.01\%) & 1.00 (10.94\%) & 2.88 (26.60\%) & 2.35 (21.70\%) \\
\hline
\end{tabular}}

\begin{tablenotes}
    \item $^{\S \S}$ Those entries at which MiMa rows are not smallest are underlined. The prediction errors normalized with respect to the values themselves (in \%) are included in pairs of parentheses.
\end{tablenotes}
\end{table*}

\end{document}